\documentclass[fleqn,usenatbib]{mnras}

\usepackage{newtxtext,newtxmath}

\usepackage[T1]{fontenc}

\DeclareRobustCommand{\VAN}[3]{#2}
\let\VANthebibliography\thebibliography
\def\thebibliography{\DeclareRobustCommand{\VAN}[3]{##3}\VANthebibliography}

\usepackage{graphicx}	
\usepackage{amsmath}	
\usepackage{subcaption}

\newcommand{\mps}{\; {\rm m\;s^{-1}}}

\newcommand{\kps}{\; {\rm Ks^{-1}}}
\newcommand{\mmps}{\; {\rm m^2\;s^{-1}}}

\newcommand{\btt}{ }
\newcommand{\bttt}{ }

\def \Kzz {K_{\mathrm{zz}}}

\def \tdrag {\tau_\mathrm{drag}}
\def \teff {T_{\rm eff}}
\def \kps {{\rm Ks^{-1}}}

\title[Global circulation and tracer mixing on brown dwarfs]{Jet streams and tracer mixing in the atmospheres of brown dwarfs and isolated young giant planets}

\author[X. Tan]{
Xianyu Tan$^{1,2}$\thanks{E-mail: \url{xianyu.tan@physics.ox.ac.uk}}
\\
$^{1}$Atmospheric Oceanic  and Planetary Physics, Department of Physics, University of Oxford, OX1 3PU, UK\\
$^{2}$Lunar and Planetary Laboratory, University of Arizona, 1629 University Boulevard, Tucson, AZ 85721, USA\\
}

\date{Accepted XXX. Received YYY; in original form ZZZ}

\pubyear{2022}

\begin{document}
\label{firstpage}
\pagerange{\pageref{firstpage}--\pageref{lastpage}}
\maketitle

\begin{abstract}
Observations of brown dwarfs  and relatively isolated young extrasolar giant planets  have provided  unprecedented details to probe atmospheric dynamics in a new  regime. Questions about mechanisms governing the global circulation and its fundamental nature remain to be completely addressed. Previous studies have shown that small-scale, randomly varying thermal perturbations  resulting from interactions between convection and the overlying stratified layers can drive zonal  jet streams, waves and turbulence.  In this work, we improve upon our previous work 
by using a  general circulation model coupled with a {\bttt two-stream} grey radiative transfer scheme to represent more realistic heating and cooling rates. We examine the formation of zonal jets and their time evolution, and  vertical mixing of passive tracers including clouds and chemical species.  Under relatively weak radiative and frictional dissipation, robust zonal jets with speeds up to a few hundred $\mps$ are typical outcomes. The off-equatorial jets tend to be pressure-independent while the equatorial jets exhibit significant vertical wind shear. On the other hand, models with strong dissipation inhibit the jet formation and leave isotropic turbulence in off-equatorial regions. Quasi-periodic oscillations of the equatorial flow with periods ranging from tens of days to months are prevalent at relatively low atmospheric temperatures. Sub-micron cloud particles can be easily transported to several scale heights above the condensation level, while larger particles form thinner layers. Cloud decks are significantly inhomogeneous near their cloud tops. Chemical tracers with chemical timescales $>10^5$ s can be driven out of equilibrium. The equivalent vertical diffusion coefficients, $\Kzz$, for the global-mean tracer transport are diagnosed  from our models and are typically on the order of $1\sim10^2\mmps$. Finally, we derive an analytic estimation of $\Kzz$ for different types of tracers under relevant conditions.
\end{abstract}

\begin{keywords}
hydrodynamics --- methods: numerical --- planets and satellites: atmospheres --- planets and satellites: gaseous planets --- brown dwarfs
\end{keywords}



\section{Introduction}
\label{ch.intro}
Our understanding of  the atmospheres of brown dwarfs (BDs) has grown sophisticated owing to the increasing range of observational constraints. Several types of observations have suggested that their atmospheres are dynamic, inhomogeneous and in disequilibrium (see a summary in \citealp{tan2021bd2}).     In particular, several recent surveys that aim at monitoring the near-infrared (IR) brightness of BDs offer perhaps the strongest support for the presence of atmospheric circulation (e.g., \citealp{vos2022,vos2020,manjavacas2019, vos2018, apai2017, yang2016, rajan2015, metchev2015, wilson2014, radigan2014,buenzli2014, heinze2013, khandrika2013}).  They revealed that low-amplitude ($ \sim 1\%$) flux variability is likely common among L and T dwarfs, while a small fraction of BDs exhibit large-amplitude flux variations (e.g., \citealp{artigau2009, radigan2012, yang2016,apai2017,lew2020,zhou2020}).  These motivate  a greater  need to better understand their global atmospheric circulation and the consequences for variability, cloud formation, and mixing of chemical species \citep{showman2020,zhang2020}. Extrasolar giant planets (EGPs) discovered by direct imaging so far are mostly young, far away from their host stars or isolated. They share similarities with BDs in terms of their spectral type, near-IR colors, inference of  clouds and chemical disequilibrium (e.g., \citealp{chauvin2017, derosa2016, skemer2016, macintosh2015, bonnefoy2014, marley2012, barman2011b, currie2011}). Near-IR brightness variability has also  been observed on these young isolated EGPs \citep{biller2015, zhou2016,biller2018,miles2019,manjavacas2019b, lew2020,zhou2020}.   

Advancements of observations for BDs and relatively isolated young EGPs provide ever unprecedented details to probe atmospheric dynamics in a new dynamical regime. This regime is typically characterized by fast rotation,  vigorous internal thermal forcing and negligible external forcing. However, these objects  receive negligible external stellar irradiation, and therefore lack the large-scale stellar heating contracts  that are responsible for driving the global circulation like those on close-in exoplanets and solar system planets \citep{showman2020}. The first-level question needed to be addressed is, what are the possible mechanisms driving the global circulation and what are the resulting properties?

Convection in the interior of BDs is vigorous and this convection is expected to perturb the overlying, stably stratified atmosphere, generating  large-scale atmospheric circulation that consist of turbulence, waves, and zonal (east–west) jet streams \citep{showman2019}.  There have been several studies along this direction. Hydrodynamic simulations in a local enclosed area by \cite{freytag2010} and \cite{allard2012} showed that gravity waves generated by interactions between the convective interior and the stratified layer can cause mixing that leads to small-scale cloud patchiness.    \cite{showman&kaspi2013} presented global convection models, and they analytically estimated the typical wind speeds and horizontal temperature differences driven by the absorption and breaking of atmospheric waves in the stably stratified atmosphere.  By injecting random forcing to a shallow-water system that represents the effect of convection perturbing the stratified atmosphere,  \cite{zhang&showman2014} showed that weak radiative dissipation and strong forcing  favor large-scale zonal jets for BDs, whereas strong dissipation and weak forcing favor transient eddies and quasi-isotropic turbulence.  Using a general circulation model coupled with parameterized thermal perturbations resulting from interactions between convective interior and the stratified atmosphere,  \cite{showman2019} showed that under conditions of relatively strong forcing and weak damping, robust zonal jets and mean meridional circulation  are  common outcomes of the dynamics. They also  revealed long-term (multiple months to years) quasi-periodic oscillations on the equatorial zonal jets, and the mechanism driving the oscillations is similar to that of the Quasi-biennial oscillation (QBO) in Earth's stratosphere. 

An alternative class of mechanisms invoking radiative feedback by clouds has been shown to be essential to induce instantaneous atmospheric variability and vigorous global circulation when the atmospheres are dusty \citep{tan2019,tan2021bd1,tan2021bd2}. These models generate large-scale cloud and temperature patchiness, vortices, waves and jet streams, as well as systematic equator-to-pole variation of cloud thickness.  Latent heat released from condensation of silicate and iron vapor also helps to  organize zonal jets and patchy storms \citep{tan2017}. Similar radiative processes associated with clouds or chemistry may also induce small-scale instability \citep{tremblin2019,tremblin2021}.

There are two main goals in this study. First, we further investigate the  general circulation of BDs and isolated young EGPs driven by  thermal perturbations using an updated general circulation model. This study is along the line of \cite{zhang&showman2014} and \cite{showman2019},  but we utilize a more realistic model to examine the findings revealed by previous studies. Our models in \cite{showman2019} implemented an idealized  relaxation scheme with a uniform radiative timescale throughout the atmosphere. {\btt While that scheme provides a good control of models over the parameter space which is essential to demonstrate dynamical mechanisms, it oversimplifies the radiative heating and cooling rates,  and the relevance of these previous results to real objects needs to be validated against a more realistic model setup.   Here, we adopt  an idealized radiative transfer scheme that calculates the three-dimensional heating and cooling rates in a more realistic and self-consistent way  which is especially important in the current context that we investigate models with different effective temperatures.  This enables us to examine the characteristic behaviors of jet formation and their long-term evolution under more relevant conditions of BDs and isolated young EGPs.}
The second goal  is to characterize the vertical mixing of passive tracers, including cloud particles and chemical species, by large-scale circulation in the stratified layers where convection is not directly responsible for mixing. This has not been investigated in the  context of BDs and isolated young EGPs. Our study will demonstrate a dynamical mechanism for tracer mixing in the stratified layers, which is relevant in understanding the dusty L dwarfs and the prevalent non-equilibrium chemistry in atmospheres of BDs and isolated EGPs.
 

This paper is organized as follows.  We introduce our general circulation model in Section \ref{ch3.model}; we then present results of zonal jet formation in Section  \ref{ch3.results} and explore vertical mixing of passive tracers  in Section \ref{ch.tracer}; finally we discuss and summarize our results in Section \ref{ch3.discussion}.

\section{Model}

\label{ch3.model}
\subsection{General Circulation Model}
We model a global three-dimensional (3D) thin atmosphere  using a general circulation model (GCM), the MITgcm (\citealp{adcroft2004}, see also \url{mitgcm.org}), which solves the hydrostatic primitive equations of dynamical meteorology in pressure coordinates.  Our model is similar to that of \cite{showman2019} but we  couple a grey radiative transfer (RT) scheme to the dynamical core instead of the idealized Newtonian cooling scheme. In brief, thermal radiative fluxes within the atmosphere are obtained by solving the plane-parallel, two-stream approximation of the RT equations using an efficient and reliable numerical tool \emph{twostr} \citep{kst1995}. Then the radiative heating rates are calculated by taking the vertical divergence of fluxes to drive dynamics.  The detailed implementation and demonstrations of our RT scheme is described in  \cite{komacek2017} and \cite{tan2019uhj} on the application of hot Jupiters and in \cite{tan2021bd1,tan2021bd2} on BDs.  The lower boundary condition is a prescribed, globally uniform temperature at the bottom  pressure which is intended to mimic a uniform-entropy  deep layer resulting from efficient convective mixing. We include gas opacity only and omit scattering. The Rosseland-mean gas opacity as a function of temperature and pressure from \citet{freedman2014} is  coupled to the RT calculation.  
Our radiative-convective equilibrium RT calculations generate  temperature-pressure (T-P) profiles with reasonable radiative-convective boundaries (RCBs) compared to the more sophisticated one-dimensional (1D) atmospheric models (e.g., \citealp{allard2001,tsuji2002,burrows2006,marley2021}). 

{\btt The deep layers in our simulation domain reach the convectively unstable regions. Small-scale convective transport of entropy and tracers that cannot be resolved in the model is parameterized using a simple column convective adjustment scheme that instantaneously adjusts the unstable region to be convectively neutral and homogenize tracers within the adjusted region. If any adjacent two  layers within a vertical atmospheric column are unstable, they are instantaneously adjusted to a convectively neutral state while conserving total enthalpy. The whole column is repeatedly scanned until convective instability is eliminated  everywhere. Tracers are also well homogenized within the adjusted domain. Similar simple schemes have been adopted in some GCMs for Earth (e.g., \citealp{cam3}), exoplanets (e.g., \citealp{deitrick2020,lee2021}) and our previous brown dwarf models \citep{tan2021bd1,tan2021bd2}. }

A Rayleigh drag is applied to horizontal winds in the deep atmosphere to crudely represent the effects of momentum mixing between the weather layer and the quiescent interior, where flows are likely to experience significant magnetohydrodynamics (MHD) drag. The drag linearly decreases with decreasing pressure and takes the form as 
$\mathcal{F}_{\rm{drag}} = - k_v(p) \mathbf{v}$,
where $\mathbf{v}$ is the horizontal velocity vector and  $k_v(p)$ is a pressure-dependent drag coefficient. $k_v(p)$ decreases from $1/\tau_{\rm{drag}}$ at the  bottom  boundary $p_{\rm{bot}}$ to zero at certain pressure $p_{\rm{drag, top}}$, where $\tau_{\rm{drag}}$ is a characteristic drag timescale. No drag is included at pressures lower than $p_{\rm{drag, top}}$.
In all simulations, we fix $p_{\rm{drag, top}}$ to 50 bars. {\btt The bottom pressure is 100 bars.} Kinetic energy dissipated by the Rayleigh drag is converted to thermal energy. As in \cite{showman2019}, the  drag timescale $\tau_{\rm{drag}}$ in this study is treated as a free parameter to explore possible circulation patterns.   As shown below, a weak drag  is essential for  jet formation.

\begin{figure*}      
\centering
\includegraphics[width=1.7\columnwidth]{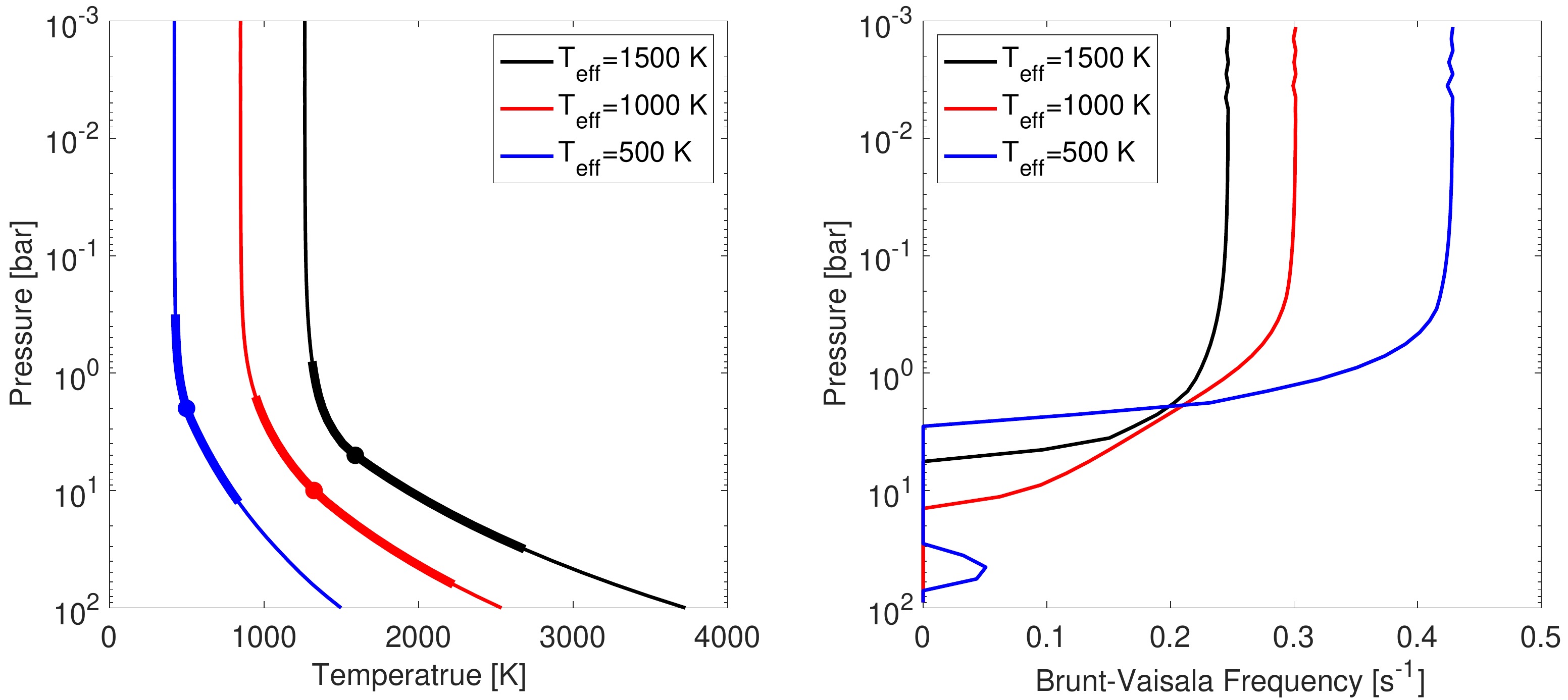}
\caption{\emph{Left:} radiative-convective equilibrium T-P profiles for three cases with effective temperatures of about 1500, 1000, 500 Kelvin, respectively. Dots are  the radiative-convective boundaries (RCBs)  and the pressure ranges within thicker lines are where the thermal perturbations apply with the  vertical forcing profile $f_v(p)$  described by Eq. (\ref{forceamp}). {\btt \emph{Right:} profiles of the Brunt-Vaisala frequency, defined as $N = \sqrt{\frac{g^2}{RT}\left(\frac{R}{c_p}-\frac{d\ln T}{d \ln p}\right)}$, for the T-P profiles shown in the left panel. }
}
\label{fig.tp}
\end{figure*}

We parameterize the global thermal perturbations resulting from interactions between convection and the overlying stratified layers  using the scheme in \cite{showman2019}. {\btt The essence of this scheme is the stochastic and isotropic nature of the forcing, such that the resulting jets and their time evolution are purely results of the internal dynamics rather than being related to the nature of the forcing. Similar types of schemes have long been used to understand turbulence and jet formation on solar-system giant planets (see reviews of \citealp{vasavada2005,galperin2008,showman2018review} and many references wherein). The work in \cite{showman2019} and here is an extension into the brown dwarf regime.  } 

The procedure is summarized as follows. The temperature field is consecutively forced by a global 3D forcing pattern $F(\lambda, \phi, p, t)$ as a function of longitude $\lambda$, latitude $\phi$, pressure $p$ and time $t$. At each time step, a forcing function is constructed by superposing a sequence of fully-normalized spherical harmonics with total wavenumbers $n=n_f-\delta n$ to $n=n_f$ and the zonal wavenumbers from 1 to $n$.  The phases of each zonal mode are randomized, and their amplitudes are statistically equal, such that their superposition  $S(\lambda,\phi)$ is spherically isotropic with a characteristic wavelength determined by $n_f$ and $\delta n$. Then the forcing function is normalized to a unit of $\rm{K~s^{-1}}$ by multiplying a forcing amplitude $s_f$. The horizontal forcing pattern $F_h(\lambda, \phi, t)$ is marched forward with time using a Markov process:
\begin{equation}
F_h(\lambda, \phi, t+\delta t) = (1-\alpha) F_h(\lambda, \phi, t) + \sqrt{2\alpha-\alpha^2} S(\lambda, \phi) s_f,
\end{equation}
where $\alpha=\delta t/\tau_s$ is a de-correlation factor,  $\delta t$ is  a dynamical time step and $\tau_s$ is a  storm timescale which characterizes the evolution timescale of the perturbation pattern. A similar horizontal forcing  was also used in  shallow-water models of \cite{zhang&showman2014}. 
In the vertical direction, the thermal perturbations are applied in regions within  two  scale heights away from the RCB of the equilibrium profile. The forcing amplitude decreases as the pressure deviates from the RCB assuming a normalized vertical forcing profile $f_v(p)$: 
\begin{equation}
 f_v(p) = \left\{ 
 \begin{array}{lr}
 (p_{\rm{rcb}}/p)^2, & \quad \ln{p_{\rm{rcb}}}+2>\ln{p} \geq \ln{p_{\rm{rcb}}} \\
 (p/p_{\rm{rcb}})^2, & \quad \ln{p_{\rm{rcb}}}-2<\ln{p}<\ln{p_{\rm{rcb}}} \\
 0,	&  \rm{else}
 \end{array}
 \right.
 \label{forceamp}
 \end{equation} 
 where $p_{\rm{rcb}}$ is the pressure at the RCB. 
 Finally, the 3D forcing pattern $F(\lambda, \phi, p, t)$ is simply $f_v(p) F_h(\lambda, \phi, t)$. 


{\btt The stochastic perturbation scheme is needed to drive the circulation given the rest of the model setup. The convective adjustment scheme is essentially a damping scheme that avoids dry superadiabatic profiles in our hydrostatic GCM. With the isotropic bottom temperature and the lack of external irradiation on brown dwarfs, the temperature-pressure profiles are relaxed towards globally isotropic, thereby damping out isobaric temperature variations and dynamics. Indeed,  experiments in which the stochastic forcing is turned off show essentially no dynamics but only radiative-convective equilibrium profiles.
 }

In this study we mainly explore atmospheric circulation of three effective temperatures, $T_{\rm eff}=1500$, 1000 and 500 K, whose equilibrium profiles are shown in the left panel of Figure \ref{fig.tp}. These represent objects with temperature of typical L, T and early Y type spectrum \citep{kirkpatrick2005,cushing2011}. The vertical forcing regions  are indicated by the thick lines in Figure \ref{fig.tp}. {\btt The right panel shows the corresponding profiles of the Brunt-Vaisala frequency defined as $N = \sqrt{\frac{g^2}{RT}\left(\frac{R}{c_p}-\frac{d\ln T}{d \ln p}\right)}$ where $g$ is gravity, $R$ is the specific gas constant and $c_p$ is the specific heat at constant pressure.  $N$ is nearly zero in the deep convectively neutral regions, then becomes positive above the RCB with a  transition to $\frac{g}{\sqrt{c_p T}}$ in the  nearly isothermal layers. Note that for the case with $T_{\rm eff}= 500$ K, a detached stratified zone appears at around 50 bars. This is not uncommon in models for slightly cooler objects (e.g., \citealp{marley2021}). } We  consider storm timescale $\tau_s = 10^4, 10^5$ and $10^6$ s, and drag timescale $\tau_{\rm{drag}} = 10^5, 10^6$ and $10^7$ s.   We assume a surface gravity of 1000 $\rm{m~s^{-2}}$, a radius of $6 \times 10^7$ meters and a rotation period of 5 hours in  simulations presented in Section \ref{ch3.results}.  
We adopt $n_f=30$ and $\delta n=1$ for most simulations in order to have sizes of perturbations much smaller than the radius of the objects in an affordable model resolution.

We solve the equations  using the cubed-sphere grids \citep{adcroft2004}. A fourth-order Shapiro filter is applied in the momentum and thermodynamics equations to maintain numerical stability.  The nominal horizontal resolution of our simulations is C96, which is  equivalent to $384\times192$ in longitude and latitude. We have tested models with higher horizontal resolution, and the large-scale circulation features are insensitive to the resolution. The simulated pressure domain is between  $10^{-3}$ to 100 bars, and is evenly discretized into 80 layers in log-pressure space.    The ideal gas law of a hydrogen-helium mixture is applied to the equation of state, with the specific heat $c_p = 1.3\times 10^4 ~\rm{Jkg^{-1}K^{-1}}$ and specific gas constant $R =  3714 ~\rm{Jkg^{-1}K^{-1}}$. {\btt With somewhat strong forcing  in many of our models, those with a strong drag (a drag timescale of $10^5$ s) equilibrate rather quickly after about 300 simulation days. Models with longer drag timescales spin up slower; for example, the ones with a drag timescale of $10^7$ s require more than 1500 simulation days to equilibrate. After the models reached statistical equilibrium, we ran an additional 500 to 1000 simulations days for time averaging. Because our models are somewhat strongly forced-dissipated, their spinup time is considerably shorter than models in the regime of solar-system giant planets \citep{showman2019}. 
}

\subsection{Passive Tracers}
\label{numeric.tracers}
To study the effects of stratospheric large-scale circulation on the mixing of cloud particles and chemical species, we implement two types of tracers $q_{\rm cond}$ and $q_{\rm chem}$ in our models. {\btt The tracers are assumed to be passively advected by the flow and do not have feedbacks on the atmospheric structure and dynamics. In reality, cloud radiative effect and latent heating would excert feedbacks to shaping the temperature structure and dynamics \citep{tan2017,tan2019,tang2021,tan2021bd1,tan2021bd2,tremblin2021}. The sinks and sources of  tracers in our models are  idealized (as will be described below) despite that processes governing clouds and chemistry in BDs and EGPs are highly complex (e.g., \citealp{helling2008,visscher2010, helling2014,marley2015, gao2021}). In this work,  we step aside  these complexities just to focus on the passive tracer transport problem. This is a necessary step because only after we understand the passive tracer problem we would appreciate tracer transport in models with more self-consistent coupling.
These simplifications follow the merits of conceptual modeling for 3D tracer  mixing in exoplanet atmospheres \citep{parmentier2013,zhang2018a,zhang2018b,komacek2019vertical,steinrueck2021}. Calculation of tracers are performed mainly in models with $T_{\rm eff}=1000$ K.}

The first type of  tracer $q_{\rm cond}$ represents the mixing ratio of a condensable  species relative to its abundance in the deep atmosphere. A critical condensation pressure $p_{\rm cond}$ is prescribed, and at pressures higher than $p_{\rm cond}$, $q_{\rm cond}$ is in the form of vapor and subjected to replenishment by efficient convective mixing from the interior. At pressures lower than $p_{\rm cond}$, $q_{\rm cond}$ is in the form of solid particles and is subjected to gravitational settling.  This is equivalent to assuming that the scale height of the saturation vapor pressure  is much smaller than a  pressure scale height. The governing equation of $q_{\rm cond}$ is:
\begin{equation} 
\frac{dq_{\rm cond}}{dt} =   \left\{
\begin{array}{lr}
-(q_{\rm cond}-q_{\rm{deep}})/\tau_{\rm{rep}} & \quad p \geq p_{\rm cond} \\ 
\\
- \partial ( q_{\rm cond} V_s)/\partial p & \quad p < p_{\rm cond}
\end{array}
\right.
\label{tracer1}
\end{equation}
where $d/dt=\partial/\partial t + \mathbf{v}\cdot\nabla_p + \omega\partial/\partial p$ is the material derivative, $\nabla_p$ is the isobaric gradient, $\omega$ is the vertical velocity in pressure coordinates, $q_{\rm{deep}}$ is the relative mixing ratio in the deep layers and is assumed to be 1, $\tau_{\rm{rep}}$ is the replenishment timescale which is set to be $10^3$ s, and $V_s$ is  the particle settling speed in pressure coordinates as a function of particle radius, pressure and temperature (with positive pointing downward to larger pressure). We adopt the formula summarized in \cite{ackerman2001}, \cite{spiegel2009} and \cite{parmentier2013} to calculate the particle settling velocity in height coordinates, and then $V_s$ is approximated using the hydrostatic balance.  The density of cloud particles is assumed to be 3190 $\rm{kg m^{-3}}$ as that of $\rm{MgSiO_3}$, which is one of the most abundant condensates in L and T dwarfs \citep{ackerman2001}. In models with $\teff=1000$ K, the condensation pressure $p_{\rm cond}$ is assumed at 10 bars which is near the RCB. This setup ensures that any vertical transport at pressures lower than $p_{\rm cond}$ is due to resolved large-scale dynamics rather than parameterized convective transport. The particle radius is treated as  a free parameter and varies from 0.05 to 5 ${\rm \mu m}$.

The second type of  tracer $q_{\rm chem}$ represents the  mixing ratio of  hypothetical chemical species relative to its deep abundance. A chemical equilibrium profile $q_{\rm{eq}}$ is assumed for the tracer. When the chemical species is driven out of equilibrium due to dynamics, it is relaxed back to $q_{\rm{eq}}$ over a characteristic  chemical  timescale $\tau_{\rm{chem}}$. The governing equation for $q_{\rm chem}$ is simply:
\begin{equation}
\frac{dq_{\rm chem}}{dt} = - \frac{q_{\rm chem} - q_{\rm{eq}}}{\tau_{\rm{chem}}}.
\label{eq.tracer2}
\end{equation}
$q_{\rm{eq}}$ is assumed to vary with pressure alone which is reasonable in conditions relevant to BDs and isolated EGPs. We define a piecewise-continuous  function for such a transitional behavior: $q_{\rm{eq}}$ is assumed to be $q_{\rm{eq, top}}$ at pressures lower than $p_{\rm{top}}$ and   $q_{\rm{eq, bot}}$ at pressures higher than $p_{\rm{bot}}$; then   $\log q_{\rm{eq}}$ varies linearly with $\log p$ in between $p_{\rm{top}}$ and $p_{\rm{bot}}$. In this work, we fix the following parameters as $q_{\rm{eq, bot}}=1$, $q_{\rm{eq, top}} = 10^{-6}$,  $p_{\rm{bot}}=10$ bars and $p_{\rm{top}}=10^{-3}$ bar. Following \cite{zhang2018a}, in a main suite of models, the chemical timescale $\tau_{\rm chem}$ is assumed independent of pressure and vary from $10^4$ to $10^7$ s.  For many species  relevant to BDs and isolated EGPs, the chemical timescale $\tau_{\rm chem}$ is expected to be short in the deep hot atmosphere and long in the cooler upper atmosphere. Therefore, in another suite of models, $\tau_{\rm chem}$ is assumed to be a function like that of $q_{\rm eq}$ with $\tau_{\rm{chem,bot}}=10^3$ s and   $\tau_{\rm{chem,top}}$ varying from $10^4$ to $10^7$ s.

Small-scale gravity waves that cannot be resolved in global models can be potentially important in transport of  tracers and angular momentum  via small-scale turbulence and interactions with mean flows. These effects can only be parameterized, a procedure common in many GCMs, albeit fraught with uncertainties associated with poorly constrained free parameters related to generation and properties of the waves,  assumptions about their dissipation and interactions between the mean flow. In this study we omit  gravity-wave parameterizations for two reasons. The first is to ensure a clean environment to understand the mixing by resolved large-scale advection. Secondly, the sub-grid gravity wave parameterization is yet highly uncertain in conditions appropriate for BDs and EGPs. The mixing of tracers from our current models may thus be considered as a baseline in real conditions.

\section{Results: Global Circulation}
\label{ch3.results}

\subsection{Zonal jet streams}

Kinetic energy injected by thermal perturbations organize itself towards large-scale, while imposed isotropic radiation and bottom frictional drag dissipate the kinetic energy. The organization of the large-scale flow is determined by the relative strengths of the forcing and damping processes.  In this section, we examine the formation of zonal jets spanning a wide range of parameter space and have the following key findings. First, although the forcing and damping are horizontally isotropic, robust zonal  jet streams can emerge from the interaction of the waves and turbulence with the planetary rotation under relatively weak radiative and frictional dissipation.   At a particular atmospheric temperature with a fixed thermal perturbation rate expected for real objects, a weak bottom frictional drag favours jet formation at all latitudes, while models with strong bottom drag exhibit isotropic turbulence at mid-to-high latitudes and jets only near the equator at altitudes above the drag domain. Then, with a fixed thermal perturbation rate and a relatively weak frictional drag, low atmospheric temperature (weak radiative damping) promotes an overall stronger circulation and zonal jets, while high temperature (strong radiative damping) leads to a much weaker circulation. With a weak frictional drag and a low atmospheric temperature, the jet speed increases and the number of jets decreases with increasing thermal perturbation rates. Finally, in models that form robust zonal jets, the off-equatorial jet structures have significant pressure independent components but the equatorial jets usually exhibit  significant vertical jet shears. Because the atmosphere is rotation dominated  at mid-to-high latitudes, the mean meridional temperature variations   associated with the off-equatorial jets are small compared to small-scale variations. Most of our findings are qualitatively consistent with those found in \cite{showman2019}. Below we describe and discuss these results in detail.

\begin{figure*}

    \begin{subfigure}[t]{0.45\textwidth}
    \centering
    \includegraphics[width=0.7\columnwidth]{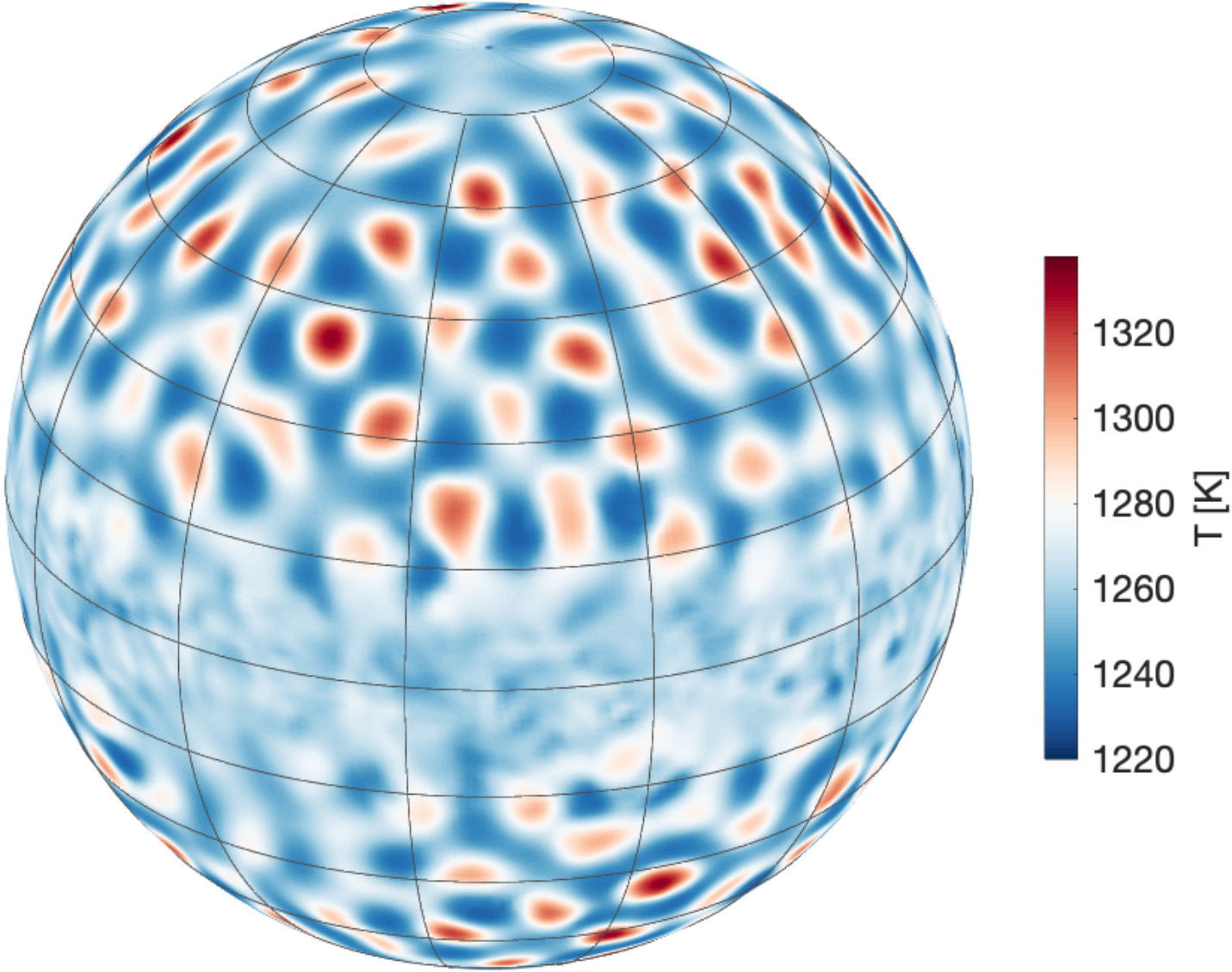}
    \caption{$\tdrag=10^5$ s.}
    \end{subfigure}
    \begin{subfigure}[t]{0.45\textwidth}
    \centering
    \includegraphics[width=0.7\columnwidth]{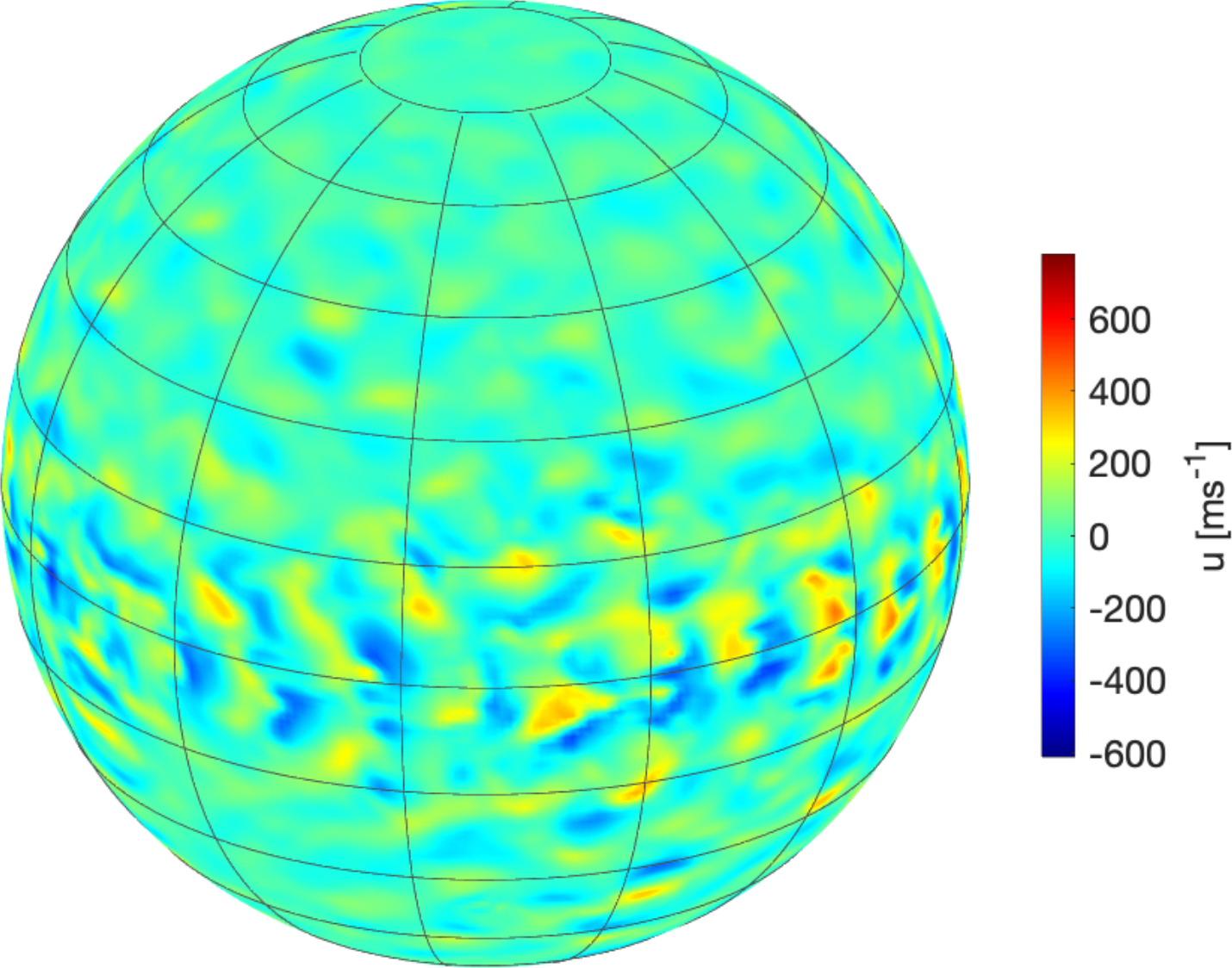}
    \caption{$\tdrag=10^5$ s.}
    \end{subfigure}
 
    \medskip
    
    \begin{subfigure}[t]{0.45\textwidth}
    \centering
    \includegraphics[width=0.7\columnwidth]{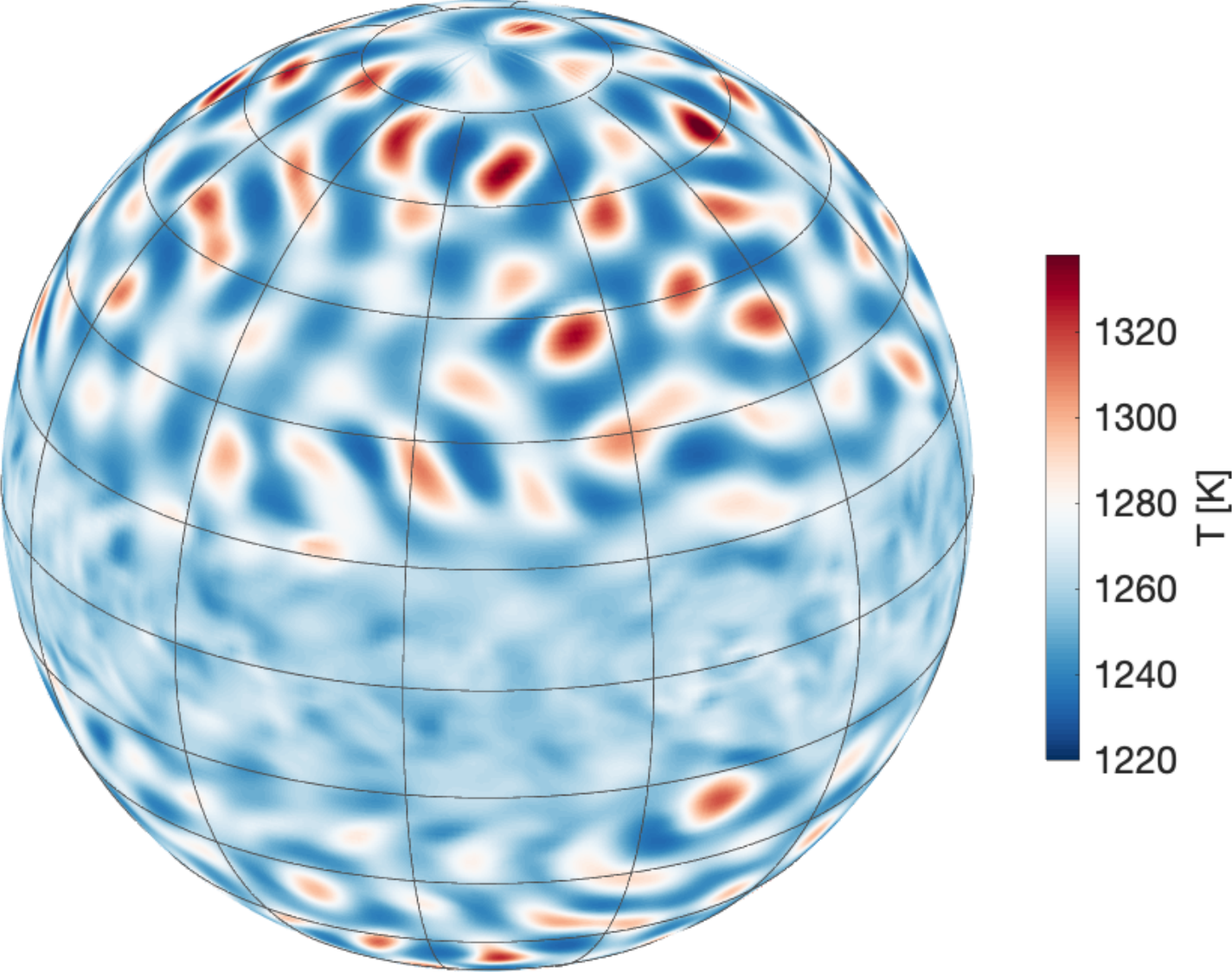}
    \caption{$\tdrag=10^6$ s.}
    \end{subfigure}
    \begin{subfigure}[t]{0.45\textwidth}
    \centering
    \includegraphics[width=0.7\columnwidth]{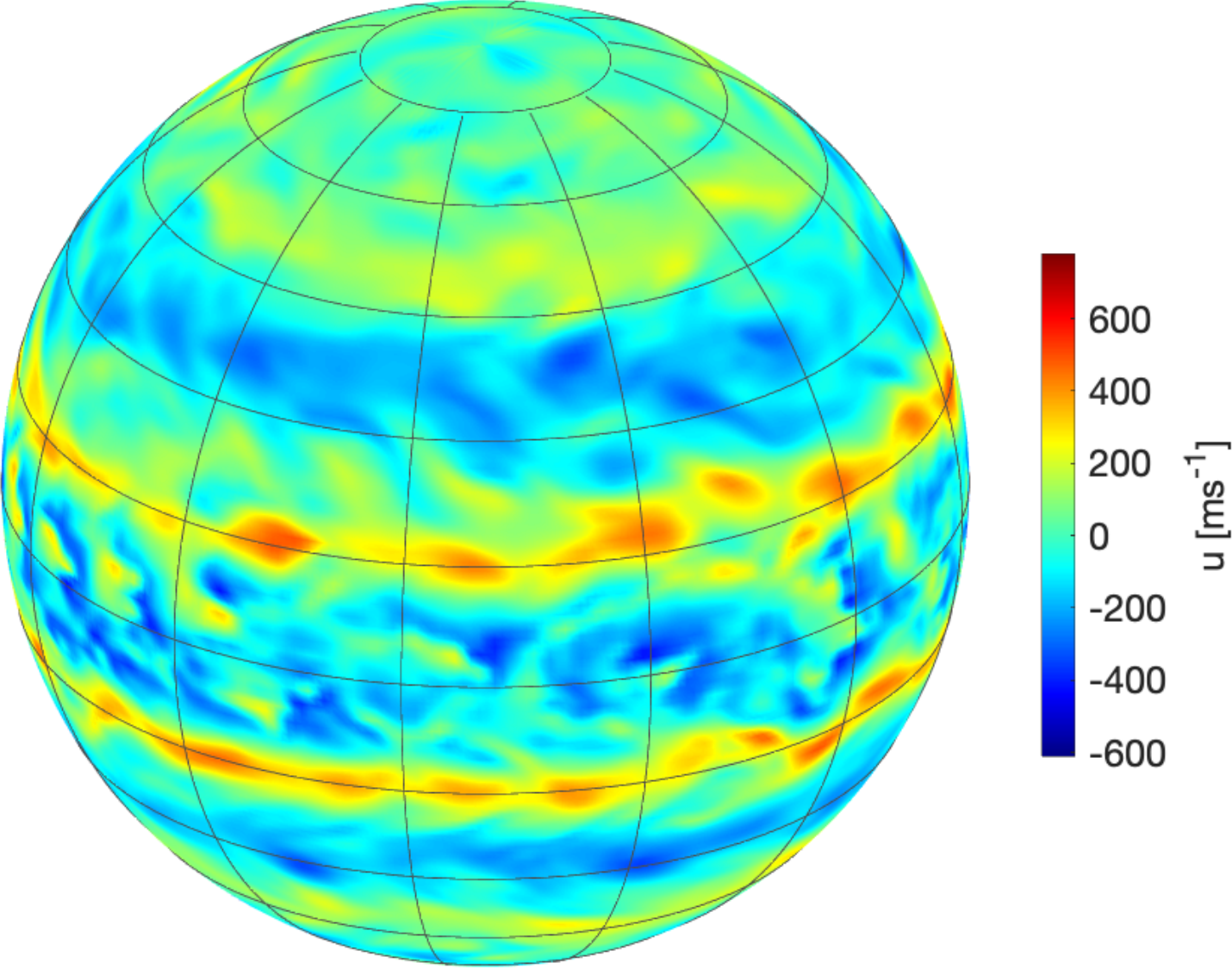}
    \caption{$\tdrag=10^6$ s.}
    \end{subfigure}
    
    \medskip
    
    \begin{subfigure}[t]{0.45\textwidth}
    \centering
    \includegraphics[width=0.7\columnwidth]{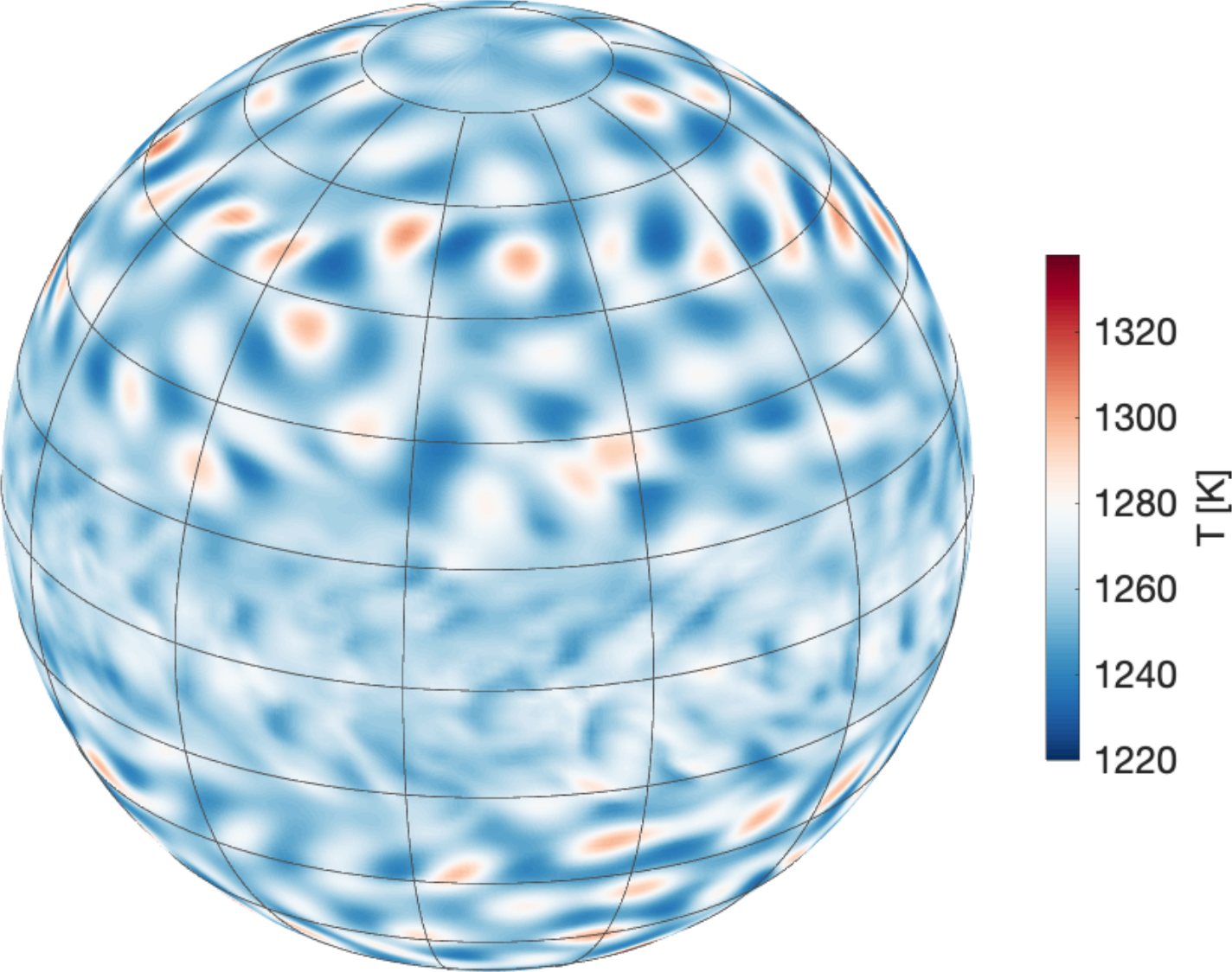}
    \caption{$\tdrag=10^7$ s.}
    \end{subfigure}
    \begin{subfigure}[t]{0.45\textwidth}
    \centering
    \includegraphics[width=0.7\columnwidth]{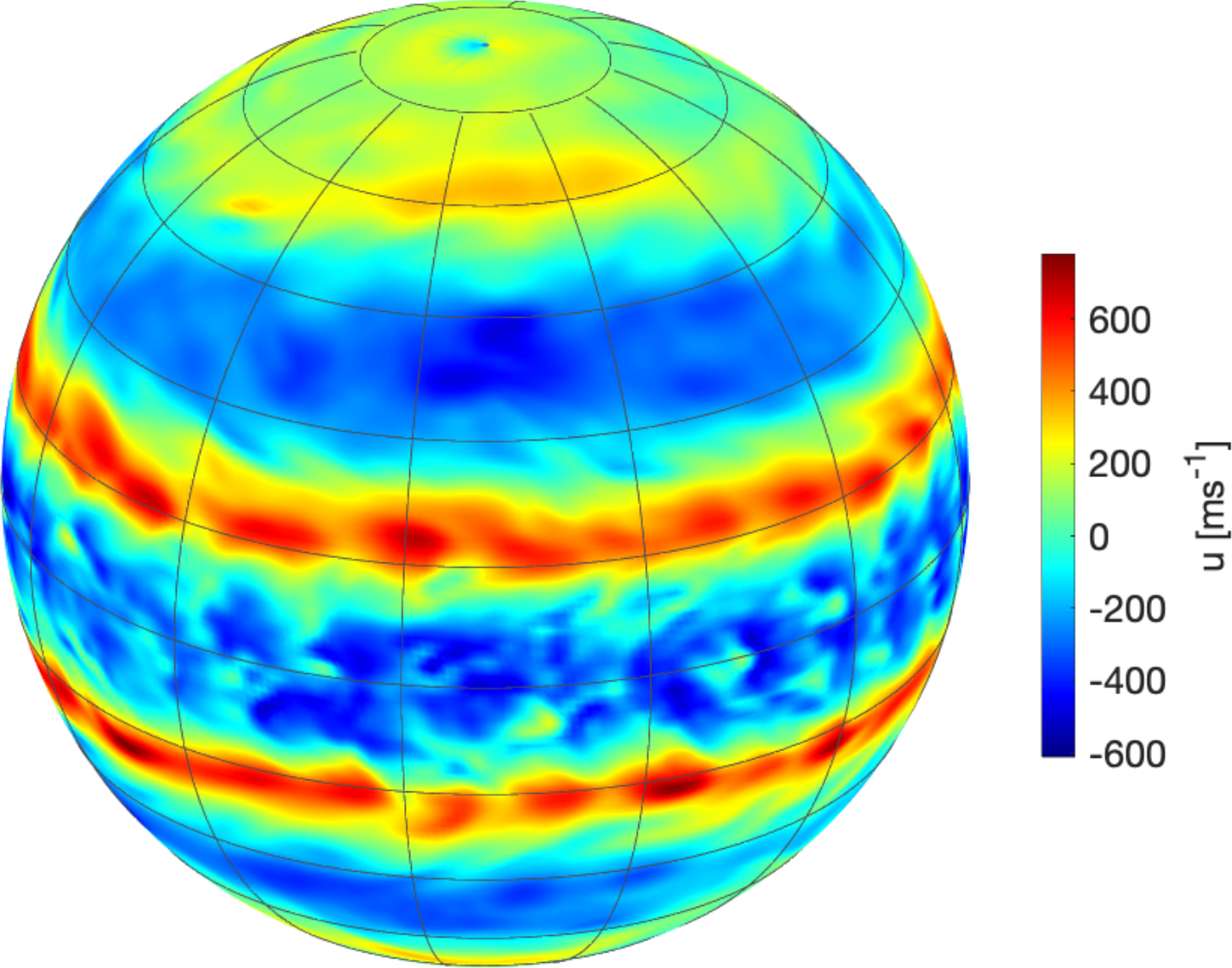}
    \caption{$\tdrag=10^7$ s.}
    \end{subfigure}
    
    \medskip

    \caption{Snapshots of temperature (left) and zonal winds (right) for models with three different bottom drag timescales of $\tdrag=10^5$ s (top), $10^6$ s (middle) and $10^7$ s (bottom). The quantities are shown at 8.2 bars.  These models have an effective temperature $T_{\rm eff}=1000$ K, a perturbation amplitude $s_f=5\times10^{-4}\;{\rm Ks^{-1}}$, a rotation period of 5 hours, surface gravity of $1000\;{\rm ms^{-2}}$, and a storm timescale $\tau_s=10^5$ s. Results are taken after the models reaching statistical equilibrium. {\btt The instantaneous fields with $\tdrag=10^5$ s were obtained at about 800 simulation days; those with $\tdrag=10^6$ s were obtained at about 1000 simulation days; and those with $\tdrag=10^7$ s were obtained at about 2500 simulation days. } }
    \label{fig.tu_drag}
\end{figure*}

\begin{figure}

\begin{subfigure}{0.45\textwidth}
    \centering
    \includegraphics[width=0.95\columnwidth]{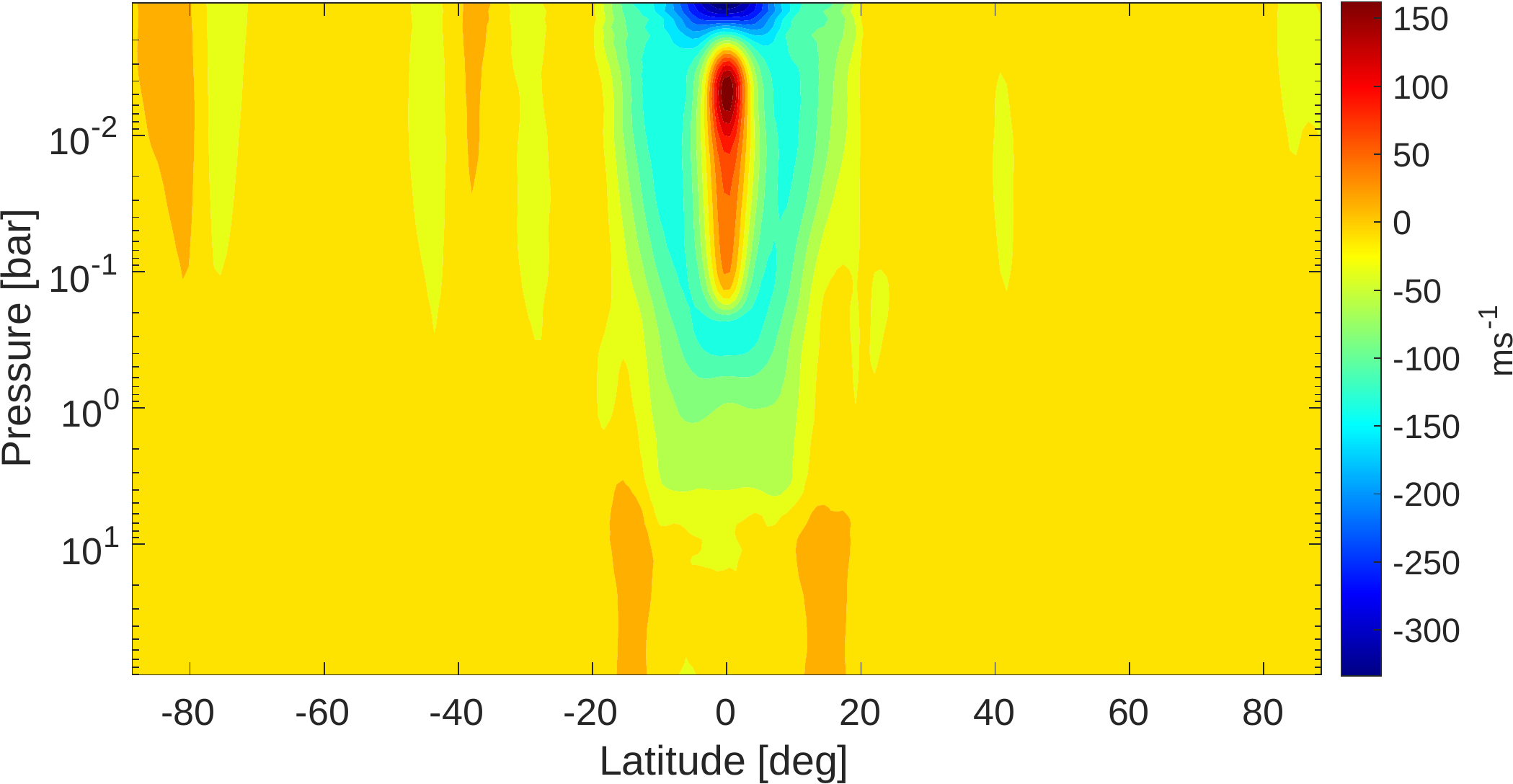}
    \caption{$\tdrag=10^5\;\mps$.}
\end{subfigure}

\medskip
\begin{subfigure}{0.45\textwidth}
    \centering
    \includegraphics[width=0.95\columnwidth]{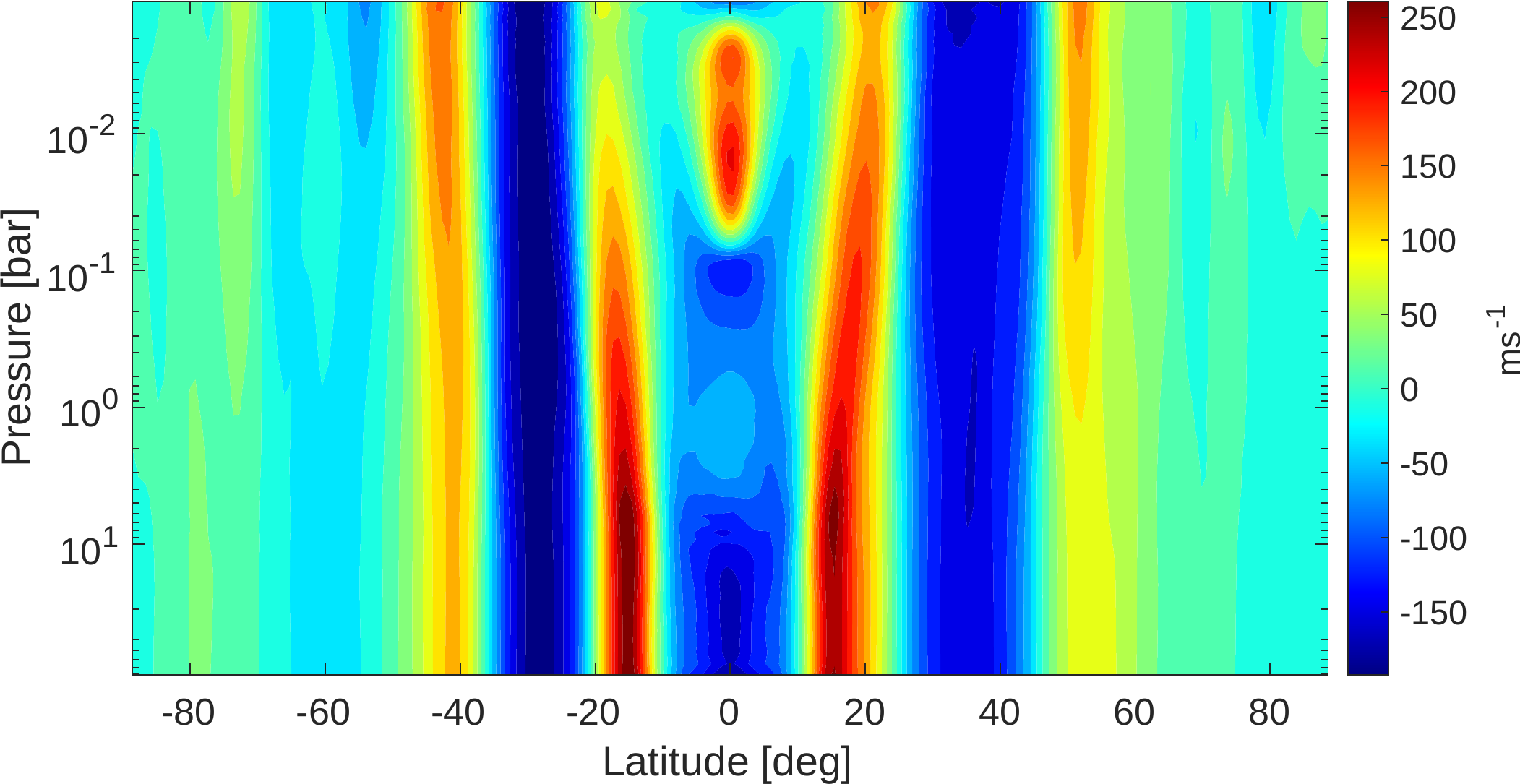}
    \caption{$\tdrag=10^6\;\mps$.}
\end{subfigure}

\medskip
\begin{subfigure}{0.45\textwidth}
    \centering
    \includegraphics[width=0.95\columnwidth]{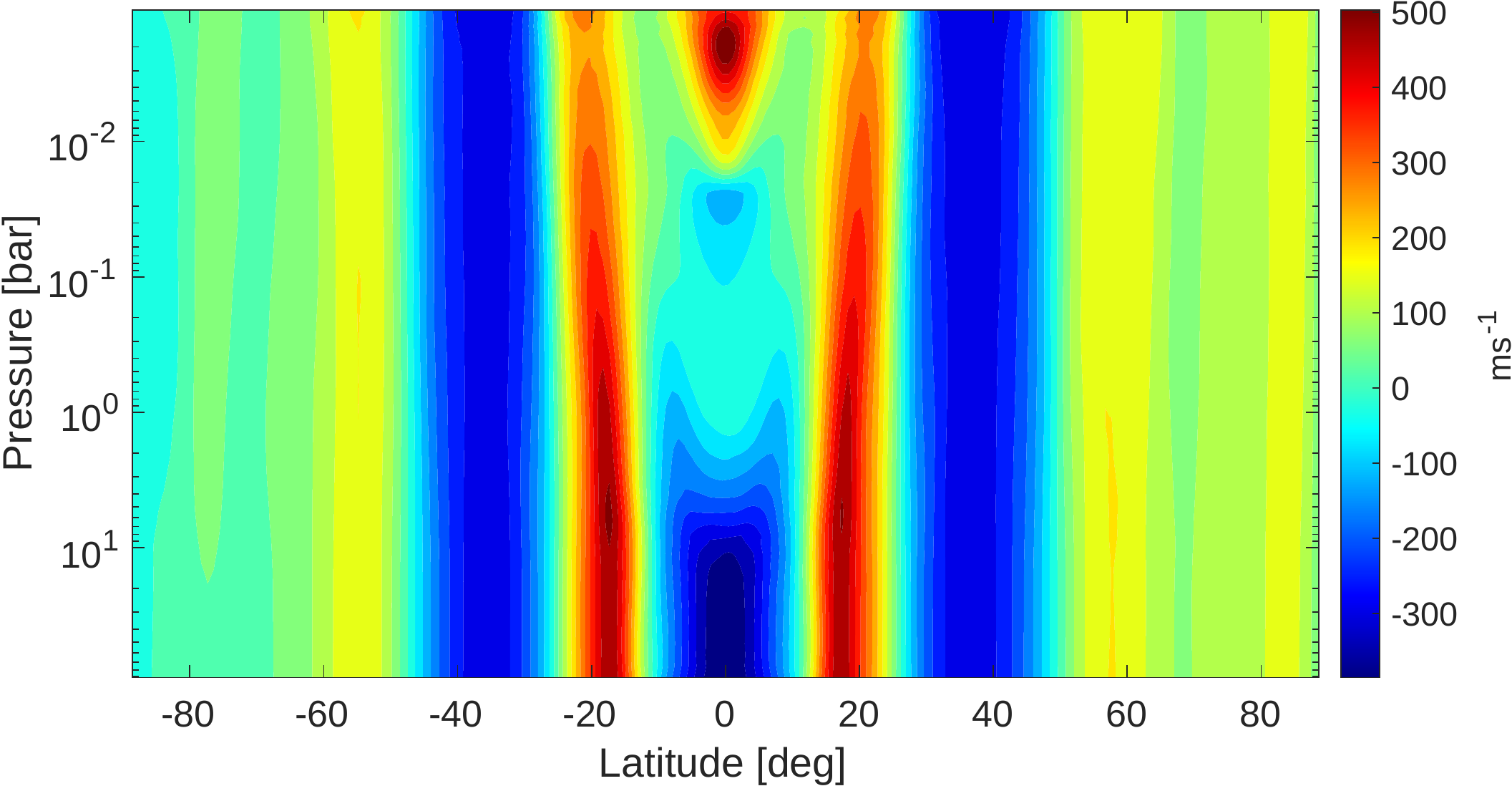}
    \caption{$\tdrag=10^7\;\mps$.}
\end{subfigure}

    \caption{Time-averaged zonal-mean zonal  wind as a function of latitude and pressure for models with $\tdrag=10^5$ s in the top panel, $\tdrag=10^6$ s in the middle panel, and $\tdrag=10^7$ s in the bottom panel. These results are from the same set of models shown in Figure \ref{fig.tu_drag}. {\btt The model with $\tdrag=10^5$ and $\tdrag=10^6$ s reached equilibrium after only about 300 and 600 simulation days, respectively, and we ran additional 500 days for time averaging. The model with $\tdrag=10^7$ s reached equilibrium after about 1600 simulation days and we ran additional 1000 days for time averaging.  } }
    \label{fig.uzonal_t2550}
\end{figure}

\begin{figure}

\begin{subfigure}{0.45\textwidth}
    \centering
    \includegraphics[width=0.7\columnwidth]{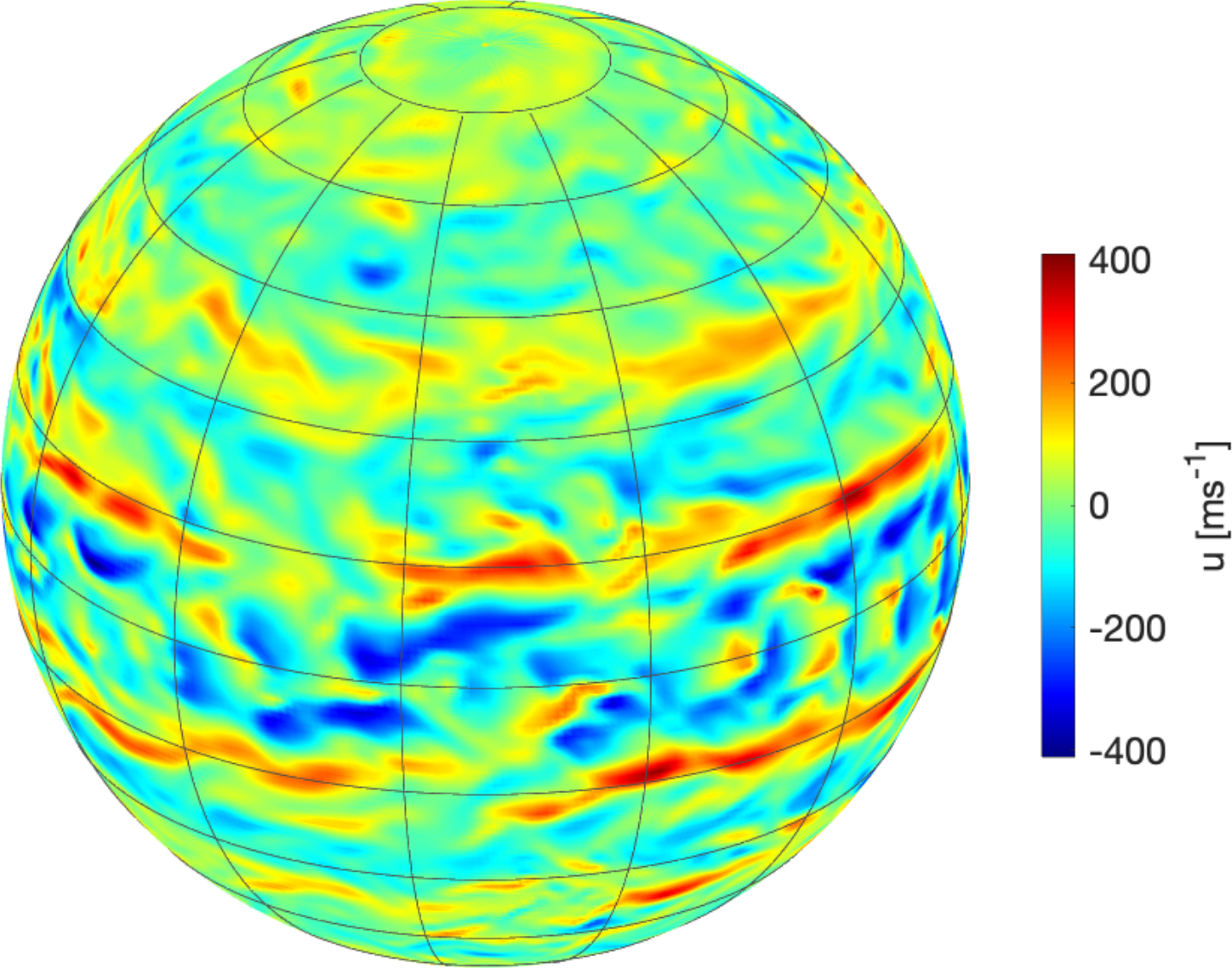}
    \caption{$\teff=500$ K.}
\end{subfigure}

\medskip
\begin{subfigure}{0.45\textwidth}
    \centering
    \includegraphics[width=0.7\columnwidth]{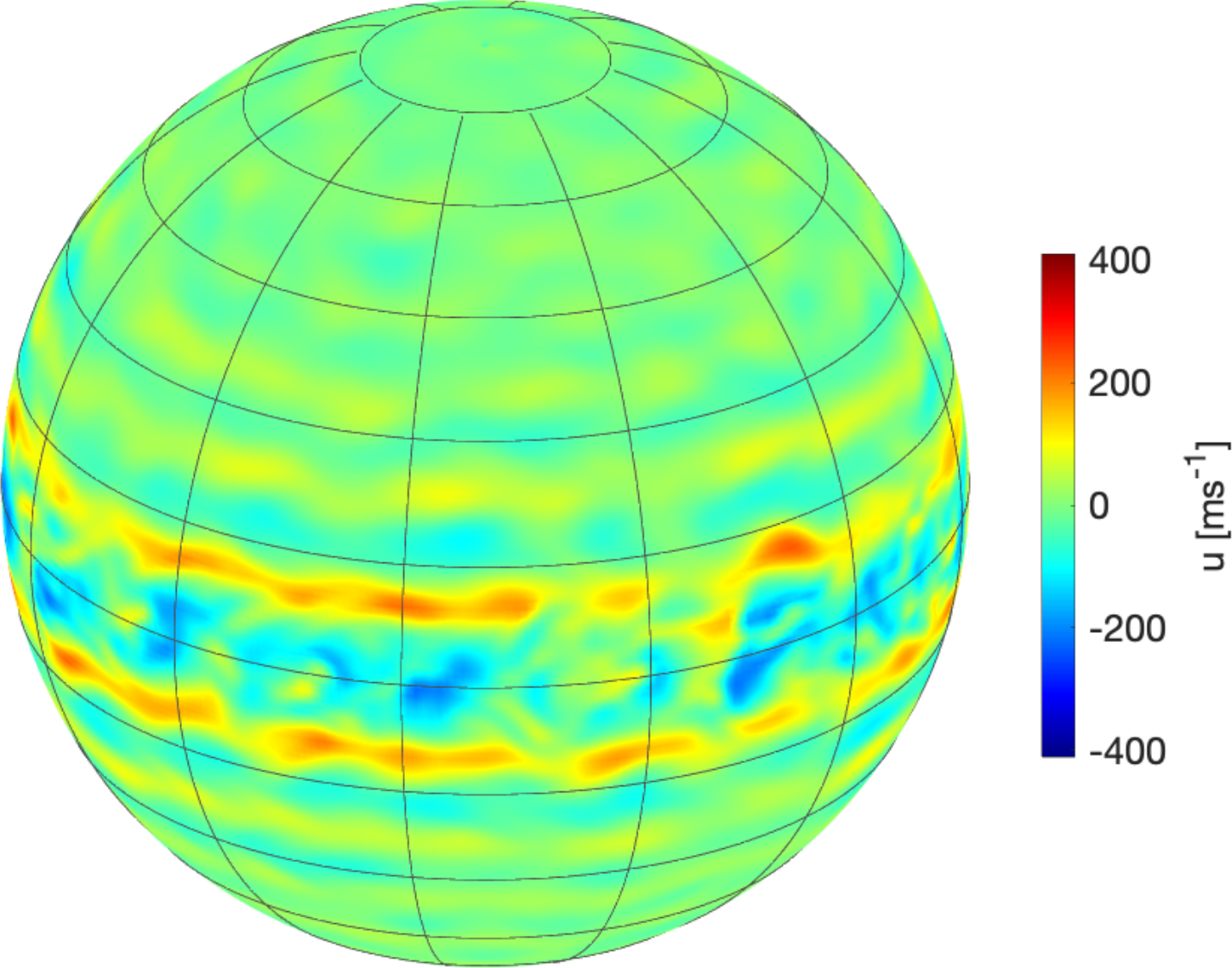}
    \caption{$\teff=1000$ K.}
\end{subfigure}

\medskip
\begin{subfigure}{0.45\textwidth}
    \centering
    \includegraphics[width=0.7\columnwidth]{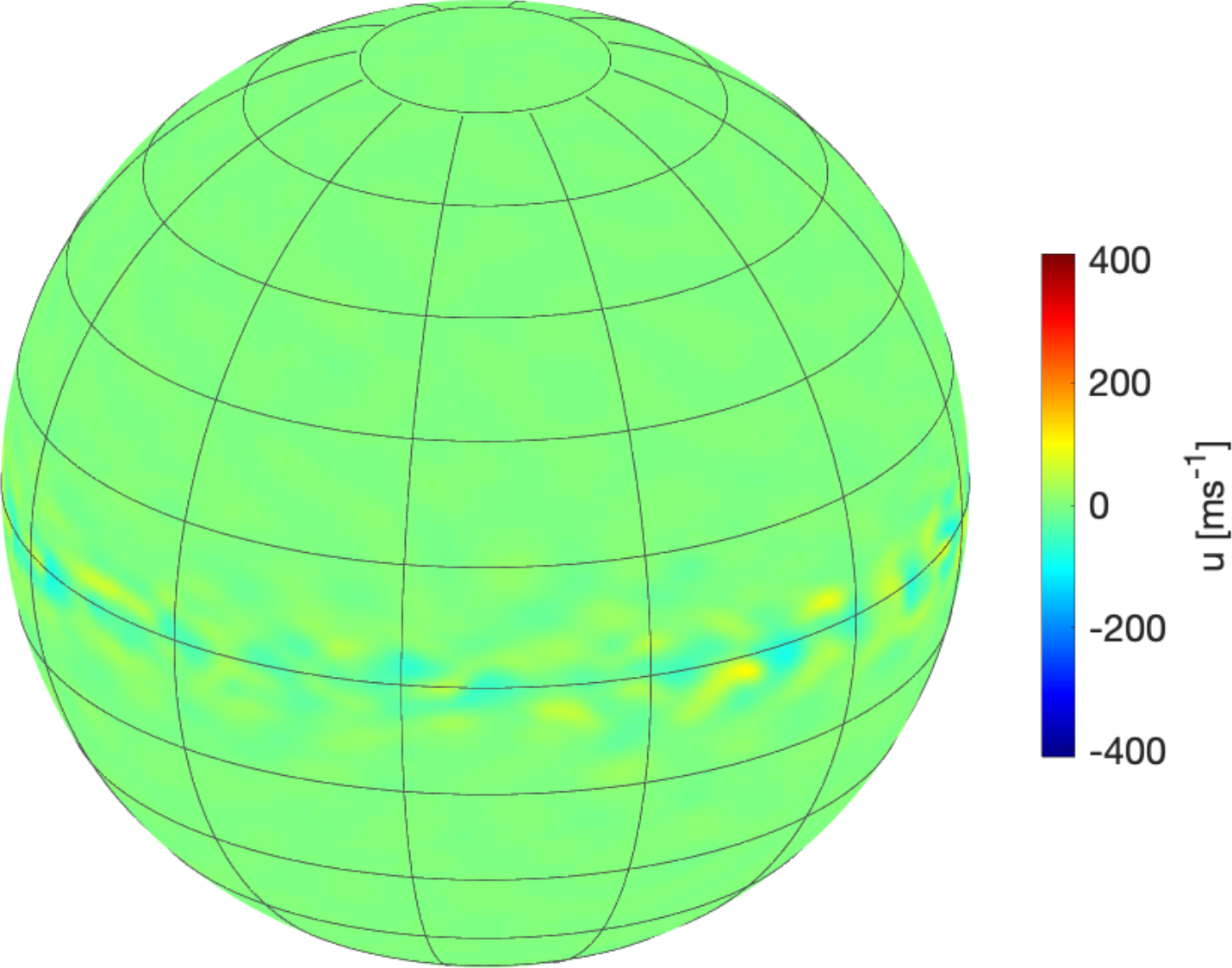}
    \caption{$\teff=1500$ K.}
\end{subfigure}

    \caption{Instantaneous zonal wind of models with $\teff=500$ K taken at 1.8 bar in the top panel, $\teff=1000$ K taken at 8.2 bars in the middle panel, and $\teff=1500$ K taken at 4.1 bars in the bottom panel. These chosen levels are slightly above their RCBs.  These results are taken after the models are fully equilibrated. All models adopt a forcing amplitude $1.5\times10^{-4}\;\kps$ and $\tdrag=10^6$ s. }
    \label{fig.u_teff}
\end{figure}

We first present a set of models with an effective temperature $T_{\rm eff}=1000$ K and a thermal forcing amplitude $s_f=5\times10^{-4}\;{\rm Ks^{-1}}$, but with three bottom drag timescales of $\tdrag=10^5$, $10^6$ and $10^7$ s. At statistical equilibrium, this $s_f$ results in typical isobaric temperature variation of $80 \sim100$ K near the RCB, consistent with the estimated values of 50 to 200 K for brown dwarf conditions (see appendix in \citealp{showman2019}). Figure \ref{fig.tu_drag} shows the instantaneous temperature at 8 bars (a level slightly above the RCB) on the left column and zonal velocity at 8 bars on the right column. Quantities at  lower pressures are qualitatively similar to those shown in Figure \ref{fig.tu_drag}. Figure \ref{fig.uzonal_t2550} shows the time-averaged zonal-mean zonal winds as a function of latitude and pressure. These results are obtained after the models reaching statistical equilibrium. 

The model with a  strong bottom drag of $\tdrag=10^5$ s does not form jets at  pressures higher than a few bars as well as throughout mid-to-high latitudes (top panel of Figure \ref{fig.uzonal_t2550}). At pressures lower than about 1 bar, alternating eastward and westward zonal flows with amplitude $\sim150\mps$ emerge at low latitudes. At pressures around 8 bars, the circulation is mainly consisted of quasi-isotropic turbulence.   
Temperature anomalies  are nearly isotropic and their spacial distributions are not very different to the forcing pattern at mid-to-high latitudes.  At low latitudes within about $\pm 10^{\circ}$,  amplitude of the temperature anomalies appears to be much  smaller than that at mid-to-high latitudes, and the shape is  distorted from the forcing pattern.   The zonal velocity at 8 bars exhibits a larger magnitude at low latitudes than that  at high latitudes. 

Dynamics at mid-to-high latitudes is rotationally dominated on rapidly rotating BDs and giant planets \citep{showman&kaspi2013}.  The importance of rotation is characterized by the Rossby number $Ro = U/fL$, where $U$ is a characteristic horizontal wind speed, $f = 2\Omega \sin\phi$ is the Coriolis parameter, $\Omega$ is the rotation rate and $L$ is a characteristic length scale of the flow.  In the   regime with $Ro \ll 1$, the Coriolis force mainly  balances the horizontal pressure gradient force, and large horizontal temperature difference can be supported \citep{charney1963}. Near the equator where advection is significant in the horizontal momentum balance, $Ro \gtrsim 1$ and we expect a ``tropical'' regime   in which gravity waves  efficiently remove temperature anomalies, leading to a weak horizontal temperature gradient \citep{sobel2001}.   Adopting the RMS of horizontal wind speed $U\sim 300 ~\rm{ms^{-1}}$, $L\sim 10^7$ m from our models, the Rossby number is $\sim$ 0.5, 0.25 and 0.12 at $\pm 5^{\circ}$, $\pm 10^{\circ}$ and $\pm 20^{\circ}$ latitudes, respectively.   This dynamical regime transition corresponds to the qualitative difference of temperature field between low  and mid-to-high latitudes.

With weaker bottom drags of $\tdrag=10^6$ and $10^7$ s,  zonal jets emerge, and the jet speed increases with increasing $\tdrag$ (see Figures \ref{fig.tu_drag} and \ref{fig.uzonal_t2550}). In the case  with $\tdrag=10^6$ s, eddy wind speeds are still comparable to the jet speeds of about $250\mps$, and the jets show significant distortions as seen in panel (d) of Figure \ref{fig.tu_drag}. Kinetic energy associated the eddies is comparable to that associated with the jets.  In the model with  $\tdrag=10^7$ s,  zonal jets are more robust and stable with a peak speed reaching up to  $500\;\mps$,  exceeding the RMS eddy speed. Kinetic energy is mostly contained in the jets. Vigorous eddy activities are still clearly seen along with the jets, and these are likely waves triggered by the thermal perturbations and breaking near the flank of jets. No stable long-lived vortex is maintained.  The equatorial jets exhibit  significant vertical wind shears, similar to that in the strong-drag model. However, the off-equatorial jets develop significant pressure-independent components as shown in Figure \ref{fig.uzonal_t2550}.  

The isobaric temperature anomalies in the weak-drag models are qualitatively similar to those in strong-drag models. There is still subtle difference in the case with $\tdrag=10^7$ s: the morphology of temperature anomalies is sheared and their amplitude is weakened by the strong zonal jets compared to those with a strong drag.  The temperature fields at off-equatorial regions do not exhibit the same zonal banded structure as in the zonal velocity field.  This is a property of the thermal wind balance which holds well in the regime of $Ro\ll1$ (e.g., \citealp{holton2012}, Chapter 3):
\begin{equation}
\frac{\partial \overline{u}}{\partial \ln p} =  \frac{R}{f} \frac{\partial \overline{T}}{\partial y},
\label{thermalwind}
\end{equation}
where $\overline{u}$ and $\overline{T}$ are zonal-mean zonal wind and temperature, and $y$ is northward distance. The largely pressure-independent off-equatorial jets are in line with the small  meridional temperature gradient  at off-equatorial regions in our models.

\begin{figure}
    \centering
    \includegraphics[width=1.\columnwidth]{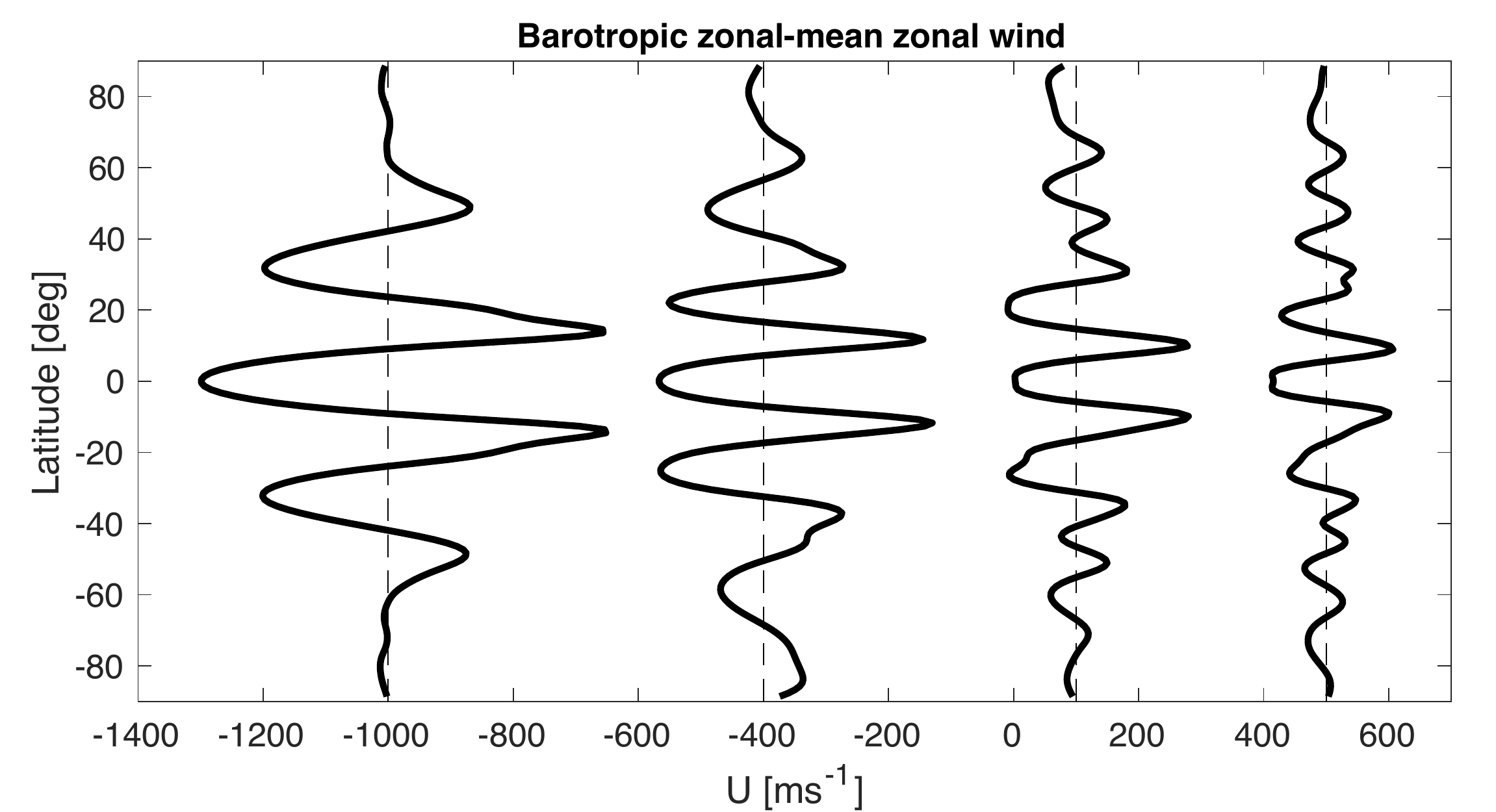}
    \caption{Time-averaged, mass-weighted zonal-mean zonal winds as a function  of latitude for models with an effective temperature of 500 K, a bottom drag timescale $\tdrag=10^7$ s and a rotation period of 5 hours. The thermal perturbation rates in these models (from left to right) are $1.5\times10^{-4}$, $6\times10^{-5}$,  $3\times10^{-5}$ and $1.5\times10^{-5}\;{\rm Ks^{-1}}$, respectively. The jet profiles are shifted with dashed lines corresponding to their zero values.  These results are obtained after the models reaching statistically equilibrium.}
    \label{fig.zonalu_t1500}
\end{figure}

Next, we  examine the effect of different atmospheric temperature (therefore different radiative damping rate) on the circulation pattern. We perform a set of models with a thermal forcing amplitude $s_f=1.5\times10^{-4}\;\kps$, a moderate drag timescale of $\tdrag=10^6$ s, and with three effective temperatures $T_{\rm eff}=500$, 1000 and 1500 K. The radiative damping timescales near their photospheres can be approximated as $\tau_{\rm rad}\sim \frac{P}{g}\frac{c_p}{4\sigma T^3}$ (e.g., \citealp{showman2002}) where $P$ is the pressure near the photosphere, $g$ is gravity and $\sigma$ is the Stefan-Boltzmann constant. We adopt pressures near their photospheres and $T\sim T_{\rm eff}$, yielding $\tau_{\rm rad}\sim 10^5$, $\sim 1.5\times10^4$ and $\sim 6\times10^3$ s for models with $T_{\rm eff}=500$, 1000 and 1500 K, respectively.  The equilibrium isobaric temperature variations reach about 60 K near the RCB in the model with $T_{\rm eff}=500$ K, but drastically decrease with increasing $T_{\rm eff}$ due to the rapidly increasing radiative damping.
Figure \ref{fig.u_teff} shows the instantaneous isobaric zonal velocity of these models  after the they fully equilibrate. The overall wind speed decreases significantly with increasing $\teff$. Zonal jets show a speed up to $\sim200\mps$  in the model with $\teff=500$ K, and only up to several tens of $\mps$ in the model with $\teff=1000$ K. In the model with  $\teff=1500$ K, no zonal jet forms but only weak eddies with local maximum speeds up to a few tens of $\mps$ are present at low latitudes. In the latter case, radiative damping sufficiently dissipates energy injected by the thermal perturbations before they can organize into zonal jets.

Lastly, we show a set of experiments with a fixed effective temperature, a weak drag, but varying amplitude of the thermal perturbation amplitudes. We choose  $\teff=500$ K, $\tdrag=10^7$ s, and $s_f=1.5\times10^{-4}$, $6\times10^{-5}$, $3\times10^{-5}$ and $1.5\times10^{-5}\;\kps$. All models exhibit robust zonal jets due to the weak drag and low temperature. These jets also show strong pressure-independent components at off-equatorial regions and vertical shears at low latitudes.  Figure \ref{fig.zonalu_t1500} shows the mass-weighted, time-averaged zonal-mean zonal wind profiles for these models, and these profiles are shifted with dashed lines corresponding to their zero values. The zonal jet speed decreases and the number of zonal jets increases with  decreasing $s_f$: there are two eastward jets with speed up to more than $350\mps$ in each hemisphere in the model with $s_f=1.5\times10^{-4}\;\kps$, but 4 to 5 eastward jets with speed up to $100\mps$ in each hemisphere with $s_f=1.5\times10^{-5}\;\kps$. These demonstrate that, with the same drag timescale and radiative damping rate, the equilibrium between energy injection and frictional dissipation is obtained at a lower jet speed when the energy injection rate is low. 
Classical two-dimensional turbulence theory predicts that the meridional jet width is correlated to an anisotropic lengthscale, the Rhines scale, $L_{\beta} \sim \pi \sqrt{2U /\beta}$ where $\beta=df/dy$ (\citealp{rhines1975}, and see a review by \citealp{vasavada2005}).  The number of jets on a sphere can be expected as $N_{\rm jet}\sim \pi a/L_{\beta}\sim \sqrt{\Omega a/U}$, where $\beta\sim 2\Omega/a$ and $a$ is the radius. With a rotation period of 5 hours and $a=6\times10^7$ m, the characteristic number of jets is expected to be about 7 for a wind speed $U=400\mps$ and 14 for a wind speed $U=100\mps$. Although the trend that higher wind speed corresponds to  less number of jets is confirmed in our models, the number of jets in our models is obviously smaller than that predicted by the theory given the wind speeds. This is related to the vorticity structure of our simulated flows as will be explained below.

The vorticity structure provides additional information on jet formation. Absolute vorticity, defined as $\zeta+f$, where $\zeta=\mathbf{k}\cdot\nabla\times\mathbf{v}$ is the relative vorticity and $\mathbf{k}$ is the local upward vector on the sphere, is  an approximation to the potential vorticity (PV), a quantity  materially conserved under frictionless, adiabatic motions \citep{vallis2006}. In models with robust zonal jets, the absolute vorticity at a given level tends to be organized into zonal strips with large meridional gradient of  absolute vorticity at the edges of adjacent strips but much smaller gradient within the strips. The relationship between the zonal-mean vorticity and zonal-mean zonal winds suggests that the boundaries between strips correspond to the eastward  jets, whereas the interior of the strips  correspond to the westward jets (e.g., \citealp{mcintyre1982,dritschel2008,dunkerton2008}).  If the absolute vorticity is well homogenized within the strip (the so-called absolute vorticity staircases), it naturally results in jets  whose meridional spacing and jet speeds are related by the Rhines scale (e.g., \citealp{scott2012}). However, in our cases, the homogenization of absolute vorticity within the zonal strips is not perfect,   and the corresponding jet structure is less sharp and the meridional jet widths are broader than those predicted from flows with perfect  absolute vorticity staircases. This explains that the number of jets (Figure \ref{fig.zonalu_t1500}) are less than that expected from the Rhines scale argument. The imperfect PV staircases seem to be the common outcome of the strongly forced and strongly dissipated systems \citep{scott2012}. In such situations, magnitude of the eddy absolute vorticity can be comparable to the difference between two adjacent zonal absolute vorticity strips and their random motions are able to smooth out the boundaries between the strips, preventing perfect staircases. Indeed, such significant turbulent, filamentary absolute vorticity structures  due to turbulence and Rossby wave breaking are present in our models. {\btt Note that such zonal-mean PV structures in our models naturally lead to a {\bttt non-violation} of the barotropic stability criteria which is in contrast to jets in the Jovian atmosphere \citep{ingersoll2004}. }

Rossby wave breaking and its  eddy mixing on PV play a crucial role in jet formation of our models that are under  isotropic forcing and damping.  Rossby waves break more easily  in regions of weaker meridional PV gradient, and the resulting turbulent mixing tends to homogenize the PV in the wave breaking regions and therefore further decreases the meridional PV gradient. This  is a positive feedback mechanism to induce zonal jets. Given an initial meridional PV structure that has variations to the planetary vorticity $f$, Rossby-wave breaking  preferentially occurs at regions with weaker PV gradient, and the associated PV mixing weakens the meridional PV gradient even more, promoting further wave breaking in these regions. In the equilibrium state, zonal jets form as a result of PV zonal striping, and the balance in the jets is achieved in between the dissipation (mainly the frictional drag in some cases) and the PV mixing; for reviews of this mechanism, see \cite{dritschel2008} and \cite{showman2013}. The diagnosis of our models and interpretations are qualitatively very similar to those shown in \cite{showman2019}, and we shall not repeat them here.

The formation of pressure-independent (the so-called barotropic) off-equatorial jets is likely a natural outcome of the tendency that kinetic energy  is stored to the gravest scale in rapidly rotating systems. Such a tendency has also been observed in idealized terrestrial GCMs (e.g., \citealp{chemke2015}) and gas giant planet GCMs (e.g., \citealp{lian2008,schneider2009,young2019,spiga2020}).   Early theoretical work using two-layer quasi-geostrophic (QG) models  \citep{rhines1977, salmon1980} and multiple-layer QG models \citep{smith2002} have shown that energy can  be converted efficiently from the smaller-scale pressure-dependent (the so-called baroclinic) modes  to the barotropic mode.  In our weak-drag models,   the emergence of barotropic off-equatorial  jets    is not surprising as the  QG approximation holds well at mid-to-high latitudes. 
Baroclinic potential energy is injected and is partially converted to baroclinic kinetic energy;  at least part of it subsequently cascades to the barotropic mode, driving the barotropic jets.

Another  mechanism contributing to the formation of  barotropic jets  is  the development of vertically in-phase  jets by  Coriolis force associated with the mean meridional circulation. This is the so-called downward control  in Earth's stratosphere  \citep{haynes1991} and has  been proposed for the Jovian atmosphere \citep{showman2006}.   The principle is the following: under rotation-dominated conditions, the zonal acceleration of zonal winds confined within a certain pressure range  causes a responsive mean meridional circulation with a Coriolis  force counteracting the jet forcing. The  overturning meridional circulation penetrates to layers without zonal forcing, causing a  reverse meridional velocity. The corresponding Coriolis force drives the zonal winds to the same direction as the jets in the forced layers.     Indeed,  diagnosis (not shown) in our models suggest that zonal jets are forced by eddy convergence and balanced by Coriolis force in the perturbed layers,  while in deeper layers, the Coriolis force associated with the mean meridional circulation helps to drive the deep jets to the same direction. 

{\btt The deep layers are nearly convectively neutral (Figure \ref{fig.tp}). The weak stratification  greatly helps with driving the strong barotropic component of the off-equatorial jets by either the downward control mechanism \citep{showman2006} or  turbulent energy transport \citep{smith2002}. In general, in many GCMs for giant planets in which the deep layers tend to be neutral, the development of strong barotropic components of the off-equatorial zonal jets is very common (e.g., \citealp{lian2008,schneider2009,showman2019,young2019,spiga2020}). 
 }

There is still a subtlety regarding the mean meridional temperature structure. If the atmosphere is externally forced by equator-to-pole irradiation difference as in the Earth, there will be a pressure-dependent jet component at mid latitudes regardless how strong the pressure-independent component is. In our cases, the imposed homogeneous  temperature at the bottom boundary acts against this configuration. The question is, will dynamics organize itself to maintain a systematic meridional temperature variation (and thus a significant vertical jet shear at mid latitudes) against radiative damping? Such a picture was proposed by \cite{showman&kaspi2013}: when alternating jets develop, some waves  dissipate preferentially at different latitudes. This drives overturning meridional circulation cells  that  advect high-entropy air down and low-entropy air upward, creating isobaric meridional temperature gradients. Indeed, there are obvious mean meridional temperature variations in some very weakly radiatively damped models in \cite{showman2019} (see their Figure 2). In our models, the off-equatorial mean meridional temperature variation is much weaker than that of the local variations, presumably because of the strong radiative damping in conditions appropriate for BDs and isolated young EGPs.

Finally, we turn our attention to the equatorial jets. In almost all models including those with a strong bottom drag, there exist  local equatorial eastward jets at altitudes well above the thermal perturbation regions (see Figure \ref{fig.uzonal_t2550} for examples). These correspond to local superrotations, and they emerge from interactions between the vertically propagating equatorial waves and the mean flows. Diagnoses (not shown) of the cases in Figure \ref{fig.uzonal_t2550} suggest that vertical eddy angular momentum transport exerts eastward torques in the superrotating region, while horizontal eddy  momentum transport counteracts it, {\btt i.e., $\frac{1}{a\cos^2\phi}\frac{\partial(\overline{u'v'}\cos^2\phi)}{\partial \phi} \sim - \frac{\partial(\overline{u'\omega'})}{\partial p}$ where $\phi$ is latitude, $a$ is radius, and $u',v'$ and $\omega'$ are  deviations of zonal, meridional and vertical velocities from the zonal average, respectively.  Momentum transport by the mean flow also contributes  to the balance.} Jets in cases shown in Figure \ref{fig.uzonal_t2550} are stable overtime. But in some other cases, especially those with a lower temperature,  equatorial jets can evolve overtime in a quasi periodic manner, which will be described in next section.

However,  all our models with relatively weak drags  form equatorial westward jets at deep layers (see examples in Figure \ref{fig.uzonal_t2550}). Despite having local superrotations at low pressures, all these models exhibit equatorial sub-rotation when their jets are mass-weighted. Part of these models are shown in Figure \ref{fig.zonalu_t1500}.  The common outcome with westward equatorial flow is similar to previous investigations using shallow-water models with  small-scale random forcing (e.g., \citealp{scott2007, showman2007, zhang&showman2014}) and  our previous study using the same forcing scheme but a different radiative damping scheme \citep{showman2019}. On the other hand,  the strong equatorial eastward jets in the Jovian and Saturnian atmospheres penetrate to levels much deeper than their cloud decks. 

As yet, there is no universal theory to predict whether  westward or eastward equatorial flow should occur under conditions appropriate for gas giants. Multiple mechanisms have been proposed, but the resulting equatorial flows are sometimes  sensitive to the detailed model setup (see a recent review by \citealp{showman2018review}). Regardless the details, in a shallow atmosphere framework (as opposed to the deep models that consider the fluid interior), the competition between the equatorially  driven and off-equatorial driven forces largely determines the direction of the equatorial flow:  (e.g., \citealp{schneider2009,liu2011, showman2015}). 
\cite{schneider2009} and \cite{liu2011} argue that if the convective  forcing is isotropic horizontally, one might expect equatorial superrotation. {\btt This is because 
convective heating fluctuations induce fluctuations in the large-scale horizontal divergence, and these represent a source of vorticity fluctuations and thus a source of Rossby waves. The Rossby wave source can be expected to be large in the equatorial region because the Rossby number there is order one or greater and large-scale horizontal flow fluctuations induced by convective heating fluctuations are divergent at leading order. On the other hand, the Rossby number at high latitudes is small, and large-scale flows are nondivergent at leading order.	}
Eddy fluxes due to waves propagating out of the equatorial region are expected to be more than those propagating into the equatorial region from mid latitudes. The former induces eastward acceleration while the latter exerts westward acceleration to the equatorial flow. Therefore, the net eddy angular momentum convergence tends to drive an equatorial superrotation. 

Surprisingly, this picture does not describe our results despite that our forcing is globally isotropic. Even when our models do not impose the drag at the equatorial region as those in \cite{schneider2009}, there is still no mass-weighted equatorial superrotation. 
One possibility may be that  waves triggered at the equator remain a baroclinic structure. In the presence  of a finite equatorial deformation radius for the baroclinic flow, these waves tend  to be trapped in the equatorial region \citep{matsuno1966} and few of them propagate out meridionally. Instead, they are transmitted in the vertical direction and drive vertically shearing  equatorial jets.  However, off-equatorial Rossby waves can freely propagate  to the equatorial region, dissipating there and  driving  the  equatorial westward flow.   Indeed, our diagnoses suggest that the mass-weighted equatorial westward flow is maintained by both the horizontal eddy and mean-flow momentum transport while  the frictional drag balances the westward forcing.  It is interesting to note that even in the absence of meridional propagation of equatorial waves, based upon a shallow-water system, \cite{warneford2017} show that dissipation of these waves can still lead to a net torque on the equatorial flow and the jet direction is sensitive to the form of the dissipation. It is yet unclear how these shallow-water theories can be transformed into understanding of the continuously stratified atmospheres.

\subsection{Quasi-Periodic Oscillations of the Equatorial Jets}

\begin{figure*}

    \begin{subfigure}[t]{0.33\textwidth}
    \centering
    \includegraphics[width=0.95\columnwidth]{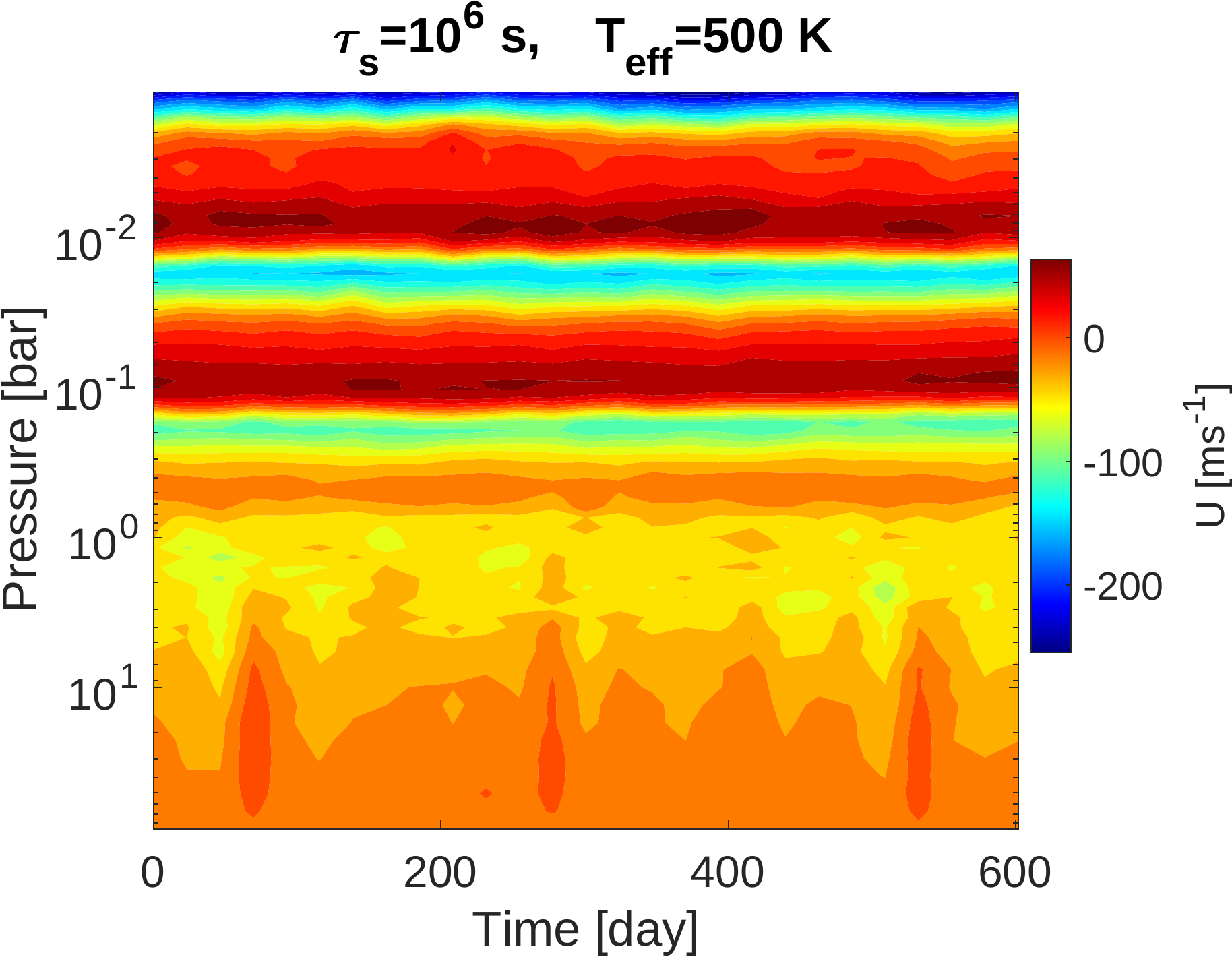}
    \end{subfigure}
    \begin{subfigure}[t]{0.33\textwidth}
    \centering
    \includegraphics[width=0.95\columnwidth]{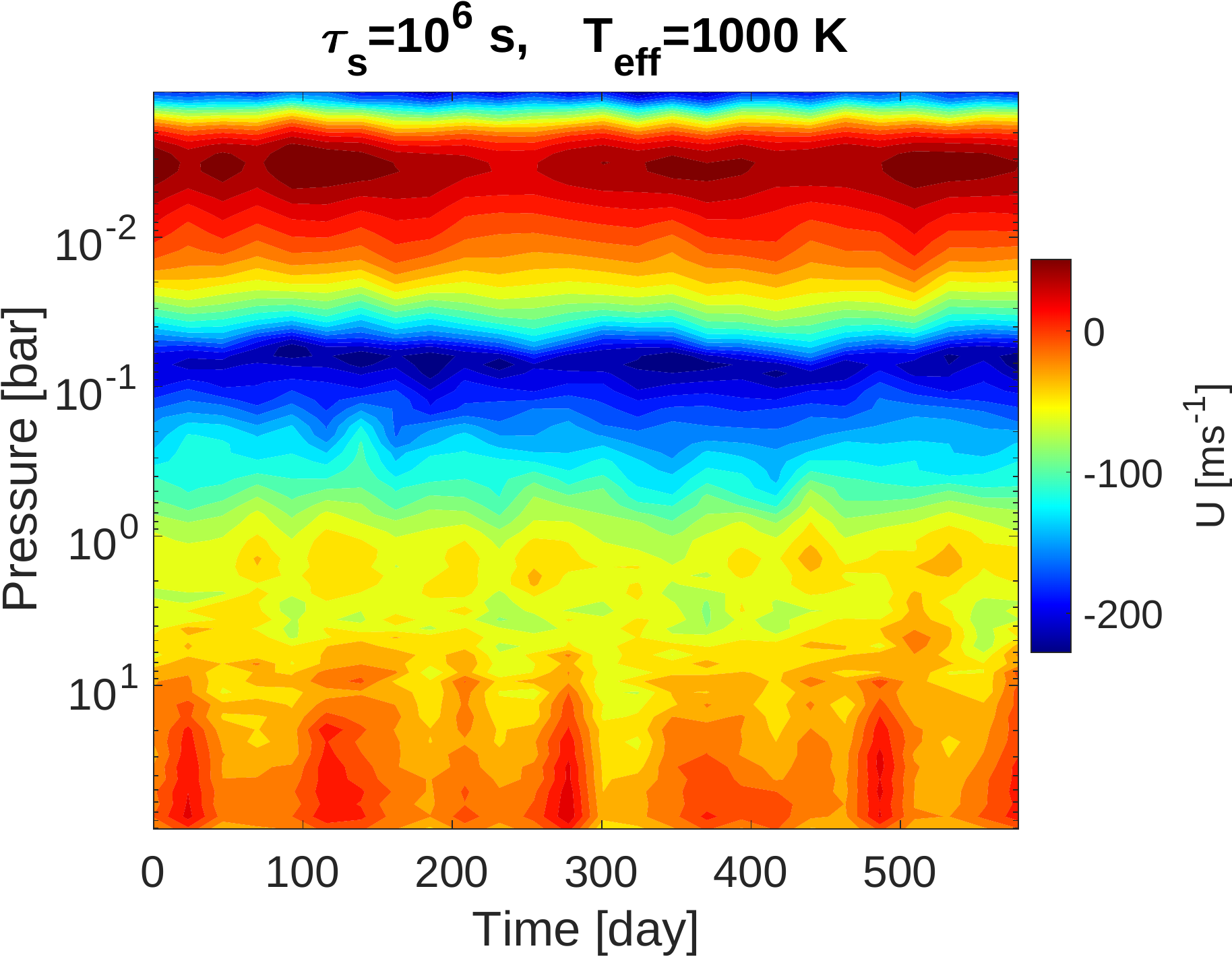}
    \end{subfigure}
    \begin{subfigure}[t]{0.33\textwidth}
    \centering
    \includegraphics[width=0.95\columnwidth]{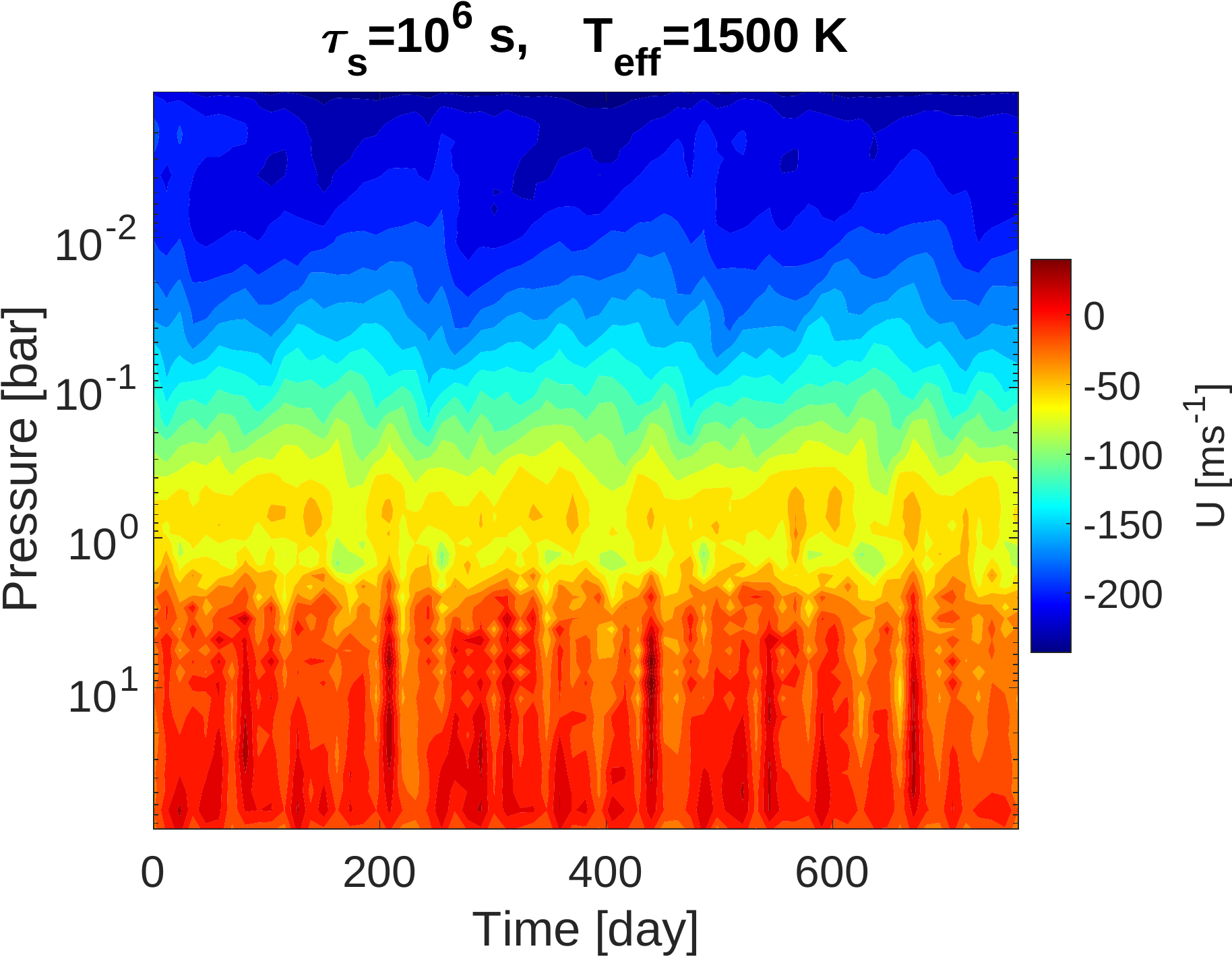}
    \end{subfigure}
 
    \medskip
    
    \begin{subfigure}[t]{0.33\textwidth}
    \centering
    \includegraphics[width=0.95\columnwidth]{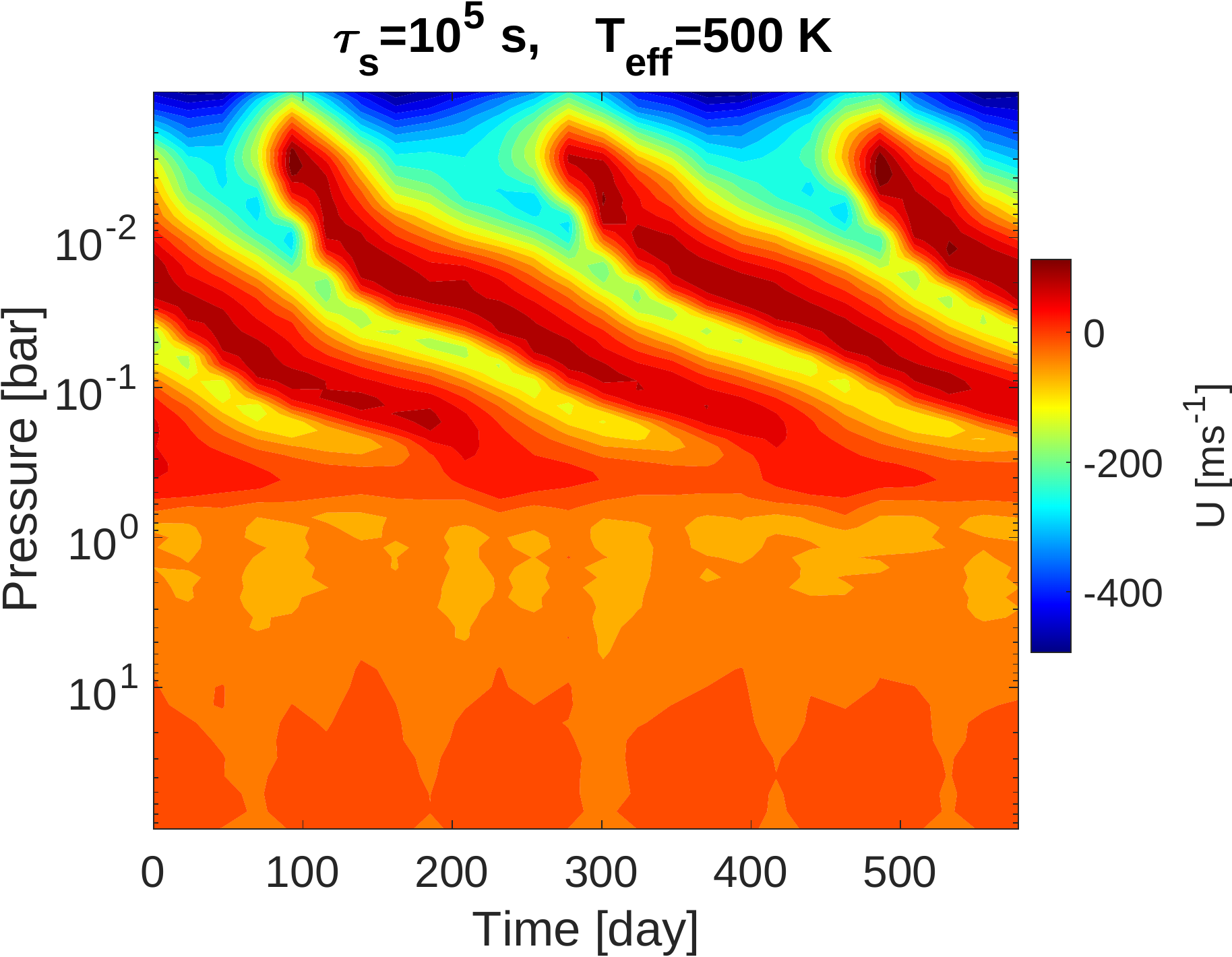}
    \end{subfigure}
    \begin{subfigure}[t]{0.33\textwidth}
    \centering
    \includegraphics[width=0.95\columnwidth]{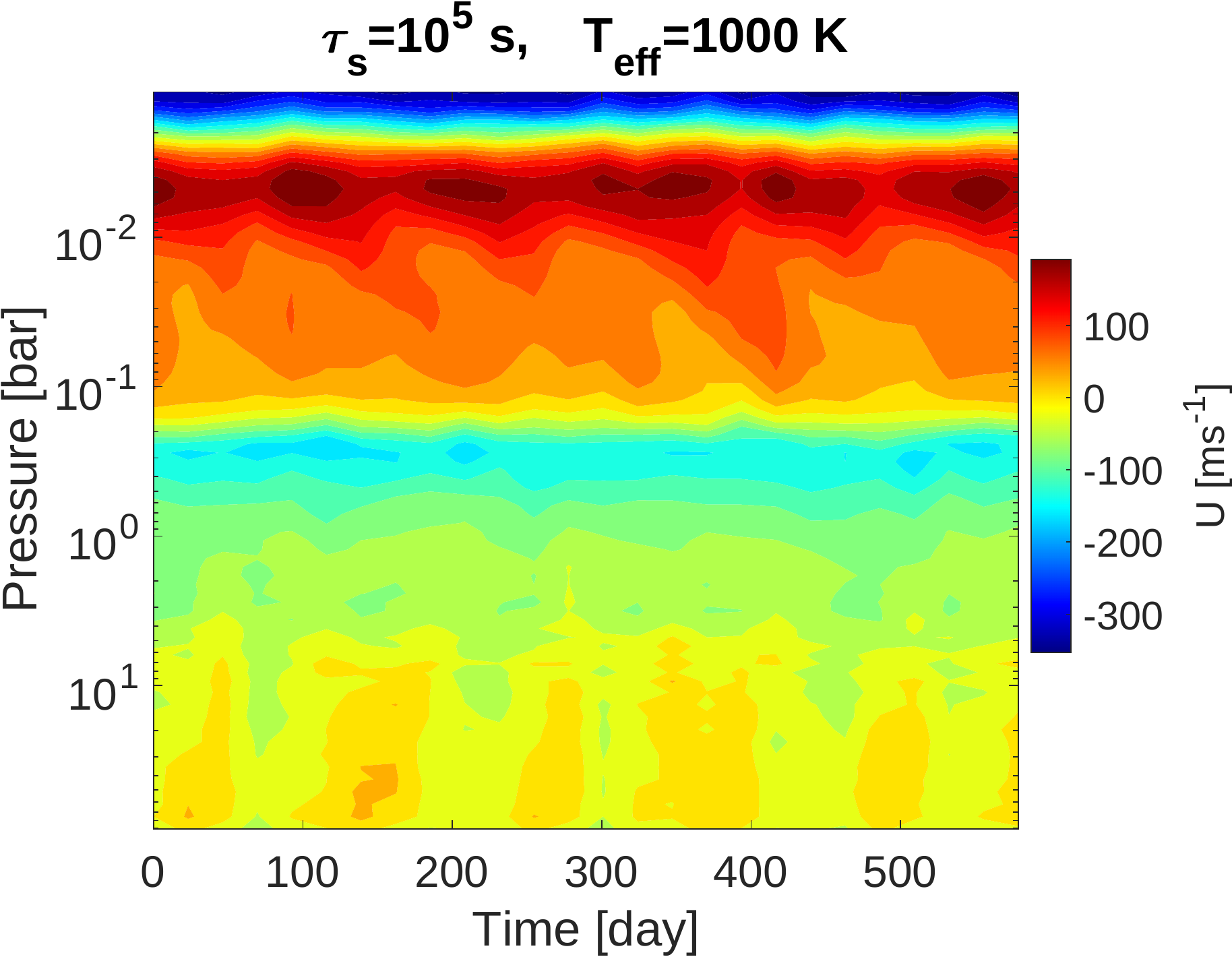}
    \end{subfigure}
    \begin{subfigure}[t]{0.33\textwidth}
    \centering
    \includegraphics[width=0.95\columnwidth]{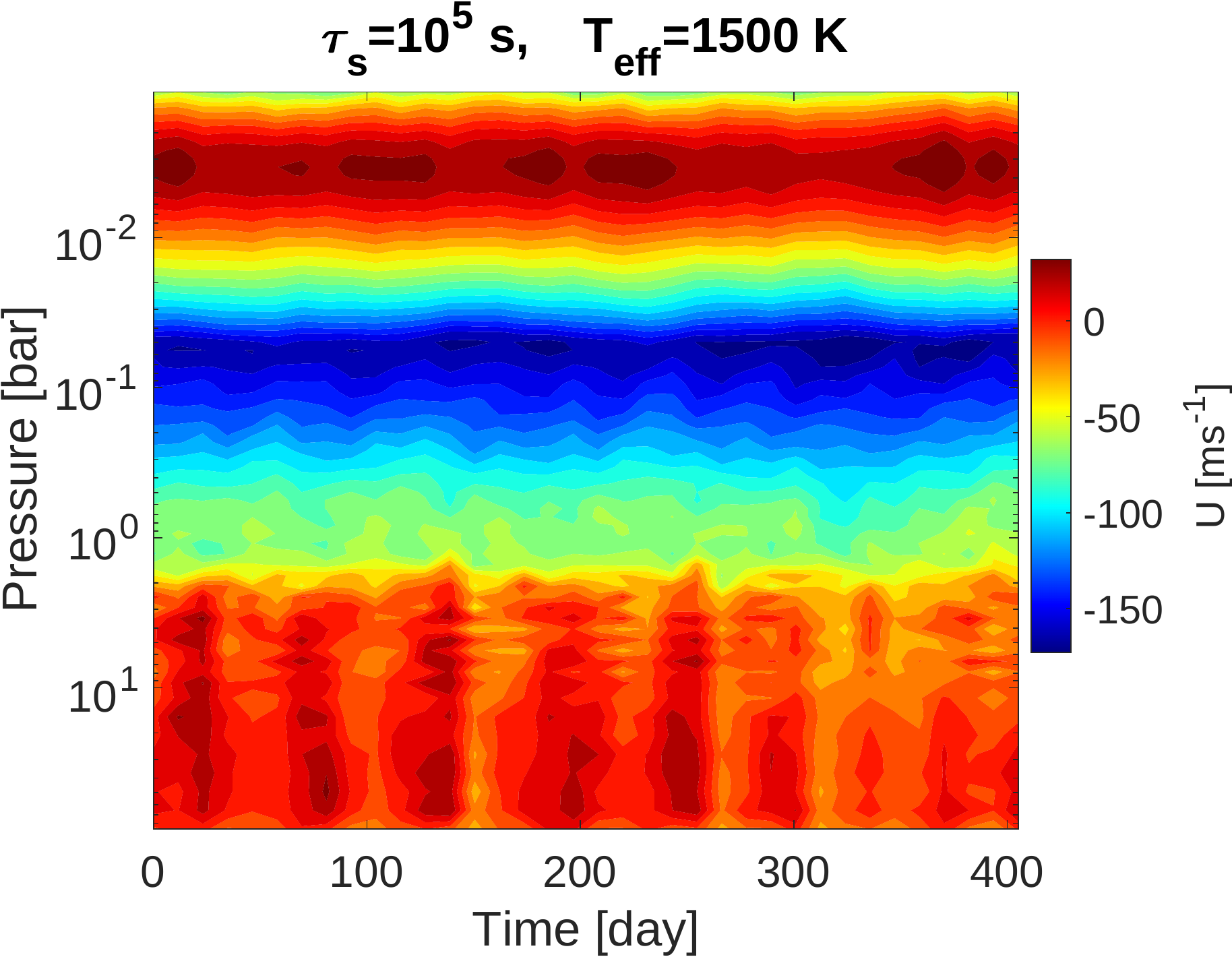}
    \end{subfigure}
    
    \medskip
    
    \begin{subfigure}[t]{0.33\textwidth}
    \centering
    \includegraphics[width=0.95\columnwidth]{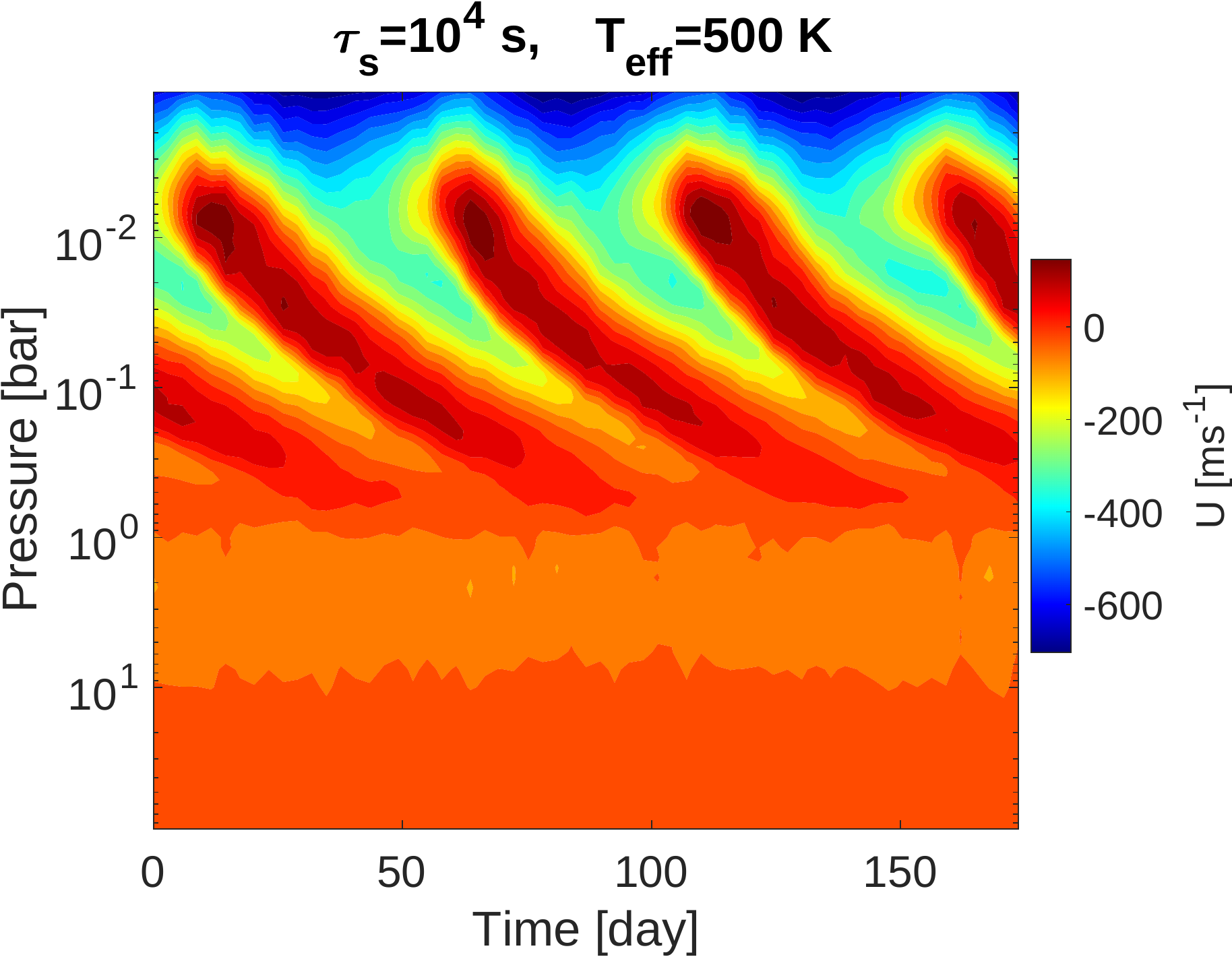}
    \end{subfigure}
    \begin{subfigure}[t]{0.33\textwidth}
    \centering
    \includegraphics[width=0.95\columnwidth]{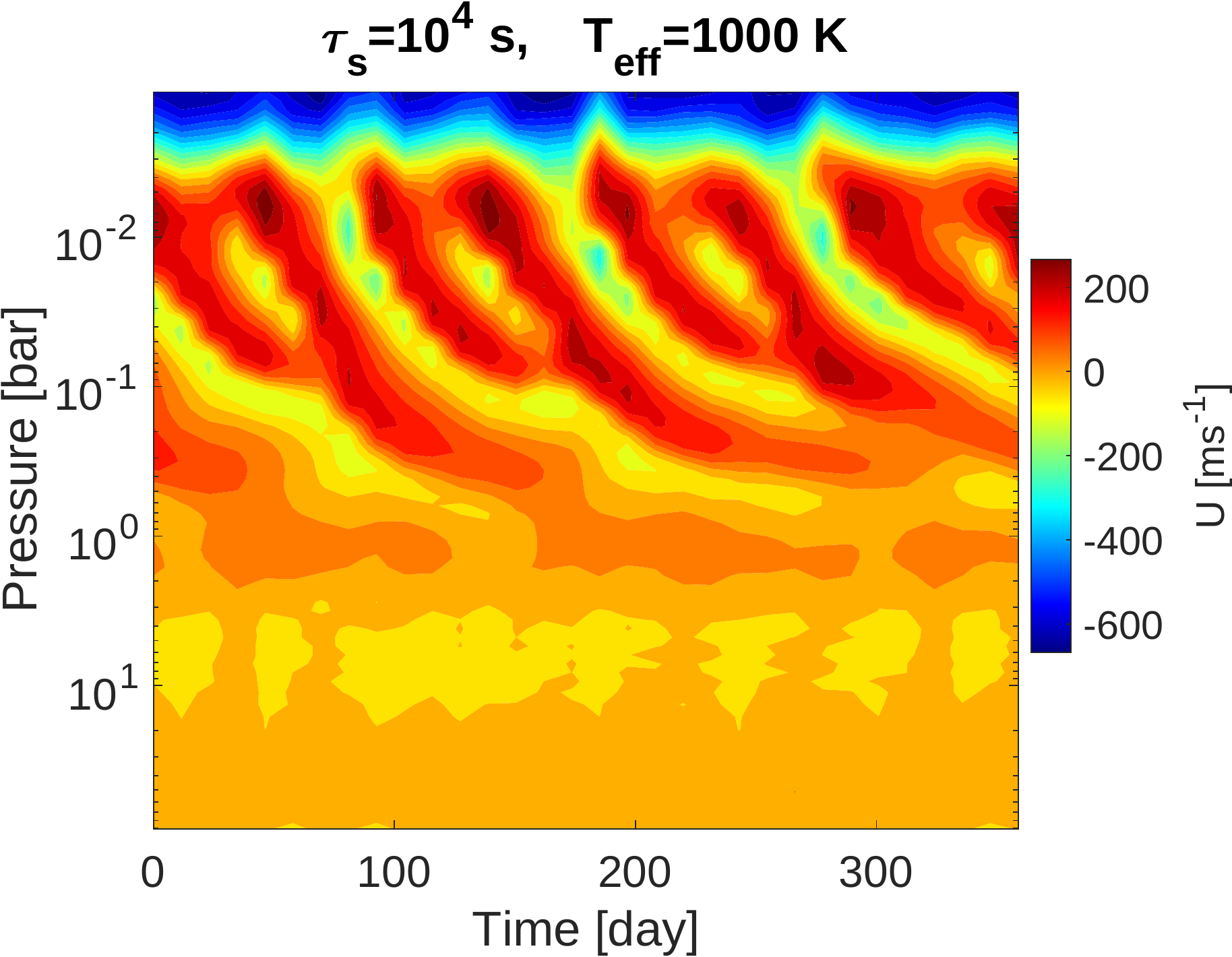}
    \end{subfigure}
    \begin{subfigure}[t]{0.33\textwidth}
    \centering
    \includegraphics[width=0.95\columnwidth]{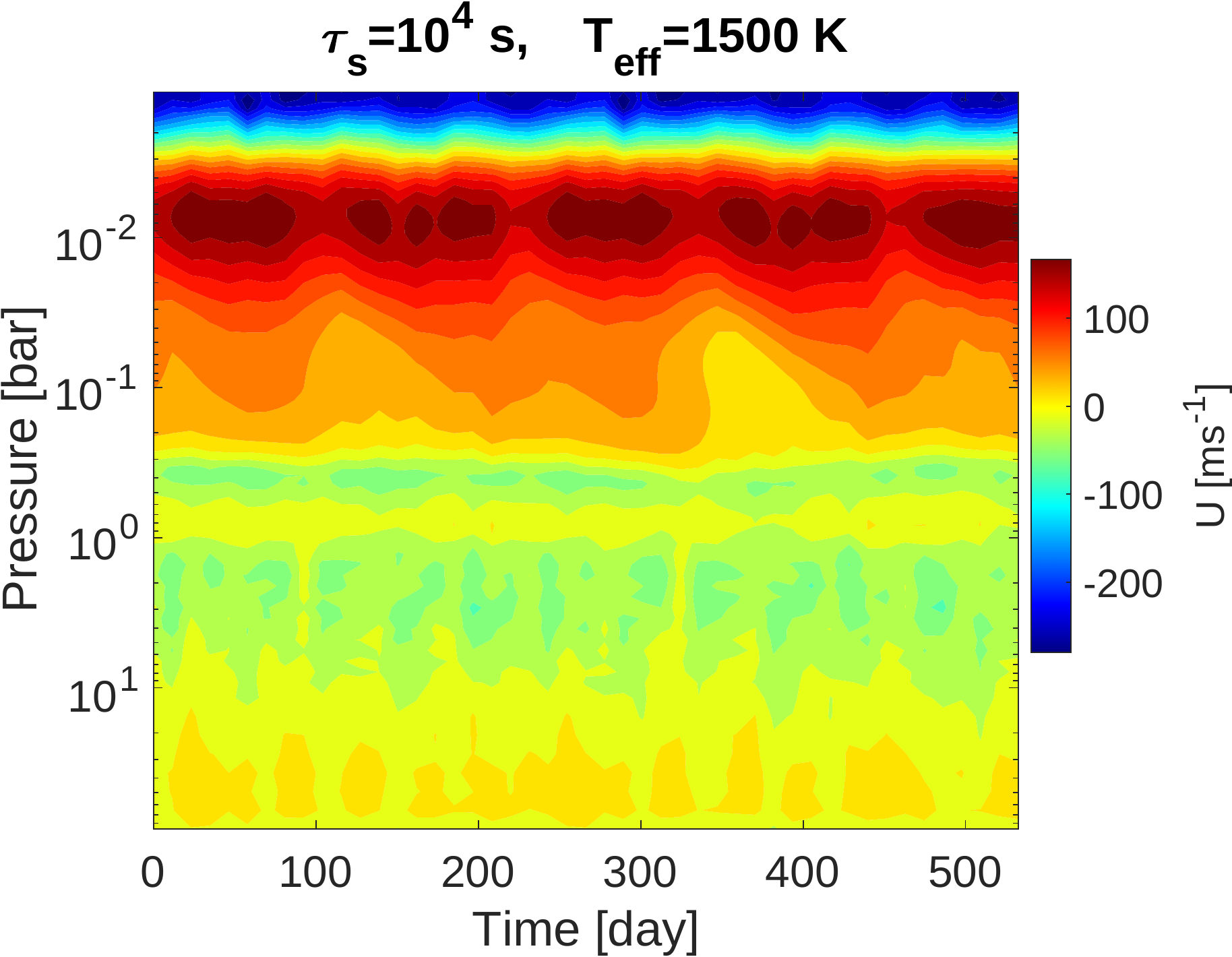}
    \end{subfigure}
    
    \medskip

    \caption{Zonal-mean zonal jet at the equator as a function of pressure and time for models in a grid with three effective temperatures of 500, 1000 and 1500 K (from the left to right) and  three storm timescales $\tau_s=10^6$, $10^5$ and $10^4$ s from the top to the bottom. The bottom drag timescale is $\tdrag=10^5$ s for all models. The time zero starts where the models have reached statistical equilibrium.}
    \label{fig.qbo}
\end{figure*}

Quasi-periodic oscillations (QPOs) in the equatorial region have long been recorded in atmospheres of solar system planets. These include the  quasi-biennial  oscillation (QBO)  in Earth's equatorial stratosphere \citep{baldwin2001}, the Quasi Quadrennial Oscillation (QQO) with a period around 4 to 5 years  in the Jovian stratosphere \citep{leovy1991, friedson1999, cosentino2017,antunano2021}, and the semiannual oscillation (SAO) with a period of $\sim$15 years in the stratosphere of Saturn \citep{fouchet2008,guerlet2011}. These QPOs  are not  related to their seasonal cycles but are  driven by   wave-mean-flow interactions involving the upward propagation of equatorial waves generated in the lower atmosphere and their damping and absorption in the middle and upper atmosphere \citep{baldwin2001,guerlet2018}. These interactions result in downward migration of the vertically stacked, alternating eastward and westward equatorial jets overtime. 
The QPOs viewed at certain levels are simply a manifest of the jet migration cycles.  Our precursor work  has shown that  QPOs may be a common outcome in  atmospheric conditions appropriate for some BDs and isolated  EGPs \citep{showman2019} . Here we further explore  properties of the QPOs in a somewhat strong forcing and damping regime over a wider parameter space using the updated GCM. 

We first conduct experiments over a grid with three effective temperatures of $\teff=500$, 1000 and 1500 K and three storm timescales $\tau_s=10^4$, $10^5$ and $10^6$ s. Note that  $\tau_s$ characterizes the typical evolution timescale of the injected thermal perturbation pattern. The former affects the radiative damping on  dynamics and the latter affects the wave properties. A bottom drag with $\tdrag=10^5$ s is applied in all models. The strong drag  suppresses formation of strong jets such that the wave properties are not influenced by the deep jets.   The thermal forcing amplitudes are $s_f=1.5\times10^{-4}\;\kps$ for $\teff=500$ K, $5\times10^{-4}\;\kps$ for $\teff=1000$ K and $2\times10^{-3}\;\kps$ for $\teff=1500$ K. The typical equilibrium isobaric temperature variations near the RCBs are between 60$\sim$100 K in these models.   Figure \ref{fig.qbo} shows the equatorial zonal-mean zonal wind as a function of time and pressure  of this grid.  All the time sequences start after the models reaching statistical equilibrium. QPOs are present in some cases. We take  the case with $T_{\rm{eff}} = 500$ K and $\tau_s = 10^5$ s as an example to illustrate its basic property. The downward migration of jets in this case has a mean period about 200 (Earth) days. The jet downward migration starts near the upper  boundary and ends slightly above 1 bar. The migrating eastward and westward jets have similar magnitudes, and have strong vertical wind shear between them. The QPO is horizontally confined within a narrow latitudinal range ($\pm\sim10^{\circ}$) around the equator.

The important overall picture  is that the QPOs occur preferentially in the low left corner of the grid---with low atmospheric temperatures and short storm timescales. In models with high atmospheric temperatures $\teff=1500$ K and models with long  storm timescale $\tau_s = 10^6$ s, the QPO never occurs. In models with $\teff=1000$ K, only the one with  $\tau_s = 10^4$ s has a QPO.  Within the same atmospheric temperature, the shorter the storm timescale, the faster the jets migrate, a phenomenon clearly illustrated in the column with $T_{\rm{eff}}=500$ K. Given a $\tau_s$, the cooler the atmosphere, the faster the jets migrate as shown in the row of $\tau_s = 10^4$ s (even considering that the hotter models  have higher forcing rates). The jet migration  in models with $T_{\rm{eff}}=500 $ K is coherent, but that in the model with $T_{\rm{eff}}=1000 $ K and $\tau_s=10^4$ s has a more complicated pattern, showing independent migration of two eastward jets in a complete cycle. 

\begin{figure}      
\centering
 \includegraphics[width=0.8\columnwidth]{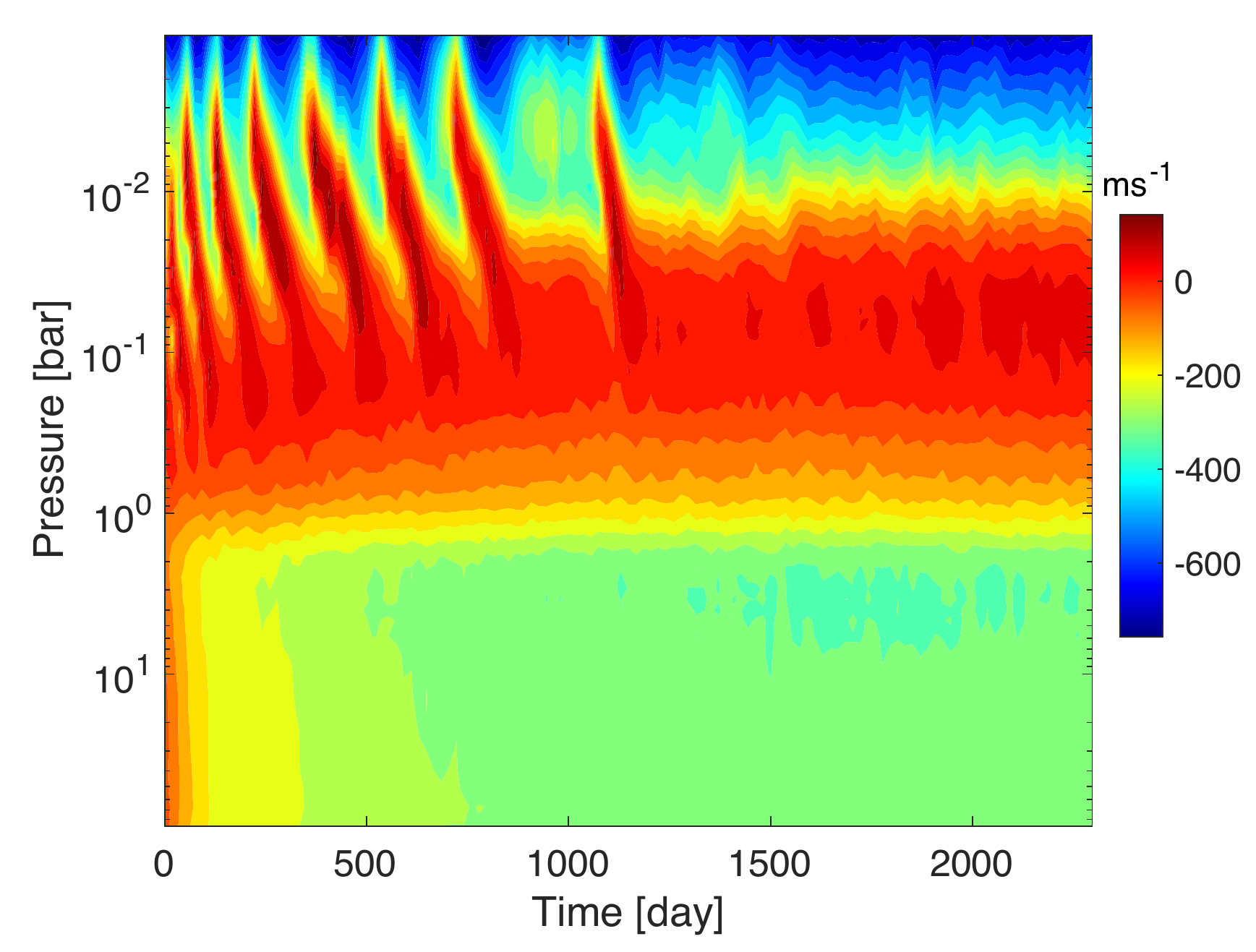}
\caption{Equatorial zonal-mean zonal wind  as a function of pressure and time for the model with an effective temperature of 500 K, a storm timescale $\tau_s=10^4$ s, a forcing amplitude $s_f=1.5\times10^{-4}\;\kps$ and a weak drag of $\tau_{\rm{drag}}=10^7$ s. }
\label{qbo1500-2}
\end{figure}

We now briefly explore the dependence of the QPOs on other parameters. A similar set of models are performed with $T_{\rm{eff}}=500$ K, $\tau_s=10^4, 10^5, 10^6$ s, $\tdrag=10^5$ s, but with thermal perturbation amplitude $s_f$ ten times smaller than those  in the left column of Figure \ref{fig.qbo}. Reducing $s_f$ simply results in less wave flux. Only the model with $\tau_s=10^4$ s exhibits a QPO, and it has a slower equatorial zonal wind speed, a longer oscillation period and a lower pressure where the jet migration   ends.   Next we conduct a similar set of  models as those in the left column of Figure \ref{fig.qbo} with $T_{\rm{eff}}=500$ K, $\tau_s=10^4, 10^5, 10^6$ s but with a longer drag timescale $\tau_{\rm{drag}}=10^7$ s that promotes  formation of strong deep equatorial jets. At the jets develop, no model can sustain the QPO. However, during the spin-up phase wherein the deep equatorial jet is still weak, the QPO can  occur in the model with $\tau_s=10^4$ s. This is illustrated  in Figure \ref{qbo1500-2}.   In the first few hundred simulation days, the QPO appears in a similar fashion as the strong-drag models. As the deep jet grows increasingly westerly,  the  QPO  period increases and the pressure at which the jet migration ends decreases over time. Finally after about 1500 days, the QPO is completely ceased when the deep westward jet approaches a steady state with a speed $\sim300\mps$.

To qualitatively understand varieties of the QPO with different parameters, we briefly recap the fundamental dynamical mechanisms  developed to understand the QBO on Earth.
The classic framework developed by  \cite{lindzen1968},  \cite{holton1972} and \cite{plumb1977} considers a 1-D height-dependent model for equatorial zonal-mean zonal winds with parameterized wave-mean-flow interactions.  Waves propagate upward through  a  background with vertical jet shear, and preferentially dissipate  and break near their critical levels (the level where the  jet speed is equal to wave zonal phase speeds). Angular momentum of waves is deposited to the jet at where they dissipate.  In the absence of large viscosity and counteracting force, the jet is accelerated. Because the peak zonal acceleration of the jet occurs at altitudes below the jet peak, the jet migrates downward over time.  As this occurs, the wave-dissipation level moves downward further accompanied with the migration of jet, and this feedback ensures a continuous jet migration.  The essence of the QBO can be understood with  a simple and elegant framework that  considers two discrete upward propagating gravity waves with identical amplitudes and equal but opposite phase speed \citep{plumb1977}. Radiative damping is applied to wave amplitudes as they propagate vertically, and a minimal viscosity is applied to the jets. 
Suppose that initially a eastward jet is at a lower altitude than a westward jet, and the eastward propagating wave is trapped below the eastward jet, but the westward propagating wave can propagate freely until reaching the westward jet above. The two waves  drive the downward migration of their corresponding jets. At some point a jet  approaches the bottom and create a strong shear,  which then leads to viscous destruction of that jet. The corresponding wave then freely propagate to higher levels, where wave damping gradually build up a new  jet. The process repeats, leading to a periodic downward migration of alternating jets. The period of the oscillation crucially depends on the  wave flux. Although latitudinal complications must be considered to explain some features of the QBO \citep{andrews1987}, the essence of this simple picture generally applies. 


The rate of radiative damping may have two crucial effects in controlling the QPO behaviors  in our models. First, radiative damping may affect the  wave momentum flux that can reach the critical levels by damping the wave amplitudes during their propagation. Strong radiative damping with a rate approaching the Doppler-shifted wave frequency may reduce the sensitivity of wave dissipation to the background mean flow.\footnote{The assumption that the radiative damping rate is much smaller than the Doppler-shifted wave frequency is generally applied in the wave-mean-flow theories for the QBO (e.g., \citealp{plumb1977}).} Radiative damping also influences the type of the waves that actually drive QPOs by preferentially damping out slowly propagating waves and preserving fast waves. 
Second, radiative damping serves as a source of  diffusion on the zonal jets. To see this,  we write the thermal wind equation Eq. (\ref{thermalwind}) in the equatorial $\beta$ plane {\bttt (\citealp{holton2012}, Chapter 12.6)}: 
\begin{equation}
\frac{\partial \overline{u}}{\partial \ln p} =  \frac{R}{\beta} \frac{\partial^2 \overline{T}}{\partial y ^2}.
\label{thermalwind2}
\end{equation}
In the thermal-wind limit which holds reasonably well for zonal-mean flows, significant vertical shear  requires substantial meridional temperature differences. If the radiative damping is strong, this meridional temperature anomaly can be difficult to maintain over a timescale comparable to the jet migration timescale. In this case, the radiative damping will tend to trigger a residual circulation  to oppose the driving force of the QPO.  Putting the above arguments together, in hotter atmospheres, the eddy flux responsible for the QPOs may be sufficiently reduced, and the ``radiative diffusion" on the zonal jets is so strong that it may statistically balance the vertical wave forcing. Thus, the equatorial regions in hot atmospheres can  easily  achieve steady states compared to cooler atmospheres regardless the storm timescale. If wave flux is small,  a moderate rate of radiative damping can diffuse away the QPO forcing, which is why models with much smaller thermal perturbations amplitude lack a QPO even with  a low temperature of $T_{\rm{eff}}=500$ K.

The storm timescale $\tau_s$ likely has significant impacts on the wave properties. Despite that  limited discretized modes are in the thermal perturbations,   waves spanning a wide range of frequency and wavelength are triggered (see Figure 14 in \citealp{showman2019}).  This indicates that processes generating these waves include nonlinear effects.  It may be possible that these processes with a shorter $\tau_s$ tend to trigger   waves whose spectra spans a wider range of frequency and therefore the wave flux may be greater. On the other hand, the excited wave spectra may be more limited by a longer $\tau_s$.  A shorter $\tau_s$ is then expected to result in a shorter oscillation periods than those with a longer $\tau_s$ due to a larger excited wave flux, as shown in Figure \ref{fig.qbo}.  Quantitative demonstration of the above speculation is out of our scope and merits further studies.

\begin{figure*}      
\includegraphics[width=1.6\columnwidth]{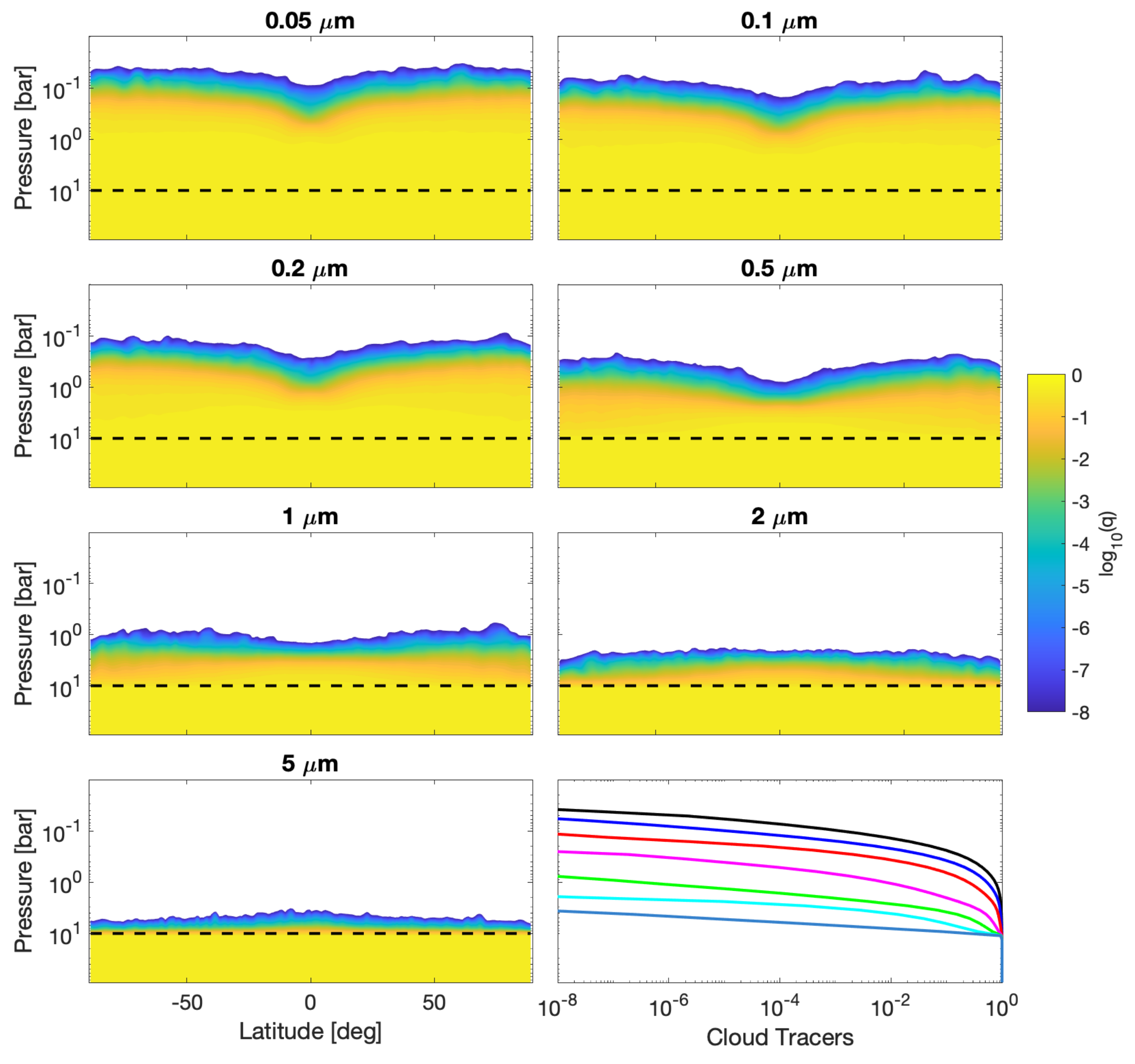}
\caption{\emph{Colored panels}: Time averaged, zonal-mean   profiles of relative cloud mixing ratio (in logarithm scale) as a function of latitude and pressure for different particle sizes. Only mixing ratio larger than $10^{-8}$ is plotted, and the blank regions have mixing ratio less than $10^{-8}$. The black dash lines are the condensation level, below which the tracers are in the form of vapor. The model has an effective temperature of  1000 K, a rotation period 15 hours, a surface gravity 1000 $\rm{ms^{-2}}$, a storm timescale $\tau_s = 10^5$ s, a drag timescale $\tau_{\rm{drag}}=10^5$ s,  and a forcing amplitude $5\times10^{-4}$ $\rm{K~s^{-1}}$. \emph{Lower right panel}: time and globally averaged relative cloud mixing ratios as a function of pressure for particles with different size. {\bttt Solid lines with different colors (from black to blue) correspond to particles with different size from 0.05 to 5 $\mu{\rm m}$.}   }
\label{cloud1}
\end{figure*}  

The development of a strong  jet near the wave-generation level affects wave propagation. Because perturbations are applied in the rotational frame, waves are triggered in the rotational frame but immediately experience the jet. When the jet is strong, waves with zonal phase velocities the same sign as the jet encounter a critical layer right after they form and never propagate out. This leaves only waves in the other phase direction propagating freely to the upper atmosphere. When the jet is not strong (compared to the zonal phase speeds of waves), the influence is weaker and does not totally suppress all waves in one direction. The strong deep jet breaks the east-west symmetry of wave forcing  necessary for the QPOs. In Figure \ref{qbo1500-2}, the increasing QPO period suggests that wave fluxes decrease as the deep jet increases, indicating the increasing ability of the deep jet on blocking part of waves. As the deep jet approaches the same speed as that of the upper downward migrating westward jet, almost all west-propagating waves that are associated with driving the QPO are trapped, and the QPO is then ceased.

Finally we note that details of QPOs in our models are somewhat sensitive to  vertical  resolution, similar to \cite{showman2019} and earth GCMs without gravity wave parameterization and tuning (e.g., \citealp{nissen2000}). 
Our goal here is not to precisely predict properties of QPOs that might occur in real BDs and EGPs but to explore their first-order behaviours over a wide parameter space.

\section{Transport of passive Tracers}
\label{ch.tracer}
In this section we show that the local overturning circulation generates vertical mixing of passive tracers and discuss the dependence of mixing on different tracer sources and sinks. Based on the GCM results, we estimate the vertical diffusion coefficients, $\Kzz$, for different tracers  if the global-mean vertical transport  is approximated as diffusion. Finally we discuss how to analytically estimate the $\Kzz$ at conditions relevant for BDs and isolated EGPs without running a GCM.

\subsection{Cloud tracers}
\label{tracer.cloud}

Cloud particles can be vertically transported in the stratified atmosphere against sedimentation via local overturning circulations driven  by the thermal perturbations. We present   representative  results using models  with $T_{\rm{eff}} = 1000$ K. Models  with different $T_{\rm{eff}}$ but the same forcing pattern show qualitatively similar results because the settling speeds depend primarily on assumed cloud particle size and gravity but only moderately on temperature. In a primary model, the following  setup is used: a surface gravity of 1000 $\rm{ms^{-2}}$, a storm timescale $\tau_s = 10^5$ s, a drag timescale $\tau_{\rm{drag}}=10^5$ s, and a forcing amplitude $s_f=5\times10^{-4}\;\kps$. However, here we use a longer rotation period of 15 hours and a higher horizontal resolution (C128) to better resolve the fine structures within the local Rossby deformation radius. This is because we found extra sensitivity of tracer transport to the horizontal resolution compared to jet formation, and very fine structures within a deformation radius are needed to be resolved in order to achieve convergent results in terms of tracer transport.  In the following models, the condensation pressure $p_{\rm cond}$ is set at 10 bars, below which the tracers are in a vapor form and above which they are subjected to sedimentation. We examine particle radius of $r_p=0.05$, 0.1, 0.2, 0.5, 1.0, 2.0 and 5.0 $\mu$m.

\begin{figure}

    \centering
    \includegraphics[width=0.8\columnwidth]{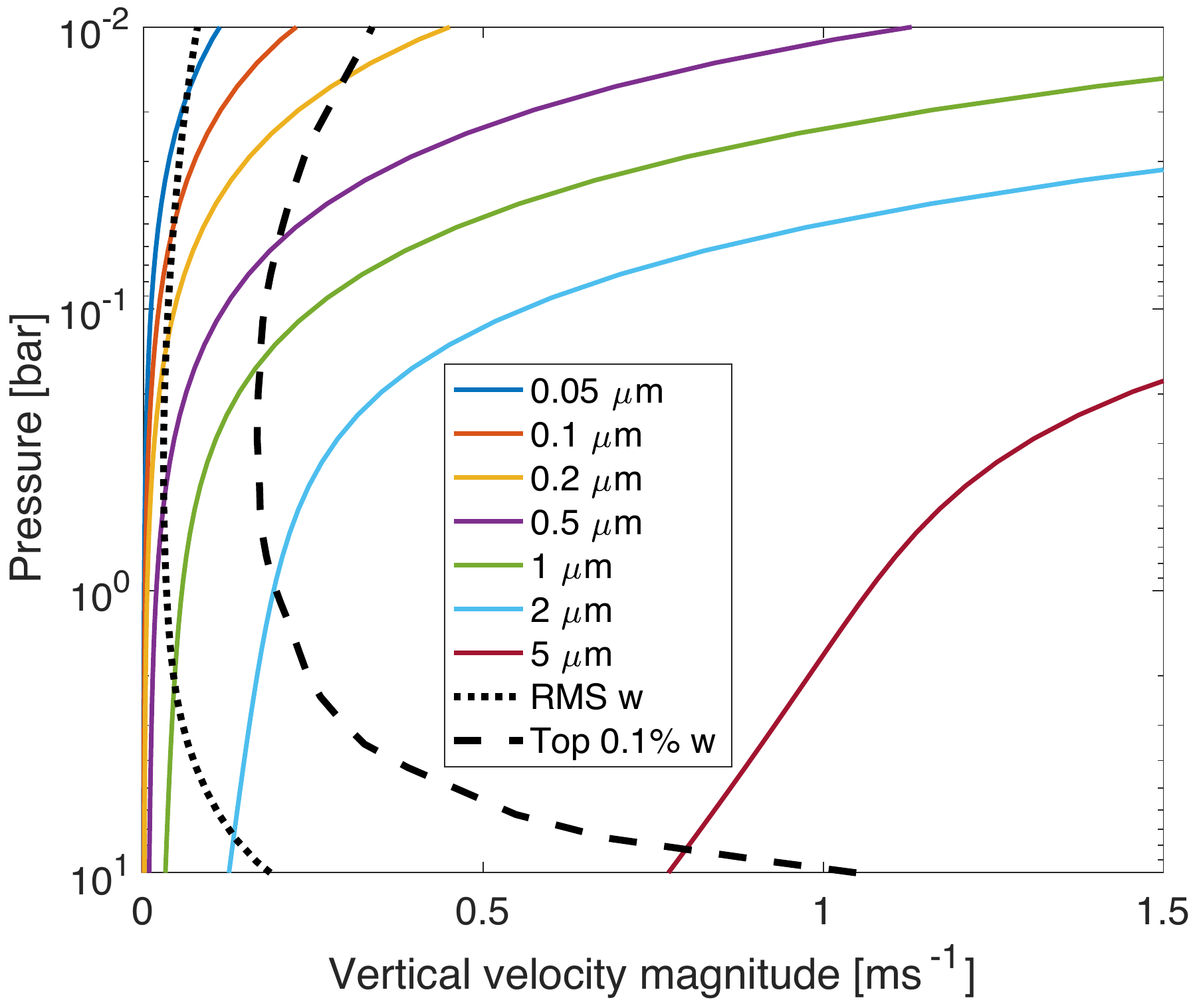}

    \caption{  Terminal settling speeds of particles with different size  as a function of pressure (solid lines), which are calculated using the equilibrium T-P profile with $\teff=1000$ K. {
    \btt Solid lines with different colors correspond to particles with different size from 0.05 to 5 $\mu{\rm m}$. } The dotted line is the isobaric RMS of vertical wind speed from the the model shown in Figure \ref{cloud1}, and the dashed line is a profile for the mean value of the top 0.1\% vertical wind speed.}
    \label{fig.settlevelocity}
\end{figure}

\begin{figure}      
\centering
\includegraphics[width=1.\columnwidth]{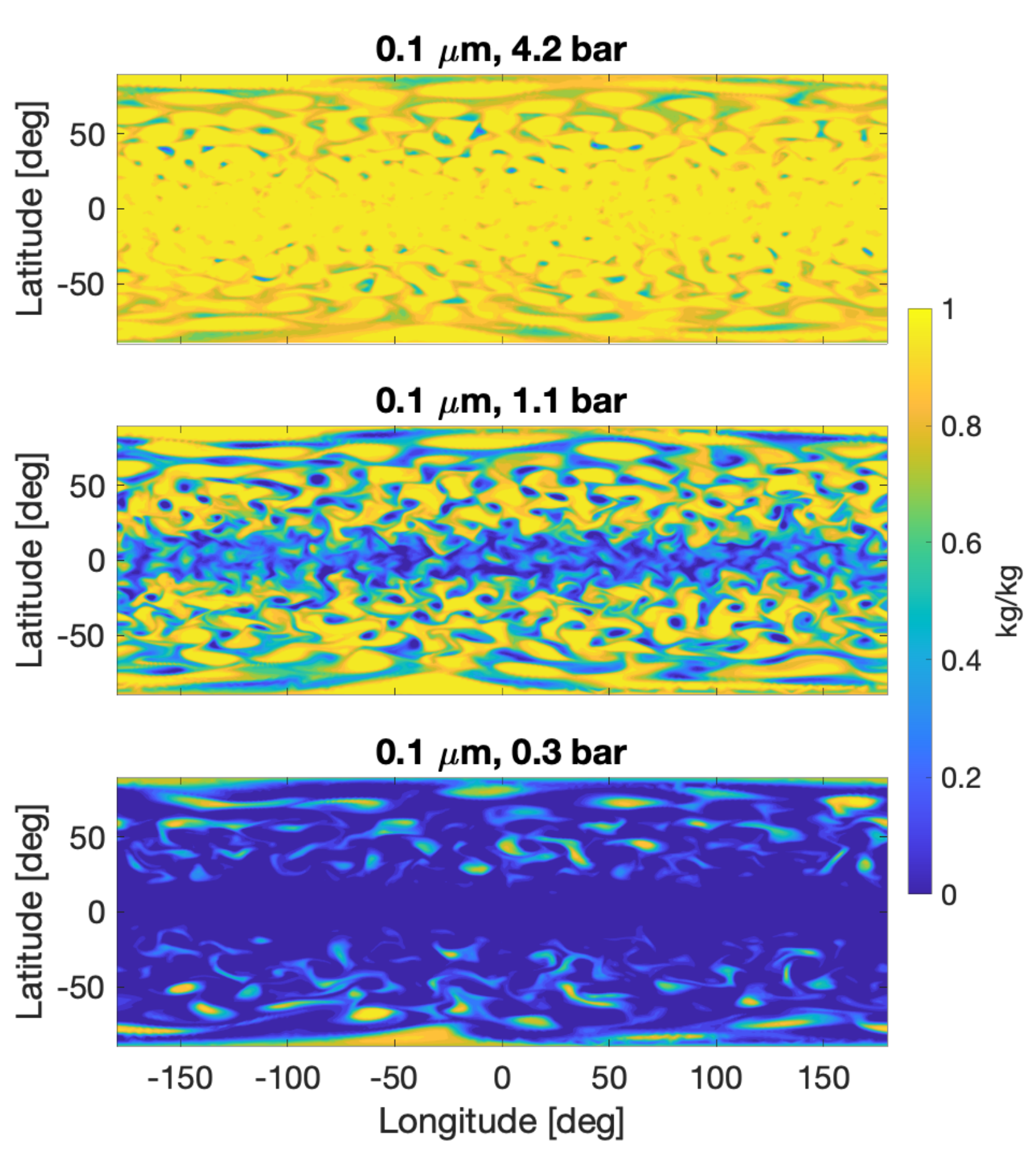}
\caption{Snapshots of cloud mixing ratio at several pressure levels indicated above each panel  for the particle size of 0.1 $\mu$m. Results are  from the  GCM shown in Figure \ref{cloud1}. }
\label{tracer2}
\end{figure}

With a given circulation, small particles are easily mixed up to   high altitudes relative to the condensation level due to low sedimentation speeds, while large particles sink down quickly and form vertically compact cloud layers. This is shown in Figure \ref{cloud1}, in which the colored contour panels show time-averaged zonal-mean mixing ratio of cloud tracers as a function of latitude and pressure (on a logarithmic scale) for particles with different size. Dashed lines are the condensation level. The global-mean mixing ratio as a function of pressure for each particle size is in the lower right panel. The 0.05-$\mu$m particle can extend roughly 4 to 5  scale heights from the condensation level, while the mixing ratio of 5-$\mu$m particle decrease significantly within about one  scale height above the condensation level. The cloud thickness smoothly decreases from small to large particle size.   Near the cloud top, cloud abundance drops rapidly with increasing altitude due to the sensitivity of settling velocity to pressure at low pressures\footnote{At low pressure the ratio of molecular mean free path to the size of cloud particles is larger than unity, and the terminal velocity  in this regime is inversely proportional to the pressure.}. The zonal-mean mixing ratio of clouds exhibits interesting meridional variation of the cloud top, in which clouds are usually mixed higher at mid-high latitudes than at low latitudes. This may be due to the weaker local overturning circulation at low latitudes, which is related to the  smaller temperature variation at low latitudes (see Figure \ref{fig.tu_drag}).

Pressures of global-mean cloud tops (here loosely defined as the pressure lower than which the global-mean tracer abundance is below $1\times10^{-8}$) roughly corresponds to where the particle settling speed exceeds the vertical wind speed. Figure \ref{fig.settlevelocity} shows the particle settling speeds as a function of pressure for different particle sizes. Profiles for the isobaric RMS vertical wind speed and the mean value  of the top 0.1\%  vertical speeds are also included for a comparison. For small particles with sizes $\lesssim0.2\;{\rm \mu m}$,  conjunctions between the RMS vertical wind speed and the particle settling speeds are very reasonable to describe the cloud top pressures (see a comparison between Figure \ref{fig.settlevelocity} the lower right panel in Figure \ref{cloud1}). For larger particles with sizes $\gtrsim2 \;{\rm \mu m}$, the the comparisons using the RMS vertical velocity obviously underestimate the cloud-top altitudes; and that  in the case with $5\;{\rm \mu m}$ even indicates that no clouds at all above 10 bars. Instead, the comparisons using the top 0.1\%  vertical speeds are more reasonable.

Cloud layers with smaller sizes are nearly horizontally homogeneous near the cloud base but show increasing horizontal inhomogeneity at altitudes closer to the cloud top. Figure \ref{tracer2} depicts  snapshots of the cloud mixing ratio at different pressure levels  for a particle size 0.1 $\mu$m.   At relative low pressures, the cloud pattern  exhibits strong patchiness with typical shape and length scales  strongly related to the regional-scale turbulence structure.  However,  larger particles ($\gtrsim 0.5~ \mu$m) show patchiness even close to the condensation level due to their large settling speed.  Despite  that the degree of patchiness differs, the shapes of cloud patterns are consistent between different pressures, indicating  vertically coherent tracer transport from the cloud base up to the cloud top.    The patchiness naturally originates from the vertical coherence of the overturning circulation. The atmosphere at sufficiently low pressures is  cloud-free because clouds must eventually settle out due to the increasing settling velocity with decreasing pressure. The overturning circulation transport the cloud-free air to high pressure with only moderate ``contamination'' from the surrounding cloudy area.  If the cloud is thick as those with small sizes, the lateral mixing has sufficient time to reduce the horizontal variation between the cloud-free downdraft and  cloudy updraft. Otherwise for those with large sizes, the inhomogeneity is easily achieved by the overturning circulation.

\begin{figure}      
\centering
\includegraphics[width=0.8\columnwidth]{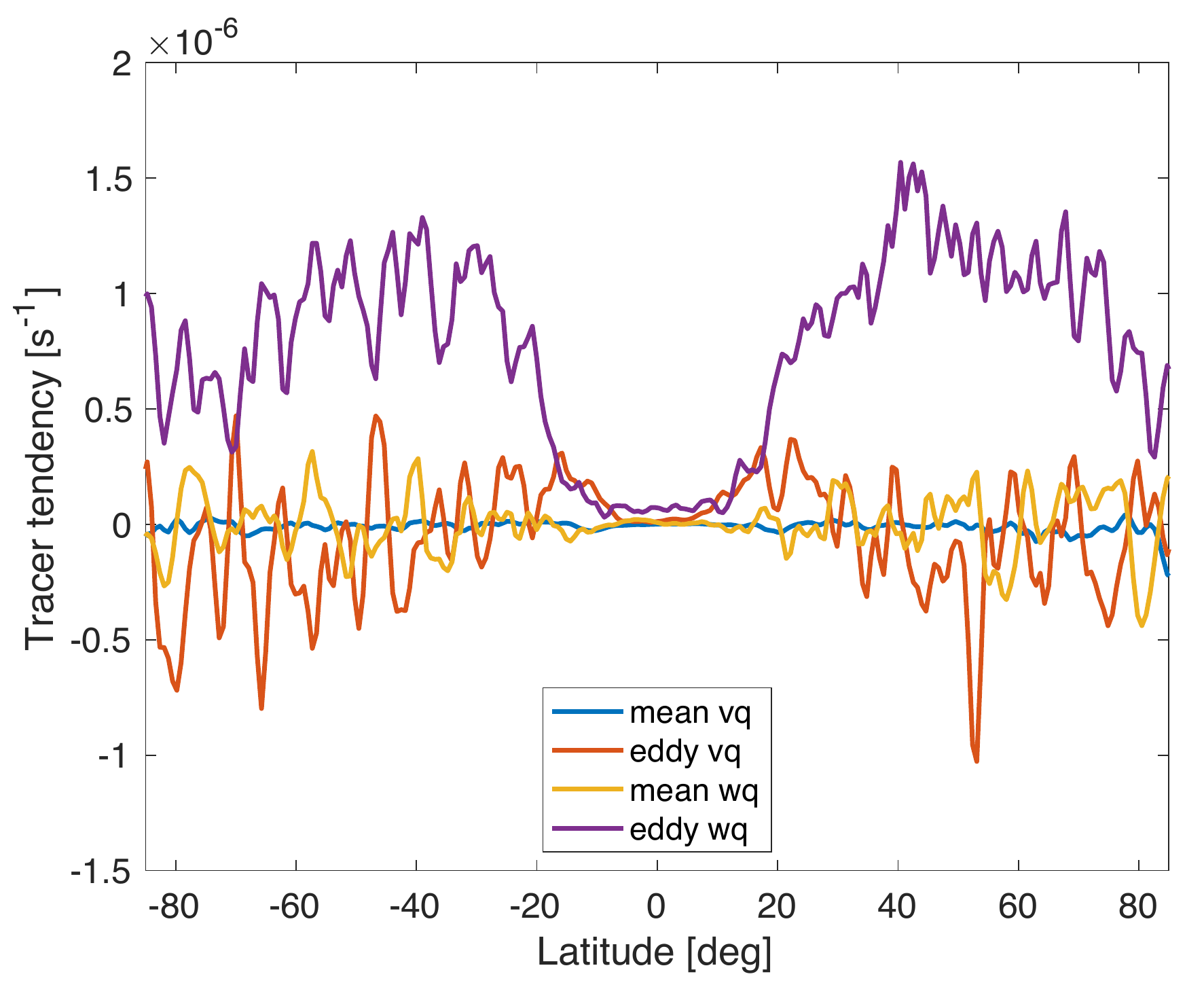}
\caption{The mean meridional transport, mean vertical transport, eddy meridional  transport and eddy vertical  transport terms for the zonal-mean tracer budget (see Eq. \ref{eq.zonalq})  as a function of latitude at  1 bar for a cloud tracer with a size of 0.2 $\mu$m. The result is time-averaged after the model reaching a statistical equilibrium. 
}
\label{fig.transport}
\end{figure}

\begin{figure}      
\centering
\includegraphics[width=1\columnwidth]{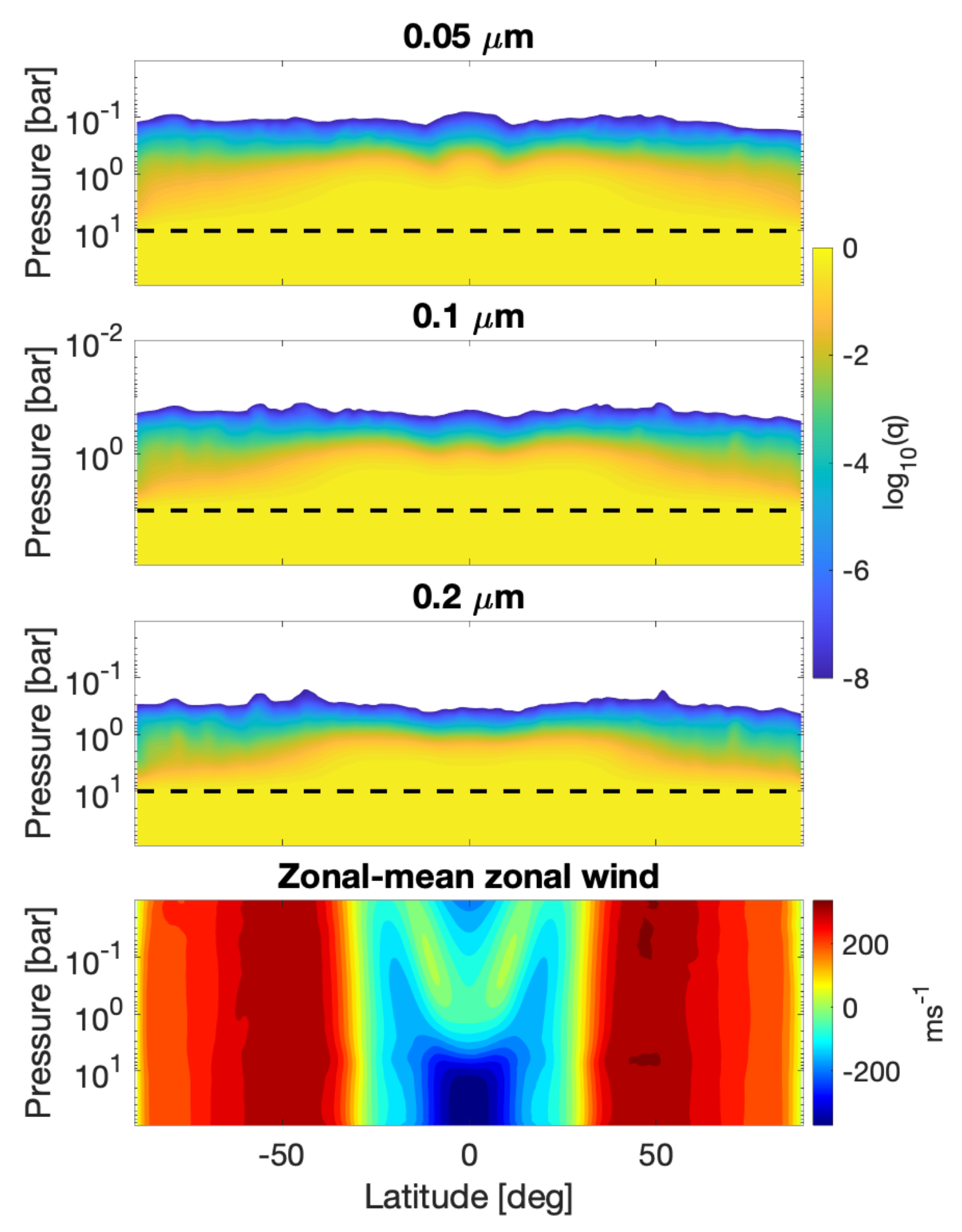}
\caption{Results from the model with a weak drag of $\tdrag=10^7$ s and other parameters the same as that shown in Figure \ref{cloud1}. \emph{Top three panels: } time-averaged zonal-mean tracer abundances for cloud particles with sizes of   0.05, 0.1 and 0.2 ${\rm \mu m}$. \emph{Bottom panel:} instantaneous zonal-mean zonal wind as a function of latitude and pressure. These results are obtained after the model reaching a statistical equilibrium.}
\label{fig.3tracer}
\end{figure}

The cloud patterns are dominated by local-scale overturning circulations rather than the zonal-mean meridional circulation. This is easily visualized in Figure \ref{tracer2} and   we quantify it here. The circulation can be decomposed to  zonal-mean and eddy (relative to the zonal mean) components, and the zonal-mean transport of a tracer $\overline{q}$ is written in pressure coordinates as follows:
\begin{equation}
    \frac{\partial \overline{q}}{\partial t} = -\frac{\overline{v}}{a}\frac{\partial\overline{q}}{\partial \phi} - \overline{\omega}\frac{\partial\overline{q}}{\partial p} 
    -\frac{1}{a\cos\phi}\frac{\partial(\overline{q'v'\cos\phi})}{\partial\phi} - \frac{\partial(\overline{q'\omega'})}{\partial p} + \overline{S}
    \label{eq.zonalq}
\end{equation}
where $\overline{A}$ represents a zonal mean of quantity $A$, $a$ is the planetary radius, $\phi$ is latitude, $\overline{v}$ is the zonal-mean meridional velocity, $\overline{\omega}$ the zonal-mean vertical velocity,  $v'$ and $\omega'$ are the eddies, and $\overline{S}$ represents the sinks. 
The vertical transport of clouds are primarily driven by the eddy vertical winds (represented by the term $\partial(\overline{q'\omega'})/\partial p$) rather than the zonal-mean circulation. Figure \ref{fig.transport} shows the above several terms as a function of latitude at  1 bar for a cloud tracer with a size of 0.2 $\mu$m. The result is time-averaged after the model reaching a statistical equilibrium.  The vertical eddy transport dominate the overall local zonal-mean tracer transport, and the mean vertical and eddy horizontal transport play secondary roles. When we integrate over latitudes to obtain the global mean vertical transport, only the vertical eddy transport is essential, which is balanced by the gravitational settling. The vertical eddy transport show minimum near the equator, corresponding to the smaller tracers profiles near the equator. 

We also perform a model with a weak-drag model of $\tau_{\rm{drag}}=10^7$ s and other parameters the same as the above.  Figure \ref{fig.3tracer} shows the time-averaged zonal-mean tracer mixing ratio for particles with sizes of 0.05, 0.1 and 0.2 $\mu$m in the upper three panels, and the instantaneous zonal-mean zonal wind in the bottom panel. Strong westward jets develop near the equator and   eastward jets are at high latitudes with speeds exceeding $\sim 300 ~\rm{ms^{-1}}$.  The cloud thickness is roughly flat in between $\sim\pm30^{\circ}$ latitudes and then decreases poleward.  The latitudinal change of cloud thickness occurs at where the zonal-mean zonal wind changes from westward to eastward. This differs from the strong-drag case, in which cloud thickness monotonically increases poleward. A careful comparison between Figure \ref{cloud1} and \ref{fig.3tracer} suggests that the overall vertical mixing is  weaker in the weak-drag model compared to the strong-drag model, and is primarily contributed by the differences poleward of $\sim\pm30^{\circ}$. The strong  barotropic mid-latitude jets do not directly generate vertical tracer transport but instead weakens the horizontal temperature variation by increasing horizontal advection.  The reduce of horizontal temperature variation  then  weakens the local overturning circulation and the vertical mixing of tracers.  {\btt We also perform a set of similar experiments in models with a cooler effective temperature of $T_{\rm eff}$ = 500 K. Still, we see an overall weaker mixing in the weak drag model where strong zonal jets develop. This is not surprising because mixing is driven by the local overturning circulation. The strong jets tend to reduce the local temperature variation by advection, and these weak-drag models show less vigorous local eddy velocities and so less vertical mixing. }

\begin{figure}      
\begin{subfigure}{0.45\textwidth}
\centering
\includegraphics[width=0.8\columnwidth]{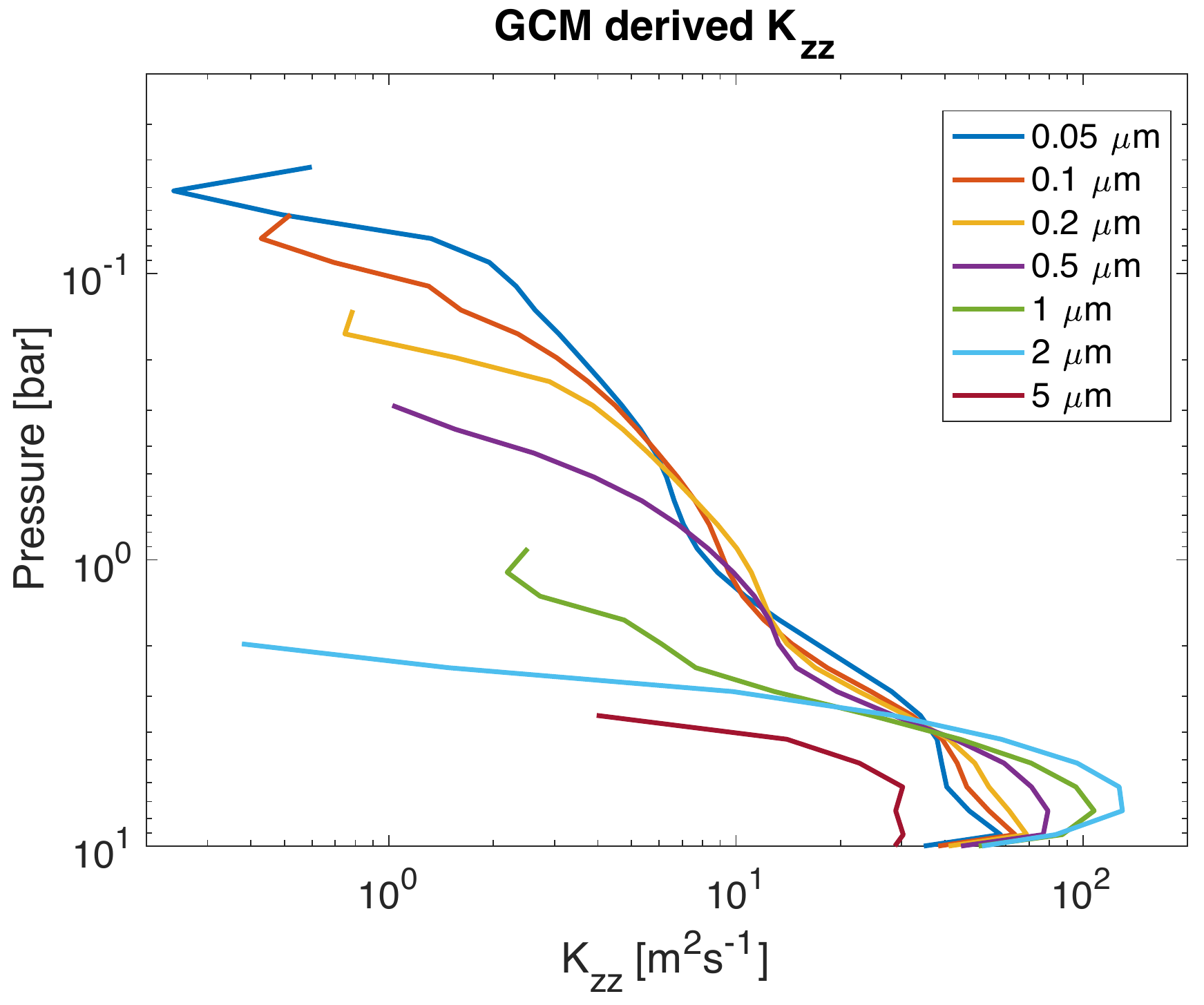}
\end{subfigure}

\medskip
\begin{subfigure}{0.45\textwidth}
    \centering
    \includegraphics[width=0.8\columnwidth]{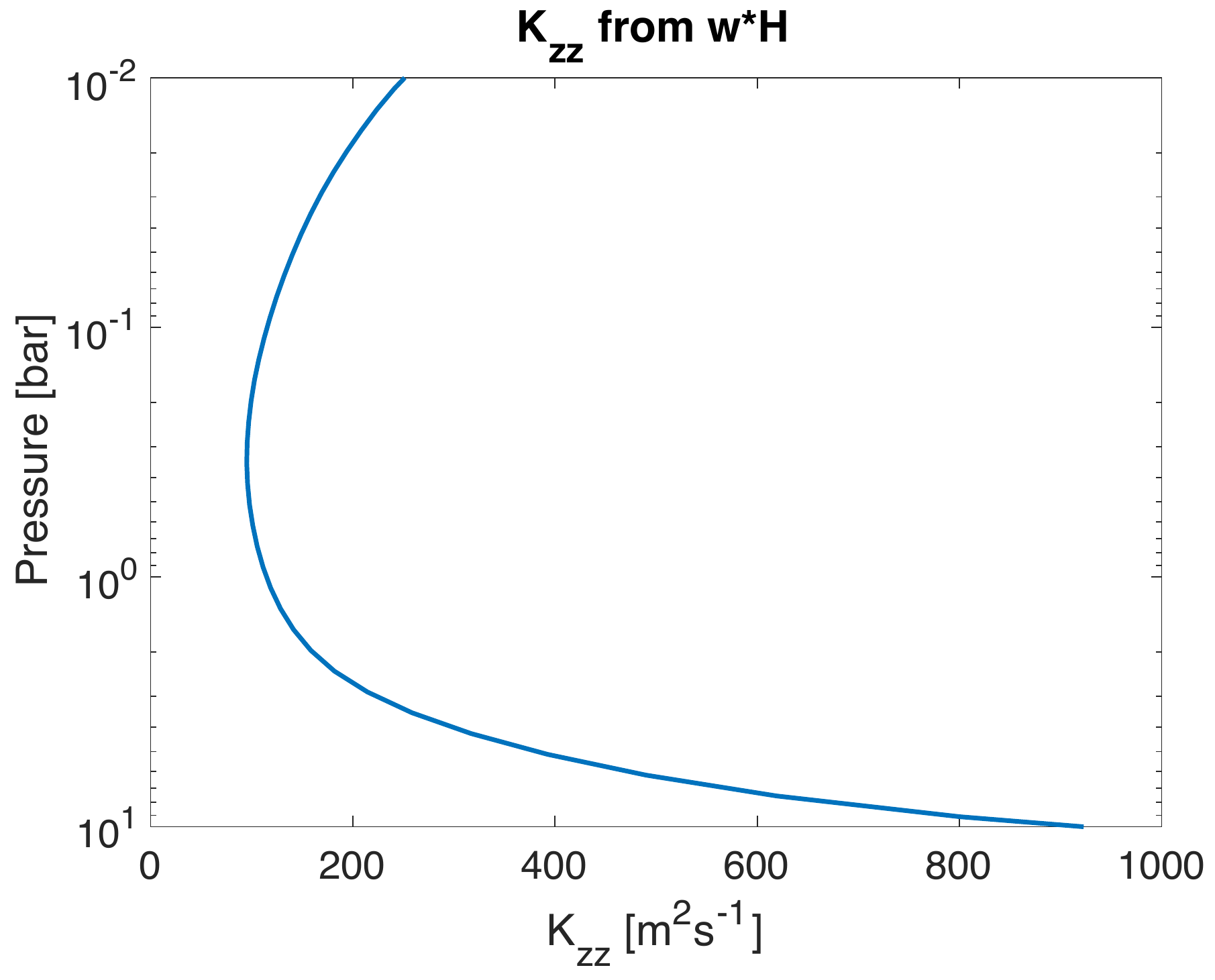}
\end{subfigure}

\caption{\emph{Top panel:} derived vertical diffusion coefficients $\Kzz$ for clouds with different sizes as a function of pressure from GCM using Eq. (\ref{eq.fluxkzz}). \emph{Bottom panel:} $\Kzz$ obtained from $wH$ where $w$ is the isobaric RMS vertical wind speed and $H$ is the scale height.
}
\label{plot.kzz}
\end{figure} 

1D atmospheric models often treat global-mean tracer transport as a diffusion process with a prescribed profile of the diffusion coefficient $\Kzz$ (e.g., \citealp{ackerman2001,moses2014,charnay2018,woitke2020}). In the stratospheres of BDs and isolated young EGPs, the source of $\Kzz$  has been interpreted as small-scale internal gravity wave mixing (e.g., \citealp{freytag2010}). In this study, we have shown that large-scale circulation serves as an alternative tracer transport mechanism. To help guiding 1D models on what reasonable $\Kzz$ profiles should be from large-scale dynamics,  we may derive  the vertical diffusion coefficients for tracers from our GCM results assuming that the global-mean transport is described as a diffusion process. The basic procedure is to diagnose the isobaric tracer flux in the GCM and relate that to the vertical mean tracer gradient \citep{parmentier2013,zhang2018a}:
\begin{equation}
\langle w q' \rangle  = -\Kzz \left\langle \frac{\partial q}{\partial z} \right\rangle,
\label{eq.fluxkzz}
\end{equation}
where the brackets represent global and time averaging, $z=-H\log (p/p_{\ast})$ is the log-pressure coordinate, $H$ is the scale height, $p_{\ast}$ is a reference pressure and $w$ is the vertical velocity in $z$-coordinates.

\begin{figure*}

    \begin{subfigure}[t]{0.33\textwidth}
    \centering
    \includegraphics[width=0.95\columnwidth]{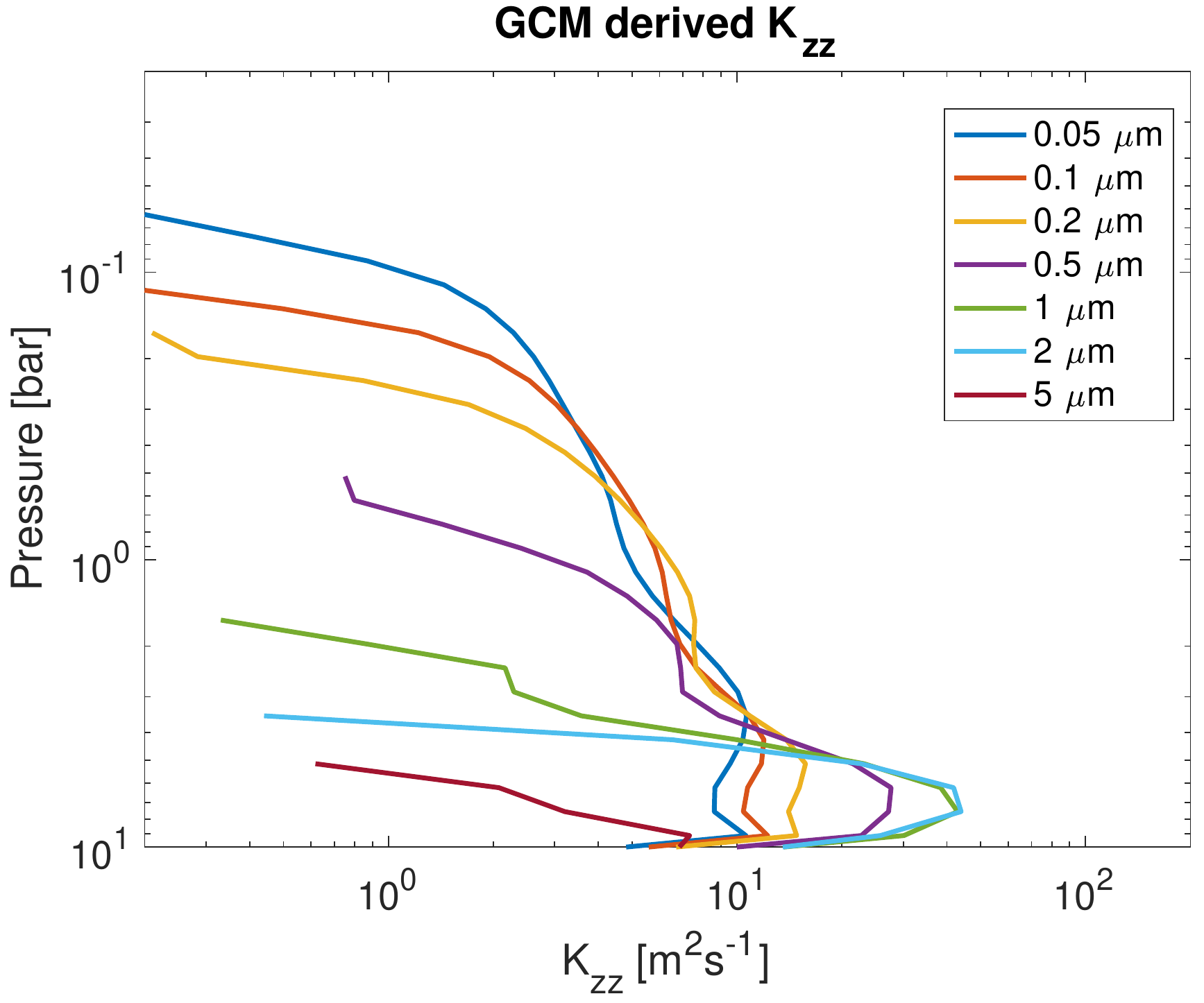}
    \end{subfigure}
    \begin{subfigure}[t]{0.33\textwidth}
    \centering
    \includegraphics[width=0.95\columnwidth]{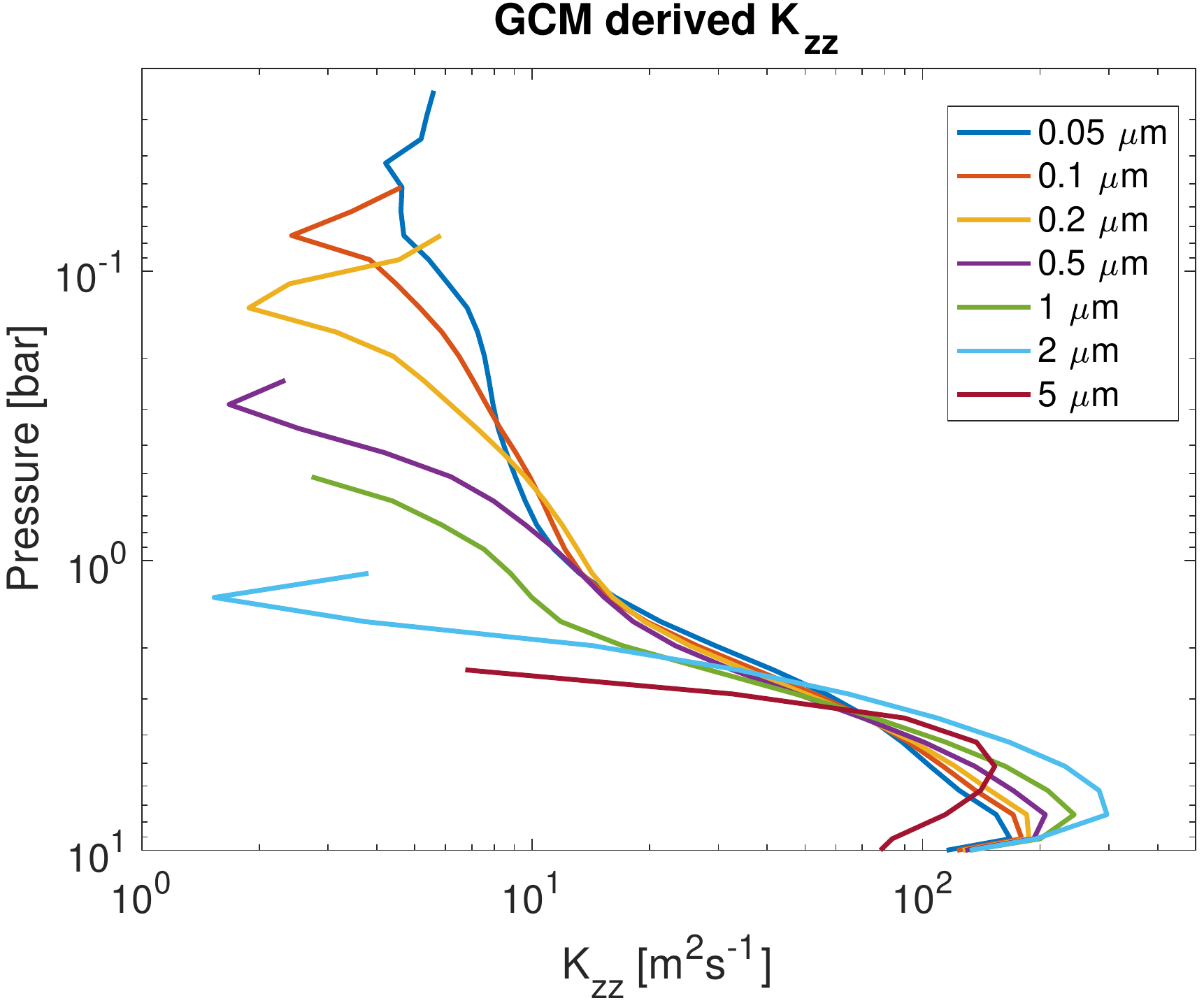}
    \end{subfigure}
    \begin{subfigure}[t]{0.33\textwidth}
    \centering
    \includegraphics[width=0.95\columnwidth]{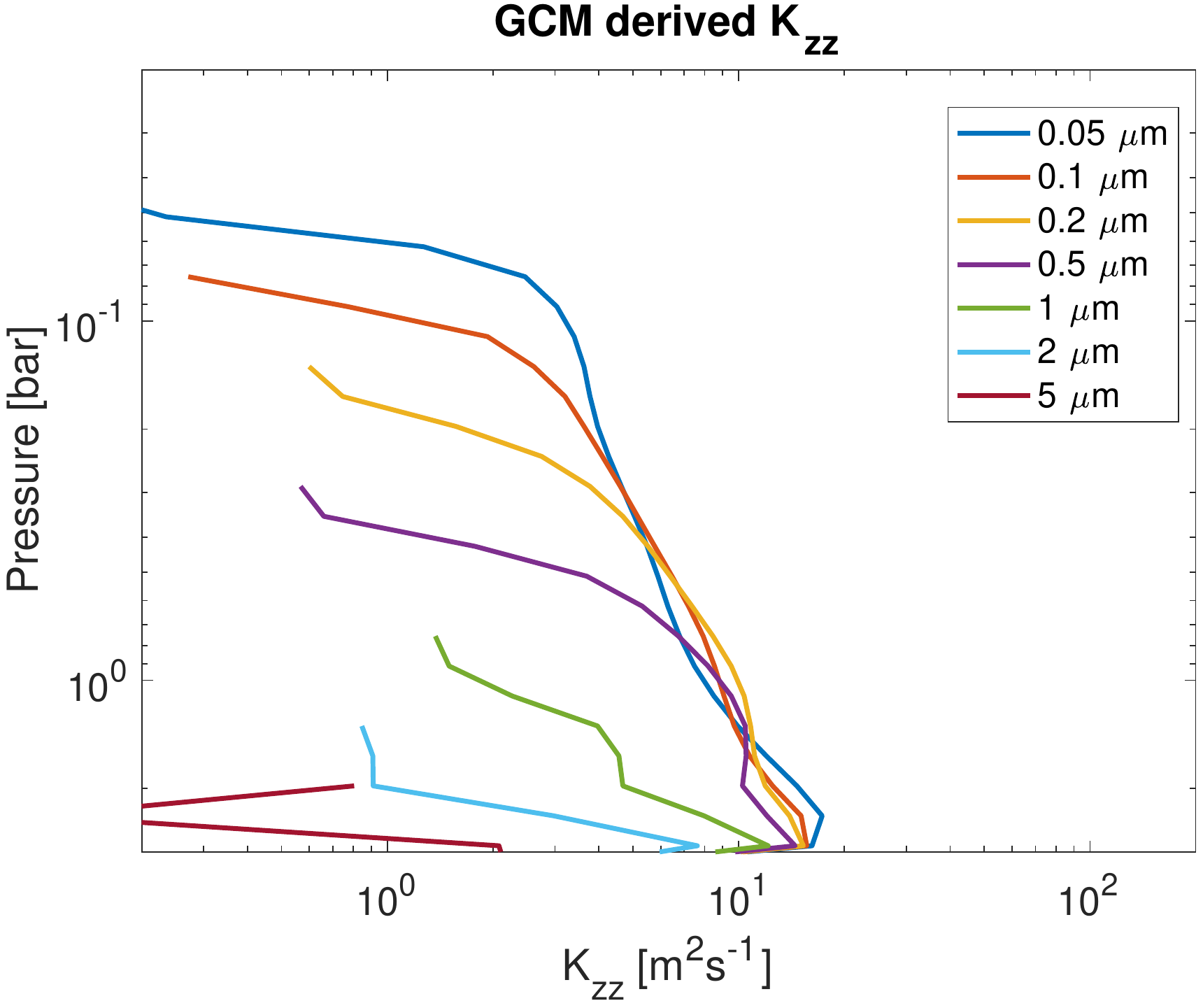}
    \end{subfigure}
    \caption{\emph{Left:} {\btt GCM derived $\Kzz$ for particles with different sizes from a model with half forcing amplitude as that shown in Figure \ref{plot.kzz}. \emph{Middle:} similar to the left panel but from a model with twice forcing amplitude as that shown in Figure \ref{plot.kzz}. \emph{Right:} GCM derived $\Kzz$ for particles with different sizes from a model with a condensation level at 3 bars but other parameters the same as that shown in Figure \ref{plot.kzz}.}
    }
    \label{fig.morekzz}
 \end{figure*}
 
The diffusion coefficients derived from our GCM as a function of pressure are shown in the top panel of Figure \ref{plot.kzz} for different particle size. The profiles at pressures with tracer abundances $<10^{-8}$ are truncated. First of all, the $\Kzz$ is different for different particle sizes. The $\Kzz$ is on the order of $10^2\;\mmps$ near the cloud base and generally decreases with decreasing pressure; for small particles the $\Kzz$ is on the order of $10\;\mmps$ at 1 bar. The decrease  is likely due to the decreasing vertical velocity with increasing altitudes (Figure \ref{fig.settlevelocity}). The $\Kzz$ rapidly become very small when approaching the cloud-top pressures. At pressures much larger than the cloud-top pressures, the $K_{\rm{zz}}$ for different particle sizes seem to be independent of particle size. This is especially the case for particles with 0.05 to 0.5 $\mu$m particles at between about 0.7 to 10 bars. The insensitivity of $\Kzz$ to particle size when settling speed is small compared to the flow vertical speed is similar to that found in hot-Jupiter GCMs \citep{parmentier2013,komacek2019vertical}. The $\Kzz$ derived based on the global-mean tracer flux using Eq. (\ref{eq.fluxkzz}) is about an order of magnitude smaller than that using a relation $\Kzz\approx wH$ where $H$ is a scale height (the latter is shown in the bottom panel of Figure \ref{plot.kzz}). The latter has been widely accepted in many 1D models or even interpretations of 3D models, but its validity has never been justified. We show that it leads to systematic overestimation of the $\Kzz$ compared to that based on true tracer fluxes. This agrees to the conclusion in \cite{parmentier2013} and \cite{komacek2019vertical} for hot Jupiter atmospheres.

{\btt We also perform models with slightly different parameters than the one shown in Figure \ref{plot.kzz}, and their derived $\Kzz$ profiles are shown in Figure \ref{fig.morekzz}. In two models, we use half and twice forcing amplitudes as the original model but with other parameters the same, and their $\Kzz$ profiles are shown in the left and middle panels of Figure \ref{fig.morekzz}.  As expected, the $\Kzz$ generally increases with increasing forcing amplitudes, but the dependence of $\Kzz$ to forcing amplitudes seems to be slightly steeper than a linear relation. In another model, we set the condensation pressure $p_{\rm cond}$ to 3 bars (instead of 10 bars in the original model) with other parameters the same. The resulting $\Kzz$ profiles are shown in the right panel, and  they generally agree well with those shown in Figure \ref{plot.kzz}. Overall, these experiments show a common property that when particle settling speed is negligible compared to atmospheric vertical speed, the $\Kzz$ seems independent of particle size.
}

\subsection{Chemical tracers}

\begin{figure}

\begin{subfigure}{0.45\textwidth}
    \centering
    \includegraphics[width=0.8\columnwidth]{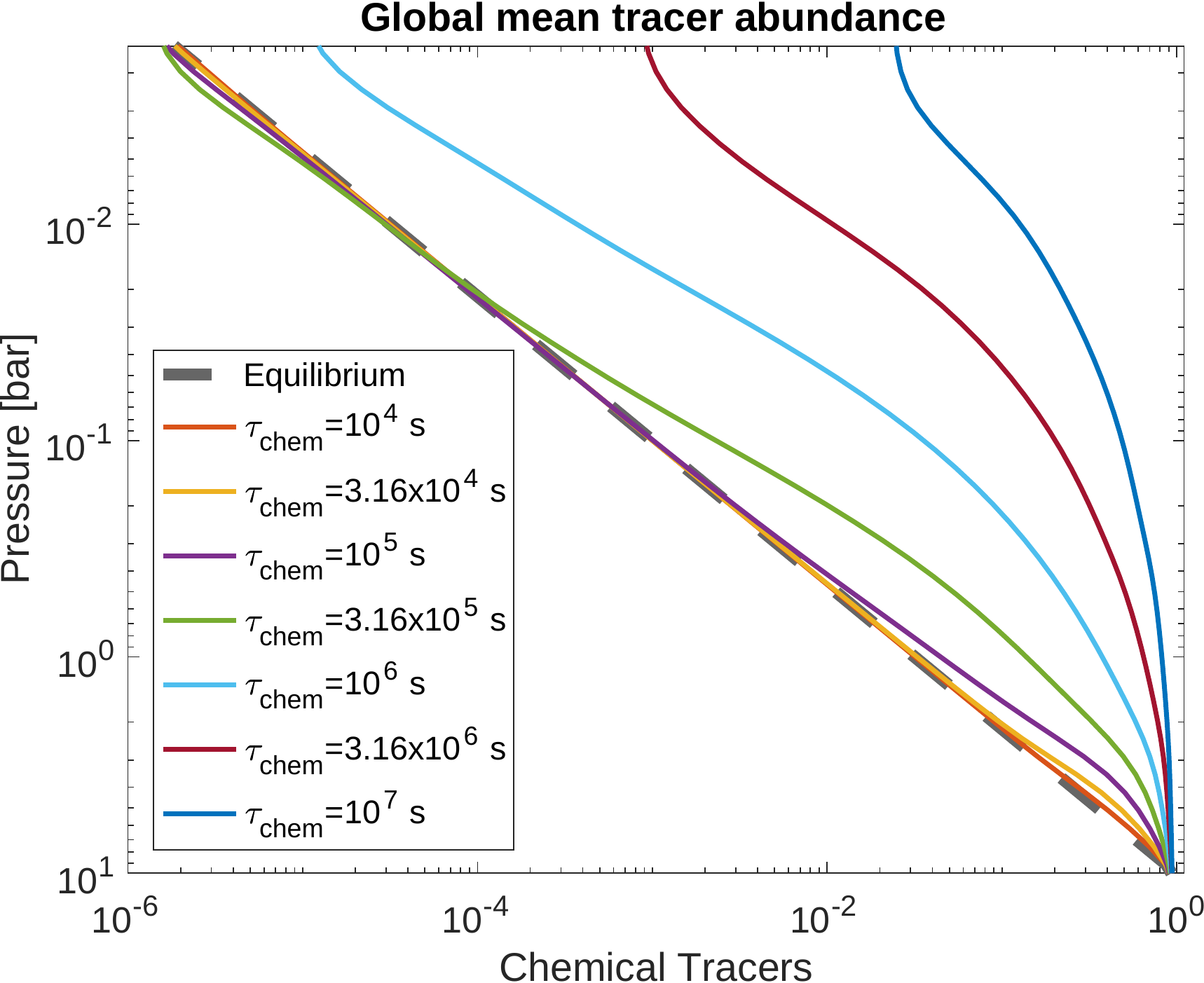}
\end{subfigure}

\medskip
\begin{subfigure}{0.45\textwidth}
    \centering
    \includegraphics[width=0.8\columnwidth]{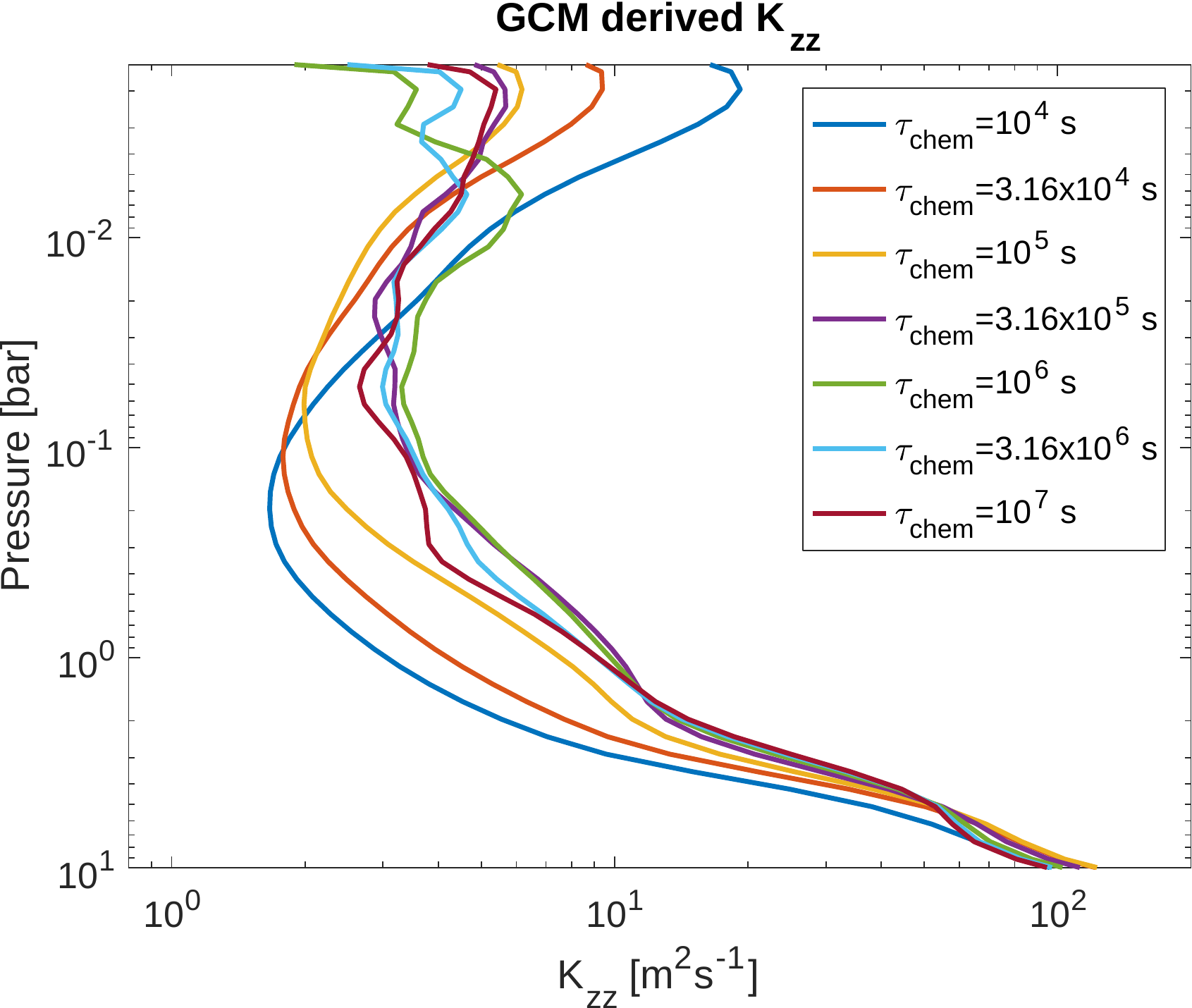}
\end{subfigure}

    \caption{\emph{Top panel:} globally averaged  abundances for tracers with different chemical timescales $\tau_{\rm chem}$ that are independent of pressure. Planetary parameters for this model are the same as those shown for Figure \ref{cloud1}. The dashed line is the  reference tracer profile. \emph{Bottom panel:} GCM-derived vertical diffusion coefficients $\Kzz$ using Eq. (\ref{eq.fluxkzz}) for tracers shown in the top panel. }
    \label{fig.chem4}
\end{figure}

\begin{figure}      
\centering
\includegraphics[width=0.8\columnwidth]{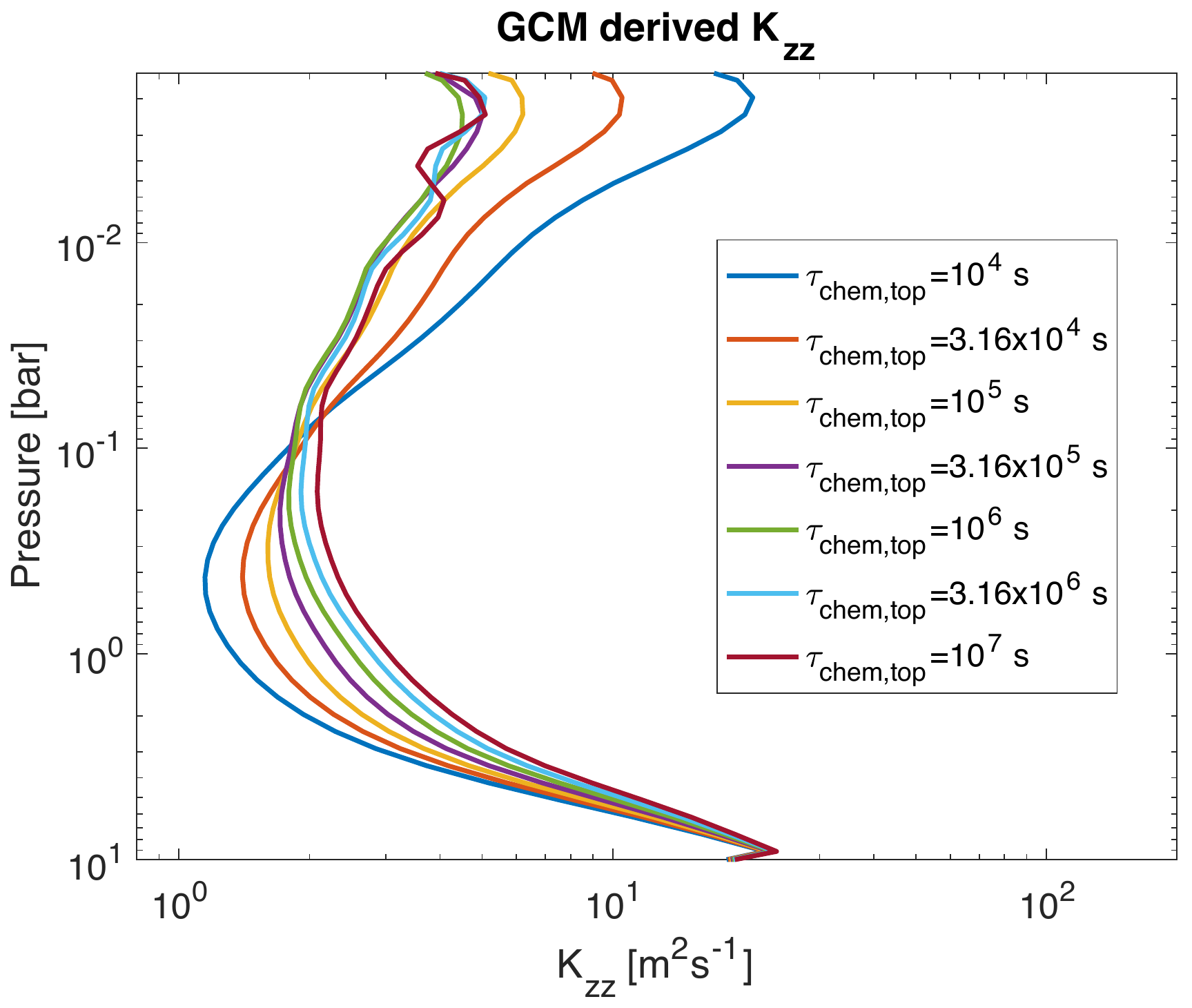}
\caption{GCM-derived vertical diffusion coefficients $\Kzz$ for chemical tracers with a  chemical timescale of $\tau_{\rm chem,bot}=10^3$ s at 10 bars  and $\tau_{\rm chem,top}$ varying from $10^4$ to $10^7$ s at the top boundary. Planetary parameters for this model are the same as those shown for Figure \ref{cloud1}. }
\label{plot.chem3}
\end{figure}

Chemical tracers with equilibrium profiles varying with pressure are subjected to vertical mixing and can  potentially be in disequilibrium depending on the chemical timescales relative to the dynamical timescales. Using the strong-drag GCM with the same parameters as that for the passive clouds, we first show the mixing of hypothetical chemical tracers with several chemical timescales of $10^4$, $3.16\times10^4$, $10^5$, $3.16\times10^5$,$10^6$, $3.16\times10^6$ and $10^7$ s that are independent of pressure. The numerical implementation of idealized chemical tracers is described in Section \ref{numeric.tracers}. The equilibrium profile is the same for all tracers and is shown as the dashed line in the top panel of Figure \ref{fig.chem4}. The time-averaged global-mean tracer profiles as a function of pressure are also shown as solid lines. 
Tracers with $\tau_{\rm chem}\lesssim10^5$ s  are close to their prescribed equilibrium at almost all pressures. Tracers with $\tau_{\rm chem}\gg 10^5$ s exhibit obvious deviation to their equilibrium profiles. However, their global-mean abundances still decrease with decreasing pressure. This is somewhat different to chemical mixing using convection resolving models in \cite{bordwell2018}, in which the mean  tracers are quenched, i.e., above a  certain level the abundance is independent of height and is determined by the value  near a level where the quenching occurs. The horizontal tracer patterns are considerably  inhomogeneous, showing isobaric tracer variations the same magnitude as the isobaric mean values. The spacial distributions of the variations are similar to those shown for cloud tracers (Figure \ref{tracer2}) and are primarily driven by local overturning circulations. Similarly, these horizontal structures are coherent at different pressures.

We performed a similar exercise to diagnose the vertical diffusion coefficient $\Kzz$ for the global-mean chemical tracers. The bottom panel of Figure \ref{fig.chem4} shows the resulting GCM-derived $\Kzz$ profiles for these tracers. The $\Kzz$ profiles with different chemical timescales show little difference at large pressures near 10 bars but moderate differences up to a factor of several at pressures lower than a few bars. The $\Kzz$ is on the order of $10^2\mmps$ near 10 bars then decreases with decreasing pressure (less than $10\mmps$ at pressures $<1$ bar) partly due to the decrease of vertical wind speed. There seems to be a trend that the $\Kzz$ is lower for tracers with shorter chemical timescales, and this is theoretically expected as will be shown below.

We also show a model with chemical timescales increases with decreasing pressure. The detailed implementation is described in Section \ref{numeric.tracers}. The chemical timescale at 10 bars is fixed at $\tau_{\rm chem,bot}=10^3$ s and that at the top boundary adopt several values as $\tau_{\rm chem,top}=10^4$, $3.16\times10^4$, $10^5$, $3.16\times10^5$,$10^6$, $3.16\times10^6$ and $10^7$ s. These tracers show much smaller deviations  to their equilibrium profiles  (not shown) compared to those with constant chemical timescales. The resulting $\Kzz$ profiles for the global-mean transport are shown in Figure \ref{plot.chem3} as a function of pressure. Their profiles are qualitative similar to but meanwhile are systematic lower than those with constant chemical timescales (bottom panel of Figure \ref{fig.chem4}).

\subsection{Analytic estimation of  $K_{zz}$}
Building upon  \cite{holton1986}, \cite{zhang2018a} presented a theory to derive the  vertical diffusion coefficient if the global-mean tracer transport is described as a diffusion process. In a regime in which the tracers are relatively short-lived with  equilibrium profiles depending on pressure along, the $K_{zz}$ can be approximated as 
\begin{equation}
    K_{zz} \approx \frac{\mathcal{W}^2}{\tau_d^{-1} + \tau_{\rm chem}^{-1}},
    \label{eq.kzz}
\end{equation}
where $\mathcal{W}$ is the magnitude of vertical wind speed. $\tau_d$ is a horizontal mixing timescale  characterized by $L_h/U$ where $L_h$ is a characteristic horizontal length scale over which distinctive tracer columns are mixed. \cite{komacek2019vertical} subsequently derived similar formula for vertical mixing in hot Jupiters' atmospheres. 

Here we apply this theory to explain the GCM-derived $\Kzz$ values, and provide  estimates of the theoretical $K_{zz}$ at conditions appropriate for BDs and young EGPs without  running GCMs. Our goal is to find appropriate scaling relations for the vertical wind speed and horizontal mixing timescale given a set of planetary parameters.  Eq. (\ref{eq.kzz}) can be applied to chemical tracers straightforwardly. However, the  diffusion coefficient for cloud tracers, which differ to chemical tracers in the sink and source, was not discussed in \cite{zhang2018a} and \cite{komacek2019vertical}. Now we  derive an expression of $\Kzz$ for cloud tracers. Following \cite{zhang2018a}, we write the tracer equation for clouds equivalent to Eq. (10) in \cite{zhang2018a} as 
\begin{equation}
    w \frac{\partial \langle q \rangle}{\partial z} + \frac{q'}{\tau_d} \approx \frac{1}{\rho}\frac{\partial(\rho q'V_{\rm term})}{\partial z} \approx q'\frac{V_{\rm term}}{H} + q'\frac{\partial V_{\rm term}}{\partial z} + V_{\rm term}\frac{\partial q'}{\partial z},
    \label{eq.cloud1}
\end{equation}
where $q'$ is the isobaric deviation to the global mean,  $V_{\rm term}$ is the terminal velocity in $z$ coordinates and $H$ is the scale height.  Multiplying Eq. (\ref{eq.cloud1}) with $w$ and globally averaging, we  obtain
\begin{equation}
    \langle wq'\rangle \left(\tau_d^{-1} - \frac{V_{\rm term}}{H}-\frac{\partial V_{\rm term}}{\partial z} \right) + \langle w^2 \rangle\frac{\partial \langle q\rangle}{\partial z}-V_{\rm term} \langle w\frac{\partial  q'}{\partial z}\rangle=0.
    \label{eq.qeddyflux_0}
\end{equation}
Our goal is to derive a relation of $\langle wq' \rangle$ to the vertical mean tracer gradient, and then $K_{zz}$ can be estimated without knowing the distribution of tracers {\it a priori}. When $V_{\rm term}\ll \mathcal{W}$, the last term in the left-hand side of Eq. (\ref{eq.qeddyflux_0}) is negligible and the relation of $\langle wq' \rangle$  and $\frac{\partial \langle q\rangle}{\partial z}$ is straightforward. When $V_{\rm term}$ is non negligible, clouds usually exhibit significant horizontal inhomogeneity due to the coherent vertical circulation. In this situation, we argue that $ \langle w\frac{\partial  q'}{\partial z}\rangle \sim \mathcal{W}\frac{\partial \langle q\rangle}{\partial z}$.\footnote{Consider two types of atmospheric columns with equal area to represent global mixing. As the correlation between vertical velocity and abundance anomaly is expected to be positive, one type of columns is upwelling with a vertical velocity $\mathcal{W}$ and an eddy abundance $q^+$ and the other is downwelling with a vertical velocity $-\mathcal{W}$ and an eddy abundance $-q^+$. The average flux is $ \langle w\frac{\partial  q'}{\partial z}\rangle = (\mathcal{W}\frac{\partial  q^+}{\partial z}-\mathcal{W}\frac{\partial  (-q^+)}{\partial z})/2$. As clouds are quite inhomogeneous, we  assume $q^+\sim \langle q\rangle$, and thus, we expect that $ \langle w\frac{\partial  q'}{\partial z}\rangle \sim \mathcal{W}\frac{\partial \langle q\rangle}{\partial z}$. }  Using this relation, Eq. (\ref{eq.qeddyflux_0}),  the equivalent eddy diffusion approximation Eq. (\ref{eq.fluxkzz}) and $\langle w^2 \rangle\approx\mathcal{W}^2$, the $K_{zz}$ for clouds is
\begin{equation}
    K_{zz}\approx \frac{\mathcal{W}(\mathcal{W}-V_{\rm term})}{\tau_d^{-1} - \frac{V_{\rm term}}{H}-\frac{\partial V_{\rm term}}{\partial z}}.
    \label{eq.kzzcloud}
\end{equation}
At levels where $V_{\rm term}\ll \mathcal{W}$ (and therefore $\frac{V_{\rm term}}{H}$ and $\frac{\partial V_{\rm term}}{\partial z}$ are also likely much smaller than $\tau_d^{-1}$), $K_{zz}$ is independent of particle size.  At levels where the settling speed approaches the vertical wind speed, $K_{zz}$ decreases rapidly and we expect that vertical cloud transport is ceased  above those levels. These properties agree well with the $\Kzz$ for clouds derived from the GCM.

Now we seek scalings for the vertical wind speed $\mathcal{W}$ and horizontal mixing timescale $\tau_d$. Vertical mixing at mid-to-high latitudes is representative for cases with either strong or weak drags, therefore we adopt the quasi-geostrophic dynamical properties there.  Thermal wind balance Eq. (\ref{thermalwind}) provides a reasonable estimate of the horizontal eddy wind speed, which reads in order of magnitude $U\sim\frac{\Delta TR\delta\ln p}{L_h\Omega}$ where $\Delta T$ is a characteristic isobaric temperature variation and $f\sim\Omega$. Temperature perturbations generated by convective penetration are expected to have vertical wavelengths on the order of the scale height, therefore eddies driven by these perturbation have $\delta\ln p\sim 1$. From the GCMs, we note that the eddy horizontal structures are mostly around the size of a deformation radius, and therefore $L_h$ can be approximated as $L_h\sim NH/\Omega$. Then, the characteristic horizontal mixing timescale is 
\begin{equation}
    \tau_d = L_h/U \sim \frac{N^2H^2}{\Omega \Delta TR}.
    \label{eq.taud}
\end{equation}
Applying the typical conditions of $\Delta T\sim 60$ to 100 K from our GCMs, we obtain $U$ of a few hundred $\mps$,  consistent with horizontal eddy wind speeds at mid latitudes (see Figure \ref{fig.tu_drag}). The resulting horizontal mixing timescale is typically on the order of $10^4$ s. For general conditions  of BDs and isolated young EGPs, the balance between convective perturbation and radiative damping may results in isobaric  temperature variation $\Delta T$ between 50 to 200 K (see the detailed argument in the appendix of \citealp{showman2019}), which is the typical range adopted in our analytic estimation.

Vertical motions are a result of horizontal convergence, and this can be decomposed into that by the ageostrophic motions and the geostrophic motions. The former scales as $U_{\rm ageo}/L\sim\mathcal{W}/H$ and the latter scales as $\beta U_{\rm geo}/f\sim U_{\rm geo}/a$, where $a$ is the planetary radius \citep{showman2019}. We may place an upper limit on the characteristic vertical wind speed due to convergence of the agostrophic wind using $\mathcal{W}/H\sim Ro U/L$, where $Ro=U/fL$ is the Rossby number and is much less than 1 at mid latitudes. The convergence due to ageostrophic motions is typically much greater than that by geostrophic motions in condition relevant here, and  the scacling for the vertical velocity can be written as
\begin{equation}
    \mathcal{W}\sim Ro\frac{UH}{L} \sim \frac{U^2H}{\Omega L^2} \sim    \frac{\Delta T^2 R^2\Omega}{N^4H^3}
    \label{eq.wmag}
\end{equation}
where we argue again that the relevant horizontal lengthscale is the deformation radius and $f\sim \Omega$. It can be shown that 
\begin{equation}
    N^2H^2 = RT\left(\kappa-\frac{\partial\ln T}{\partial\ln p}\right) \equiv \gamma RT\kappa,
    \label{eq.nh}
\end{equation}
where $\kappa=R/c_p$ and $\gamma=1$ for an isothermal atmosphere and $\gamma=0$ for an adiabatic atmosphere. For conciseness, we use a fixed value for $NH$ for a given T-P profile rather than evaluating $NH$ locally at each pressure levels. This is because eddies span certain vertical depths and a somewhat averaged $NH$ is more appropriate. Here we adopted a value $\gamma=0.8$.

\begin{figure}      
\centering
\includegraphics[width=0.8\columnwidth]{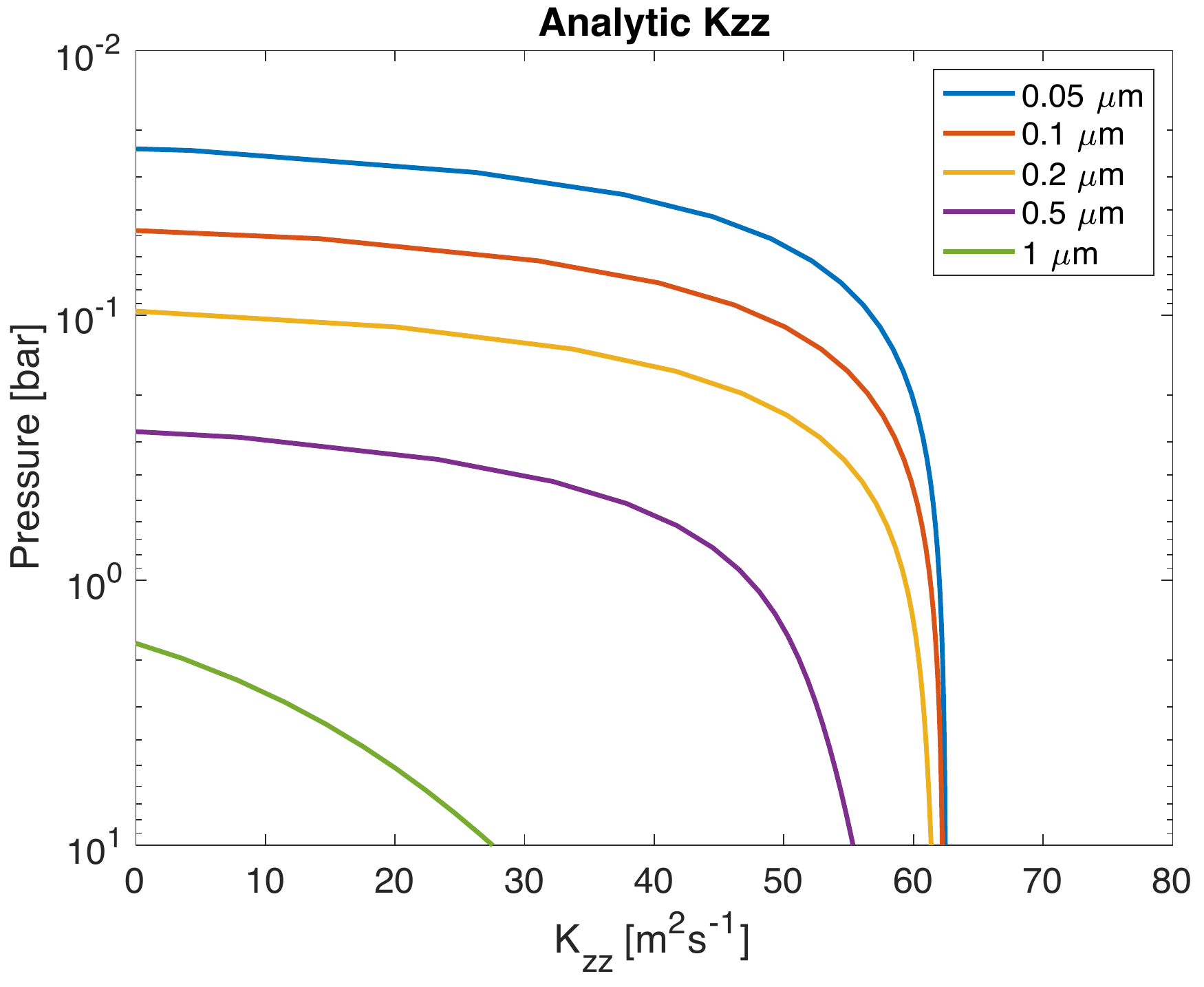}
\caption{Analytically derived $\Kzz$ profiles for clouds with different particle sizes using Eqs. (\ref{eq.kzzcloud}), (\ref{eq.taud}) and (\ref{eq.wmag}), and adopting $\Delta T=80$ K, $\Omega=1.16\times10^{-4}\;{\rm s^{-1}}$ and a mean temperature of 1100 K. The settling speed is calculated using a T-P profile of $\teff=1000$ K.
}
\label{fig.analytickzz}
\end{figure}

With the above scalings of Eqs. (\ref{eq.taud}) and (\ref{eq.wmag}), settling speed of particles given a T-P profile and the known chemical timescales, we are in the position to estimate the theoretical $K_{zz}$ for both chemical tracers using Eq. (\ref{eq.kzz}) and cloud tracers using Eq. (\ref{eq.kzzcloud}). Note that we do not intend to represent variations of vertical wind speeds and horizontal mixing as a function of height, and the resulting vertical dependence of analytic $\Kzz$ comes from the variation of tracer sink and source. Here we apply $\Delta T=80$ K, $\Omega=1.16\times10^{-4}\;{\rm s^{-1}}$ and a mean temperature of 1100 K relevant to GCMs in this section. For chemical tracers with local chemical timescales of $10^3$, $10^4$, $10^5$, $10^6$ and $10^7$ s,  the analytic $K_{zz}$ is about 2, 17, 49, 61 and 62 ${\rm m^2s^{-1}}$, respectively. The analytic $K_{zz}$ captures the order of magnitude behavior of the GCM results but cannot capture the pressure dependent variation of the GCM-derived $K_{zz}$ even in cases with  constant $\tau_{\rm chem}$. The strong dependence of $\Kzz$ on $\tau_{\rm chem}$ suggested that the variation of GCM-derived $\Kzz$ shown in Figures \ref{fig.chem4} and \ref{plot.chem3} is contributed by the variation of $\tau_{\rm chem}$.

For cloud tracers, Figure \ref{fig.analytickzz} shows the analytic $K_{zz}$ profiles as a function of pressure for particle sizes of 0.05, 0.1, 0.2, 0.5 and 1 ${\rm \mu m}$. Clouds with sizes of 2 and 5 ${\rm \mu m}$ have settling speeds greater than the estimated vertical wind speed, and therefore cannot be applied. For small particles, the $\Kzz$ is on the order between $10$ to $10^2\mmps$ at high pressures but rapidly decreases when it approaches cloud top levels.  The analytic $K_{zz}$ profiles again capture the overall behavior as well, including the agreement on the order of magnitude value and the behavior that the $\Kzz$ approaches very small values near the level where setting speed roughly match the mean vertical wind speed.  The theory of $K_{zz}$ by global circulation takes vast idealization of the actual dynamical system and its inaccuracy when comparing to GCM results  is not surprising \citep{zhang2018a}. Nevertheless, the theory captures some key properties and the order-of-magnitude estimation of $K_{zz}$ as a function of various physical parameters and  provides a guidance for parameterization of mixing for   1D models.

\section{Discussion and Summary}
\label{ch3.discussion}

\subsection{Effects of dynamics on the synthetic lightcurves}

\begin{figure}      
\includegraphics[width=1.\columnwidth]{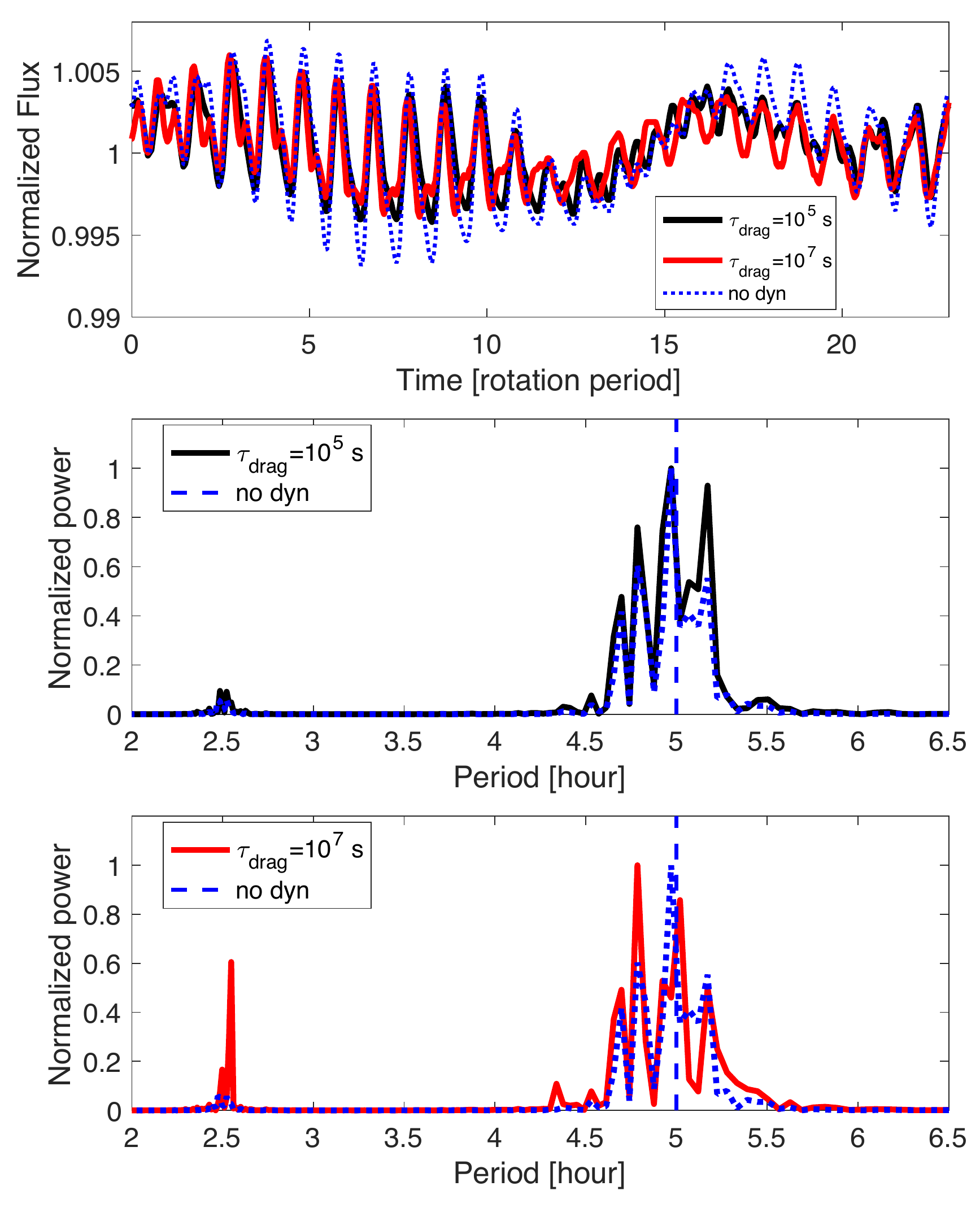}
\caption{\emph{Top panel:} normalized lightcurves from GCMs with most parameters the same as those shown in Figure \ref{fig.tu_drag}. Lightcurve in black line is from the model with $\tdrag=10^5$ s; the red line is with $\tdrag=10^7$ s; and the dotted line is that without dynamics. \emph{Middle panel:} normalized power spectrum (relative to the maximum power) of periodogram performed for the lightcurves with $\tdrag=10^5$ s and without dynamics.  \emph{Bottom panel:} same as the middle panel but with $\tdrag=10^7$ s and without dynamics. Lightcurves for the periodogram analysis span over 100 rotation periods. }
\label{fluxcurve}
\end{figure}

Recent long-term lightcurve observations of several BDs provide a valuable window to probe jets, waves, large-scale vortices and their time evolution in these atmospheres \citep{apai2017,apai2021}. The injected temperature perturbation patterns  by themselves alone  produce  lightcurve variability even without dynamics. The question is then how does dynamics shape the lightcurve on top of the forcing pattern?  Here we show simulated lightcurves viewed at the equator for models with a 5-hour rotation period, $T_{\rm{eff}} = 1000$ K, with $\tau_{\rm{drag}}=10^7$, $10^5$ s and without dynamics in the top panel of Figure \ref{fluxcurve}. The rotational modulations  are obvious in all cases, showing sinusoidal-like shapes with a period corresponding to rotation period of 5 hours.   The  variability amplitudes  are small ($\sim 1\%$), and those with dynamics show lower  amplitudes than that without dynamics.  Long-time evolution of the lightcurves  results from temporal variability in the temperature spatial patterns. We perform periodogram to the lightcurves that cover over 100 rotation periods and their nomalized power spectrum are shown in the middle and bottom panels of Figure \ref{fluxcurve}. Properties of the lightcurve from the strong-drag model show small difference to that without dynamics. This is  because with a  strong drag,  temperature patterns are close to the forcing patterns. The spectral power of the weak-drag model is overall quite similar but shows some quantitative differences to that without dynamics: the relative power of certain peaks near the rotation period is different, and perhaps more notably, the peak at 2.5 hour is exceptionally strong.  The weak-drag model  develops a global-scale, zonally propagating Rossby wave in each hemisphere and they induce certain temperature variations. The two waves have roughly the same zonal phase velocities but  their zonal phases differ. Their effect on the lightcurve is as if there is a feature with zonal wavenumber 2, which is likely responsible for the peak at 2.5 hour in the power spectrum. 


\subsection{Connections to the interior and constraints on the jets}

Large-scale winds are  weak in the metallic interiors of BDs and young EGPs where  MHD effects is important to break large-scale zonal jets, which have been suggested by theoretical arguments and numerical simulations of convection in rotating,  electrically conducting spherical shells (e.g., \citealp{busse2002,liu2008}). Application to typical BDs showing that the typical wind speed are of $\sim0.1-0.3\mps$ in the metallic regions \citep{showman&kaspi2013}. In the molecular envelope overlying the metallic region, the thermal wind balance likely holds well for rapidly rotating BDs and EGPs. Due to the rapid rotation, zonal flows tend to be organized into columns parallel to the rotation axis \citep{busse1976}. As winds are weak at the bottom of the molecular envelope, the characteristic zonal jet speeds near the top of the convective zone depend on the characteristic large-scale horizontal density differences   along the rotation axis. Scaling theories and numerical simulations suggest that this density difference is small in the fully convective region (e.g., \citealp{stevenson1979,showman2011Iscaling}). \cite{showman&kaspi2013} suggest that the typical off-equatorial zonal wind speed near the top of the convective zone is expected to be a few $\mps$ (if the jet width is close to the planetary radius) to a few tens of $\mps$ (if the jet widths are alike those of Jupiter). 
    
On the other hand, the strong off-equatorial zonal jets of Jupiter  likely penetrate to a significant depth in the molecular envelope \citep{kaspi2018}. Recent studies have suggested that the presence of stably stratified layers above the metallic region may explain the deep penetration of the off-equatorial zonal jets (e.g., \citealp{christensen2020,gastine2021}). 
These stratification may be a result of compositional separation of hydrogen and helium. For BDs and hot EGPs, their interior may be too hot for helium-hydrogen separation to occur in their interior \citep{guillot2005,marley2014interiors}.  If the the interior is fully convective (and therefore is closed to be barotropic), theoretically we would expect that off-equatorial jets near the top of convective zone would be at most a few tens of $\mps$.  Other possible mechanisms generating stratification or breaking the thermal wind balance in the interior may potentially alter this expectation. Given that the off-equatorial jet structures in our shallow models are largely barotropic, the above consideration may set a somewhat strong dissipation to the zonal jets in the observable layers of BDs and young EGPs generated by shallow mechanisms. 

The equatorial jet needs special consideration. Near the equator, cylinders concentric with the rotation axis do not intersect the layer with MHD drag, and the equatorial jet may not suffer from the drag \citep{schneider2009}. The latitudinal region of this ``drag-free'' zone depends on the depth over which the MHD drag becomes significant. The depth may be very shallow for BDs and young EGPs as their interiors are hot and thermal ionization could provide significant conductivity even before hydrogen becomes metallic. To crudely illustrate this, we  consider the ionization of potassium alone and estimate effects of the Lorentz drag on the horizontal winds in the kinematic limit as in \cite{perna2010}. Assuming a solar metallicity and a surface magnetic field strength of 1 kG (which may be appropriate for some field BDs, e.g., \citealp{kao2018}), for three models shown in Figure \ref{fig.tp}, the maximum characteristic drag timescales at 100 bars are on the order of $10^{-1}$, $10^{0}$ and $5\times 10^{3}$ s. For a Jupiter-like field strength of 3 G, the drag timescales are about $5\times 10^{3}$, $2\times 10^{5}$ and $6\times 10^{8}$ s. These estimates indicate that, with a magnetic field strength reasonable for these objects, the MHD effects extend significantly to near the surface for  L and T dwarfs.  Assuming the surface is at about 1 bar, the depth at, for example, 100 bars, is on the order of $5\times10^{-4}$ of the radius assuming a Jovian size, leading to the meridional width of the equatorial ``drag-free'' zone of about $\pm 2^{\circ}$ within the equator. Using the width of the drag-free equatorial zone and the argument of absolute vorticity homogenization (which assumes that jet is shallow and cannot violate the  barotropic stability), one may obtain an estimate of the equatorial jet speed \citep{schneider2009}. Applying this argument, we have $U\lesssim\beta L_e^2/2$ where $L_e$ is the distance between the equator and the latitude where the equatorial jet profile drops to zero. For a 5-hour rotating object with a Jovian size, $\beta=1\times10^{-11}\;{\rm m^{-1}s^{-1}}$,  and we obtain $U\approx 24$ to $500\mps$ if assuming $L_e\sim2.2\times10^6$ to $10^7$ m (a depth-to-radius ratio of $5\times10^{-4}$ to $5\times10^{-3}$).  Of course, there is no guarantee that the jet profile will follow the absolute vorticity homogenization configuration---the jet speed can be well below these values if the absolute vorticity is less homogenized or exceed those if the jet violate the  barotropic stability criteria.

Lightcurves in IR and radio wavelengths of a nearby T dwarf revealed evidence of fast eastward traveling features with a speed up to $\sim600\mps$ \citep{allers2020}. Models in \cite{tan2021bd2} showed that zonal waves triggered by cloud radiative feedbacks may explain this net eastward signal. Other possibility includes advection by eastward zonal jets. Using both types of models and careful comparisons with observations, it may be possible to disentangle between the jet and wave mechanisms.

\subsection{Other implications}
Thermal perturbations injected by interactions between convection and the overlying stratified layer provide the minimal circulation in the stratified layers that  easily advect  sub-micron cloud particles  several  scale heights above the condensation level. Cloud decks of sub-micron particles are  expected to be homogeneous near the cloud base and inhomogeneous towards the cloud top. However, the cloud decks of larger particles are vertically thin and significantly inhomogeneous even in the cloud base.   L dwarf atmospheres are characterized by thick clouds composed of mostly   sub-micron particles \citep{marocco2014,hiranaka2016, burningham2017}, while T dwarf atmospheres are characterized by  large particles and vertically thin layers (e.g., \citealp{allard2001,burrows2006,saumon2008,charnay2018}). The  emergence of significant cloud patchiness and thinning has been proposed to explain the rapid L to T transition \citep{knapp2004,marley2010}, and our circulation models and cloud transport provide a natural dynamical mechanism as to why these cloud behaviors would be expected when cloud particles become large.
Chemical disequilibrium has been inferred in many field L and T dwarfs. The needed vertical diffusion coefficient $\Kzz$ of chemistry mixing in 1D models to match the observed spectrum spans a wide range between 1 to $10^4\mmps$ ($10^4$ to $10^8\;{\rm cm^2s^{-1}}$), and most models need the mixing above the convective layers (e.g., \citealp{saumon2006,saumon2007, stephens2009,moses2016, leggett2016b,leggett2017, miles2020}). The lower end of this range of $\Kzz$ seems much smaller than that derived from direct  convective mixing (e.g., \citealp{ackerman2001}), but agrees well to our expectation from large-scale transport in the stratified layers driven by convective perturbations. In the convection resolving models by \cite{freytag2010}, the $\Kzz$ of cloud tracers can reach up to $\sim10^3\mmps$ in the stratified regions by internal gravity waves, providing the high-end mixing in the stratified layers needed to interpret observations.

\subsection{Summary}
Using an updated general circulation model with a grey radiative transfer scheme, we have investigated the formation and evolution of jet streams and tracer mixing in atmospheres  of brown dwarfs and distant young giant planets under globally isotropic  thermal perturbations which are thought to be caused by interactions between convection and the overlying stratified atmosphere \citep{showman2019}.   We have reached the following conclusions: 
\begin{itemize}
	\item Under relatively weak damping conditions, robust zonal  jet streams with speeds up to several hundred $\mps$ are a natural outcome of interactions of waves and turbulence with the planetary rotation.   At a particular atmospheric temperature with a reasonable thermal perturbation rate,  weak bottom frictional drags promote jet formation at all latitudes, while strong drags leave isotropic turbulence at mid-to-high latitudes and jet only at the equator and at altitudes above the drag domain. With a fixed thermal perturbation rate and  weak  drags, low atmospheric temperature (representing weak radiative damping) promotes an overall stronger circulation and zonal jets, while high temperature (strong radiative damping) leads to a much weaker circulation. With both weak frictional and radiative damping, the jet speed increases and the number of jets decreases with increasing thermal perturbation rates. The off-equatorial jets are largely  pressure independent  but the equatorial jets usually exhibit  significant vertical jet shears.  There is negligible systematic meridional temperature variation  associated with the off-equatorial jet structure.
	\item Similar to \cite{showman2019}, our models exhibit long-term oscillations in the equatorial zonal jets, consisting of stacked eastward and westward zonal jets that migrate downward over timescales ranging from tens of days to months. The mechanism responsible for the quasi-periodic oscillations (QPOs) is similar to those in stratospheres of Earth, Jupiter and Saturn. Conditions with moderate to low atmospheric temperatures and strong thermal perturbations favor the formation of QPOs, while high temperature (strong radiative damping) inhibit QPOs.
	\item Circulation driven by the thermal perturbations leads to efficient vertical mixing of tracers including clouds and chemical species in the stratified atmosphere. The mixing is mostly due to local-scale overturning circulation rather than the global-scale mean meridional circulation. Sub-micron cloud particles are easily mixed up to several scale heights above the condensation levels, showing horizontal homogenization near the cloud base but  patchiness toward the cloud top. Larger cloud particles show vertically thin layers with significant patchiness even near the cloud base. Chemical tracers with relatively long chemical timescales ($>10^5$ s) are driven out of equilibrium throughout the atmosphere. If the global-mean tracer transport is characterized by diffusion, the vertical diffusion coefficient, $\Kzz$, derived from the GCM results is on the order of $1$ to $10^2\mmps$ ($10^4$ to $10^6\;{\rm cm^2s^{-1}}$), and its profile differs for different types of tracers. We have derived analytic scaling relations to estimate $\Kzz$ for different types of tracers appropriate for atmospheres of BDs and young isolated EGPs. Our analytic $\Kzz$ profiles qualitatively capture some key features of and on order of magnitude agree well to those derived from GCM results.
\end{itemize}

\section*{Acknowledgements}
This work was initiated when X.T. conducted his PhD with A.P. Showman and is dedicated to Adam P. Showman (1968–2020).  The code generating thermal perturbations was obtained from A.P. Showman.  {\btt We thank constructive comments from the referee.}  X.T.  acknowledges support from the European community through the ERC advanced grant EXOCONDENSE (PI: R.T. Pierrehumbert). 

\section*{Data availability}
The data underlying this article will be shared on reasonable request to the corresponding author.



\bibliographystyle{mnras}
\bibliography{draft} 





\bsp	
\label{lastpage}
\end{document}